\documentclass{aa}  

\usepackage{graphicx}
\usepackage{txfonts}
\usepackage{natbib}
\bibpunct{(}{)}{;}{a}{}{,} 

\begin{document} 

   \title{Six new supermassive black hole mass determinations from adaptive-optics assisted SINFONI observations}

   \author{Sabine Thater\inst{1}  \and Davor Krajnovi\'{c}\inst{1} \and Michele Cappellari\inst{2} \and Timothy A. Davis\inst{3} \and P. Tim de~Zeeuw\inst{4,5} \and Richard~M.~McDermid\inst{6} \and  Marc Sarzi\inst{7,8}
}

   \institute{Leibniz-Institute for Astrophysics Potsdam (AIP), An der Sternwarte 16, D-14482 Potsdam, Germany,\email{sthater@aip.de} \and 
             Sub-Department of Astrophysics, University of Oxford, Denys Wilkinson Building, Keble Road, Oxford OX1 3RH, UK \and 
             School of Physics \& Astronomy, Cardiff University, Queens Buildings, The Parade, Cardiff CF24 3AA, UK              \and
             Sterrewacht Leiden, Leiden University, Postbus 9513, 2300 CA Leiden, The Netherlands \and
             Max Planck Institute for Extraterrestrial Physics (MPE), Giessenbachstrasse 1, D-85748 Garching b. München, Germany \and
             Department of Physics and Astronomy, Macquarie University, Sydney, NSW 2109, Australia \and
             Centre for Astrophysics Research, School of Physics, Astronomy and Mathematics, University of Hertfordshire, College Lane,
Hatfield, Hertfordshire AL10 9AB, UK \and
             Armagh Observatory and Planetarium, College Hill, Armagh, BT61 9DG, UK 
}

   \date{Received ... 2018/ Accepted ?}
 
 \abstract
{Different massive black hole mass - host galaxy scaling relations suggest that the growth of massive black holes is entangled with the evolution of their host galaxies. The number of measured black hole masses is still limited, and additional measurements are necessary to understand the underlying physics of this apparent co-evolution. We add six new black hole mass (M$_{\rm BH}$) measurements of nearby fast rotating early-type galaxies to the known black hole mass sample, namely NGC~584, NGC~2784, NGC~3640, NGC~4570, NGC~4281 and NGC~7049. Our target galaxies have effective velocity dispersions ($\sigma_{\rm e}$) between 170 and 245 km s$^{-1}$, and thus this work provides additional insight into the black hole properties of intermediate-mass early-type galaxies. We combine high-resolution adaptive-optics SINFONI data with large-scale MUSE, VIMOS and SAURON data from ATLAS$^{\textrm{3D}}$ to derive two-dimensional stellar kinematics maps. We then build both Jeans Anisotropic Models and axisymmetric Schwarzschild models to measure the central black hole masses. Our Schwarzschild models provide black hole masses of $(1.3\pm 0.5) \times 10^8 M_{\sun}$ for NGC~584, $(1.0\pm 0.6) \times 10^8 M_{\sun}$ for NGC~2784, $(7.7\pm 5) \times 10^7 M_{\sun}$ for NGC~3640, $(5.4 \pm 0.8) \times 10^8 M_{\sun}$ for NGC~4281, $(6.8\pm 2.0) \times 10^7 M_{\sun}$ for NGC~4570 and $(3.2\pm 0.8) \times 10^8 M_{\sun}$ for NGC~7049 at 3$\sigma$ confidence level, which are consistent with recent M$_{\rm BH}$ - $\sigma_{\rm e}$ scaling relations. NGC~3640 has a velocity dispersion dip and NGC~7049 a constant velocity dispersion in the center, but we can clearly constrain their lower black hole mass limit. We conclude our analysis with a test on NGC~4570 taking into account a variable mass-to-light ratio (M/L) when constructing dynamical models. When considering M/L variations linked mostly to radial changes in the stellar metallicity, we find that the dynamically determined black hole mass from NGC~4570 decreases by 30\%. Further investigations are needed in the future to account for the impact of radial M/L gradients on dynamical modeling.}


   \titlerunning{MBHs in axisymmetric galaxies with SINFONI}
   
   \authorrunning{Thater et al.}

   \maketitle

\section{Introduction}\label{s:intro}

\begin{table*}
   \caption{The sample}
   \label{t:properties}
$$
  \begin{array}{lcccccccccl}
    \hline
    \noalign{\smallskip}

$Galaxy$ &  $Type$ &  $Distance$ & $Linear scale$ & $M$_{\textrm{K}} & $R$_{\textrm{e}}  & \sigma_{\textrm{e}} & \sigma_{0} & \log $(M$_{\textrm{bulge}}$)$ & i & $Large Scale$\\
       &       &  ($Mpc$)    & ($pc arcsec$^{-1})&  ($mag$)  & ($arcsec$)     &  ($km s$^{-1})& ($km s$^{-1})&  \log($M$_{\sun})& (^{\circ}) &\\
    (1)    &  (2)    &   (3)        &    (4)     &   (5)             &   (6)   & (7)      & (8) & (9)& (10) & (11)\\
\hline
$NGC  584$ &   $S0$   &   19.1 \pm 1.0 
& 93  &  -24.19      &  33.0        &  189 \pm 5 & 216 & 10.48 &  51 & $MUSE$\\
$NGC 2784$ &   $S0$   &    9.6 \pm 1.8 & 47 &   -23.31      &  40.2     &  188\pm 8 & 243 & 10.44 & 66 & $VIMOS$\\
$NGC 3640$ &   $E3$   &   26.3 \pm 1.7 &  128 &   -24.60      &  38.5     &  176 \pm 8 & 173 & 11.00 & 68 & $SAURON$ \\
$NGC 4281$ &   $S0$   &   24.4 \pm 2.2 & 118 &   -24.01      &  24.5      &  227 \pm 11 & 314 & 10.88 &  71 & $SAURON$\\
$NGC 4570$ &   $S0$   &   17.1 \pm 1.3 & 83   &   -23.48      &  17.9     &  170 \pm 8 & 209 & 10.18 &  88  & $SAURON$\\
$NGC 7049$ &   $S0$   &   29.9 \pm 2.6  & 145  & -25.00      &  35.4        &  245 \pm 8 & 266  & 11.02 & 42  & $VIMOS$\\
       \noalign{\smallskip}
    \hline
  \end{array}
$$ 
{\textbf{Notes} -- Column 1: Galaxy name. Column 2: Morphological type \citep{deVaucouleurs1991rc3}. NGC 584 had been misclassified in the earlier work and we adopt the classification by \cite{Huang2013} here. Column 3: Distance to the galaxy (taken from \cite{Cappellari2011} for SAURON/ATLAS$^{\textrm{3D}}$ galaxies or NED for VIMOS and MUSE galaxies), the uncertainties were calculated by dividing the NED standard deviations by $\sqrt{N}$ where N is the number of measurements. Column 4: Linear scale derived from the distance.  Column 5: 2MASS total K-band magnitude \citep{Jarrett2000}. Column 6: Effective radius derived from  B-band (CGS) or r-band (ATLAS$^{\rm 3D}$) imaging data. The values were taken from \cite{Ho2011} or \cite{Cappellari2013b} for ATLAS$^{\rm 3D}$ galaxies. Column 7: Effective velocity dispersion derived by co-adding the spectra of the large-scale optical IFU data in elliptical annuli of the size of the effective radius. For ATLAS$^{\rm 3D}$ galaxies taken from \cite{Cappellari2013b}. Column 8: Central velocity dispersion derived by co-adding the spectra of the high-spatial-resolution SINFONI IFU data in elliptical annuli within one arcsec. Column 9: Bulge mass calculated by multiplying the bulge-to-total ratios from \cite{Krajnovic2013a} for ATLAS$^{\textrm{3D}}$ galaxies or \cite{Gao2018} for the remaining galaxies with the total dynamical mass from \cite{Cappellari2013b} or this paper. Column 10: Inclination from \cite{Cappellari2013b} for ATLAS$^{\textrm{3D}}$ galaxies and \cite{Ho2011} for remaining galaxies. For NGC 584 from \cite{Laurikainen2010}.  Column 11: Large scale kinematics data which is used for the Schwarzschild dynamical models. The SAURON data comes from the ATLAS$^{\textrm{3D}}$ galaxy survey.}
\end{table*}

Most massive galaxies harbor a supermassive black hole (SMBH) in their centers. While black holes are invisible by their nature, their mass can be estimated using the motion of dynamical tracers (i.e., stars or gas) in combination with sophisticated dynamical models. The literature contains more than 100 robust dynamical black hole mass determinations, slowly growing into a statistically significant sample. Relating these measured black hole masses (M$_{\rm BH}$) to different host galaxy properties (such as bulge stellar mass, bulge velocity dispersion $\sigma_e$, S\'ersic index n and star formation) 
revealed several noticeably tight correlations, i.e. M$_{\rm BH}$-L \citep{Kormendy1995,Magorrian1998}; M$_{\rm BH}$-$\sigma_e$ \citep{Ferrarese2000,Gebhardt2000}; M$_{\rm BH}$-n \citep{Graham2001}. 
Connecting vastly different scales these relations raise the question whether the growth of the black hole and the evolution of the host galaxy are entangled with each other (see recent reviews by \citealt{Kormendy2013} and \citealt{Graham2016}). Current explanations suggest that black holes grow via two main processes: self-regulation by accretion of gas onto the black hole (facilitated by galaxy merging or accretion of gas) \citep{Silk1998,Fabian1999,DiMatteo2008, Volonteri2010} and by mergers of black holes (following dry major mergers). \cite{Kulier2015} and \cite{Yoo2007} show that accretion is the main channel of black hole growth, but galaxy mergers become relevant for more massive galaxies \citep[see also][]{Graham2012,Graham2013,Krajnovic2018b}. Based on the scaling relations we can see a clear trend that the more massive the galaxy is, the more massive is usually its central black hole. The exact shape of the various scaling relations is however still under debate. While early studies suggested a single-power law \citep{Kormendy2013}, it is nowadays a question whether the fundamental relation between black hole and host galaxy properties scales as double power-law \citep{Graham2013} or has to be described by a three-parameter plane \citep{Bosch2016,Saglia2016}. Moreover, \cite{Krajnovic2018b} and \cite{Mezcua2018} recently reported an up-bending of the scaling relations with higher galaxy mass questioning the existence of one universal scaling relation. The search for a fundamental relation is made even more difficult by an increased internal scatter in both the low- and high-mass regime of the scaling relations. In order to understand and reduce the increased scatter, different observational strategies need to be developed. It is important to understand the different measurement methods with their associated systematic uncertainties by obtaining multiple M$_{\rm BH}$ measurements with different methods for individual galaxies as was done, e.g. in \cite{Walsh2010,Barth2016,Davis2017a,Davis2018,Krajnovic2018}. On the other hand, it is also important to figure out intrinsic scatter due to different galaxy formation scenarios by obtaining more and more homogeneous measurements over the complete SMBH mass range in order to strengthen current theories and ideas. 
\\ \\
Our SMASHING sample (see Section 2 for details) was created to exploit the capabilities of natural and laser guide star adaptive optics (AO) systems at 8m ground-based telescopes. Its purpose is to fill up the scaling relations with additional M$_{\rm BH}$ measurements of early-type galaxies. By the time of the creation of the project in 2009, the black hole mass measurements were almost exclusively populated by Hubble Space Telescope (HST) measurements with the exception of \cite{Nowak2008} and \cite{Krajnovic2009a} who pioneered a new method to measure M$_{\rm BH}$ by using ground-based spectroscopy in combination with AO systems, laser and natural guide stars, respectively. The SMASHING survey was planned to expand the AO method to a large range of early-type galaxies with different velocity dispersions, from the low (100 km/s) to the high ($\approx$ 300 km/s) end. First results, based on observations with NIFS and GEMINI, were published in \cite{Thater2017} and \cite{Krajnovic2018}. Unlike many other M$_{\rm BH}$ measurements in the literature, we used both small (high spatial resolution) and large-field integral field spectroscopic (IFU) data for our measurements. The high-resolution kinematics are crucial to probe the orbital structure in the vicinity of the SMBH (also outside of its sphere of influence) and the large-scale kinematics are needed for constraining the global dynamical M/L, as well as to trace the influence of the stars on radial orbits, which pass close to the SMBH, but spend most of the time at large radii. Including both data sets provides more robust M$_{\rm BH}$ measurements, especially if the sphere of influence is hardly resolved. \\ \\
This paper is the first of a series of papers based on SINFONI observations. We present adaptive optics-assisted near-infrared integral-field spectroscopic data for six fast-rotating axisymmetric early-type galaxies to study the stellar kinematics in the vicinity of their central black hole. We begin by introducing the sample and its selection in Section~\ref{s:sample}, followed by the data acquisition and reduction in Section~\ref{s:observation}. In Section 4, we describe the extraction of the stellar kinematics from the near-infrared SINFONI and optical MUSE and VIMOS integral-field spectroscopic data. In addition to the kinematics, we combine high-resolution HST and Sloan Digital Sky Survey (SDSS) data to model the stellar surface brightness and thus examine the stellar brightness density of our target galaxies. In Section 5, we present the dynamical models which
were constructed using two different and independent methods: Jeans Anisotropic Modeling \citep{Cappellari2008} and \cite{Schwarzschild1979} orbit
superposition modeling. We analyze our assumptions for the dynamical modeling with particular attention to M/L variations and discuss our results in the context of the M$_{\mathrm{BH}}$-$\sigma_{\mathrm{e}}$ relation in Section 6, and finally, conclude in Section 7. 

\section{The Sample}\label{s:sample}
The six galaxies analyzed in this paper belong to our SMASHING galaxy sample to dynamically determine black hole masses in the nearby universe. Three of our target galaxies were selected from the ATLAS$^{\textrm{3D}}$ volume-limited galaxy sample \citep{Cappellari2011}, from which one galaxy had already been observed in the SAURON project \citep{deZeeuw2002}. The three remaining galaxies were observed with the VIMOS or MUSE instruments. Additional high spatial resolution data was obtained with the near-infrared SINFONI instrument to probe the direct vicinity of the SMBH. Based on their velocity dispersion, the sample galaxies are expected to be located in the intermediate $M_{\rm BH}$ range. 
The main properties of our six sample galaxies are summarized in Table~\ref{t:properties}. \\
Our target galaxies were selected based on a number of requirements for a successful $M_{\rm BH}$ determination. An important criterion for a robust black hole mass determination is the need to resolve the sphere-of-influence (SoI) of the black hole within which the SMBH dominates the galaxy potential. The SoI depends on the mass of the black hole $M_{BH}$ and the velocity dispersion of the galaxy within an effective radius $\sigma_e$ and is defined as $r_{\textrm{SoI}}=G M_{\rm BH}/\sigma_{e}^2$ where G is the gravitational constant. We calculated an estimated value for $r_{\textrm{SoI}}$ using black hole masses based on the $M_{\rm BH}-\sigma_{\rm e}$ relation from \cite{Tremaine2002}\footnote{The data acquisition process for this project started in 2008. In that time \cite{Tremaine2002} was one of the best representations of the black hole - host galaxy scaling relations. \cite{Tremaine2002} is very similar to the scaling relations that we show in Fig.~10. The selection based on the scaling relation by \cite{Tremaine2002} was only to select galaxies that were most likely to provide robust $M_{\mathrm{BH}}$ estimates. However, the required observing time and obtaining useful data in the near-infrared with LGS AO trimmed the sample more significantly than any scaling relation.} and the ATLAS$^{\textrm{3D}}$ velocity dispersions from \cite{Cappellari2013}. Using the large set of information from both large-scale and high-resolution IFUs, we can probe sphere-of-influences which are 2-3 times lower than the spatial resolution \citep{Krajnovic2009a,Thater2017}. With the goal to gain the best possible resolution, we utilized the AO mode from the SINFONI instrument using a natural guide star (if possible) or a laser guide star to correct for unstable seeing conditions. 
\\
Furthermore, archival HST imaging was needed for the galaxies of our sample to build detailed light models of the galaxy's centers. We also ensured that the selected galaxies would not include any obvious bars or merger features indicating a non-relaxed galactic potential which would make the galaxies unsuitable for dynamical modeling with static potential models, as used here.

\section{OBSERVATIONS}\label{s:observation}
The mass measurement of massive black holes requires a large variety of data sets. Both high spatial resolution kinematic information of the central galaxy region to constrain the wide range of different stellar orbit families and large-scale IFU data to constrain the global galaxy characteristics are essential for a precise measurement. The IFU data is complemented by imaging data from HST and ground-based telescopes to construct a detailed mass model of the host galaxy. In the following section, we present the different observations from the IFUs towards the imaging data.

\begin{table*}
\caption{Details of the SINFONI observing runs}
\centering
\begin{tabular}{lccccccl}
\hline\hline
Galaxy   &  Date                          &  PID          & Pixel scale & N & N& $T_{exp}$ & AO mode\\
         &                                &               & (mas)       & of exp. & comb. exp. & (h)    &  \\
   (1)   &   (2)                          &  (3)          & (4)  & (5) & (6) & (7)    & (8)  \\
\hline
NGC 584  & 2007 Jul 23,24                 & 079.B-0402(A) & 100  &  3 &  3  &     0.75      &  NGS        \\
NGC 2784 & 2007 Dec 12,29, 2008 Jan 01,02 & 080.B-0015(A) & 100  & 9 & 9 &     4.5         & NGS          \\
NGC 3640 & 2010 Apr 08,09                 & 085.B-0221(A) & 100  & 13 &  12  &     3.16       & LGS          \\
         & 2013 May 08,09, May 09         & 085.B-0221(A) & 100  & 8 & 7 &  3.16         &  LGS         \\
         & 2013 Dec 27, 2014 Jan 07,24    & 291.B-5019(A) & 100  & 14 & 12   &   3.16        & LGS         \\
NGC 4281 & 2010 Sep 04                    & 085.B-0221(A) & 100 & 4&  3    &    2.25       &  LGS         \\
         & 2012 Mar 20                    & 085.B-0221(A) & 100  & 4&  4   &    2.25      &  LGS        \\
         & 2013 May 09,11                 & 091.B-0129(A) & 100 & 19 &   16   &     2.25      &  LGS         \\ 
NGC 4570 &  2013 May 07                   & 091.B-0129(A) & 100 & 16  &  15  &     2      &  LGS         \\
NGC 7049 &  2005 Jun 08,09,10,14,19       & 075.B-0495(A) & 100  & 20 &  16 &    6.25       &  NGS          \\
         &  2005 Jun 27, Jul 02,03,07     & 075.B-0495(A) & 100 &  12 &  9    &   6.25        &  NGS         \\
\hline
\\
\end{tabular}
\\
\tablefoot{Column 1: Galaxy name. Column 2: Dates of the observations. Column 3: Identification number of the Proposal. Column 4: Spatial pixel scaling of the observation. Column 5: Number of available single exposure frames. Column 6: Number of single exposure frames in the combined data cube. Column 7: Combined exposure time in hours. Column 8: Adaptive optics mode applied for the data, either using a natural guide star (NGS) or a laser guide star (LGS).  } 
\label{t:observ}
\end{table*}

\subsection{SINFONI IFU data}
The near-infrared portion of our IFU data was obtained between 2005 and 2013 with the Spectrograph for INtegral Field Observations in the Near Infrared (SINFONI) instrument mounted on UT4 (Yepun) of the Very Large Telescope (VLT) at Cerro Paranal, Chile. SINFONI consists of the Spectrometer for Infrared Faint Field Imaging (SPIFFI) assisted by the adaptive optics (AO) module, Multi-Application Curvature Adaptive Optics (MACAO) \citep{Eisenhauer2003,Bonnet2004}. We observed each galaxy at K-band grating (1.94 - 2.45 $\mu$m) providing a spectral resolution of $R \sim 4000$ and a pixel scaling of 100 mas leading to a total field-of-view (FOV) on the sky of about $3.2 \times 3.2 \arcsec$ per pointing. 
Details of the observing runs for each galaxy are provided in Table~\ref{t:observ}.
For each of our observations, we made use of the AO mode, either using a natural guide star (NGS) or an artificial sodium laser guide star (LGS) in order to correct for the ground-layer turbulence and achieve the highest spatial resolution possible. In the ideal case, the LGS mode still requires a natural guide star to correct for the tip-tilt disturbances in the wavefront, which are not sensed by the LGS. However, we often did not have a suitable tip-tilt star close to the galaxy and tip-tilt on the nucleus was not always possible, such that we applied the SINFONI Seeing Enhancer mode which provided a slight improvement to the natural seeing. Our observations show typical Strehl ratios of about 10 \% (see Table~\ref{t:spatialres}). The observations were performed using the object-sky-object nodding scheme. At the beginning and end of each observing block, the respective standard star was observed at a similar airmass and with the same optical setup in order to correct for the telluric features at similar atmospheric and instrumental conditions. 
We used the SINFONI reduction pipeline to reduce the data and reconstruct the data cubes of the individual observations. This science frame contains spatial information in the X and Y directions and spectral information in the Z direction. 
As the data reduction was quite extensive, we mention a number of steps individually in the next subsections.

\subsubsection{Data reduction and sky correction}
The data reduction followed mostly the steps that are described in the SINFONI instrument handbook. The observations were reduced using the ESO SINFONI pipeline
\citep[version 2.4.8, ][]{Modigliani2007} in combination with additional external corrections to optimize the resulting data cubes. The ESO pipeline includes the bias-correction, dark-subtraction, flat-fielding, non-linearity correction, distortion correction and wavelength calibration (using a Neon arc lamp frame) for each observation of target and standard star. The nearest sky exposure was used to remove the night-sky OH airglow emission using the method described by \cite{Davies2008}. In the last step of the data reduction, each observation was reconstructed into a three-dimensional data cube.

\subsubsection{Telluric and heliocentric velocity correction}
A significant part of the data correction in the near-infrared regime is the correction for telluric absorption which originates in the Earth's atmosphere (mainly ozone, gaseous oxygen, and water vapor).
Telluric absorption lines are exceptionally deep at the blue end of the K-band and may vary over the time of the observation.
Therefore, it is necessary to correct each science frame individually.
Standard stars with known spectra are typically used to remove these atmospheric absorption features from science cubes.\\
For the telluric correction of the near-infrared spectra, we wrote a Python script to apply the same method as described in \citet{Krajnovic2009a}. In most of the observation nights, two telluric stars were observed which gave us the opportunity to choose the telluric stars with a similar airmass to our science target. The telluric stars were either solar-like G0-2V stars or hotter B2-5V stars in an unsystematic order. 
We used the Python version 6.06 of the penalized Pixel fitting software\footnote{\url{http://purl.org/cappellari/software}} \citep[pPXF, ][]{Cappellari2004} as upgraded in \cite{Cappellari2017} to fit a stellar template showing the characteristic features of the telluric star. For the solar-like G-type stars, we use a high-resolution solar template \citep{Livingston1991}\footnote{\url{ftp://nsokp.nso.edu/pub/atlas/photatl/}} and in the case of spectrally almost feature-less B-type stars we fitted a blackbody spectrum.\\
The telluric absorption corrected spectra were then corrected for the Doppler shifts due to the motion of the earth around the sun, commonly known as heliocentric correction. As some of our targets were observed at different times of the year, the velocity shifts in the observed spectra could be between 10-40 km/s, which is of the order of the LSF. We used our own python routine to correct the wavelength into the heliocentric frame of reference. The corrected wavelength is defined by $\lambda_{\textrm{corrected}}=(1+v_{\textrm{helio}}/c) \times \lambda_{\textrm{uncorrected}}$ where c is the speed of light and $v_{\textrm{helio}}$ is the projected heliocentric velocity which was calculated from the ESO pipeline for each data frame.  The heliocentric correction was necessary for NGC 3640, NGC 4281 and NGC 7049, as in these cases the different observing blocks were spread widely throughout the year. We had to apply the heliocentric correction to each spaxel of each of our science frames individually.

\subsubsection{Merging of the data cube}
The individual frames of the observations from the different observing blocks were then combined spatially using the position of the center of the galaxy as reference. This position changed for each of our science frames as a dithering pattern of a few pixels was applied for each subsequent observation to ensure that the galaxy would not always fall onto the same pixels of the detector and thus adding systematic uncertainties. It was possible to identify the center of the galaxy with an accuracy of about one pixel (=0.05$\arcsec$) by comparing the isophotes of the reconstructed images and re-center them. In this step, we also excluded science frames with a bad PSF. Bad PSFs can be the result of poor seeing or an insufficient AO correction. In Table~\ref{t:observ} we specify how many science frames were excluded for each galaxy.
After the re-centering, we applied a sigma-clipping pixel reject algorithm to align the single science frames and created the final data cubes as in \cite{Krajnovic2009a} and \cite{Thater2017}. The algorithm defines a new square pixel grid and interpolates the science frame to this grid. Flux values of the final data cubes were calculated as median flux values of the single data frames.
Finally, we obtained $3 \times 3 \arcsec$ data cubes with 0.05$\arcsec$ pixel scale for the 100 mas SINFONI observations.

\subsubsection{Correction of line-spread function inhomogeneities}
\begin{figure}
  \centering
    \includegraphics[width=0.5\textwidth]{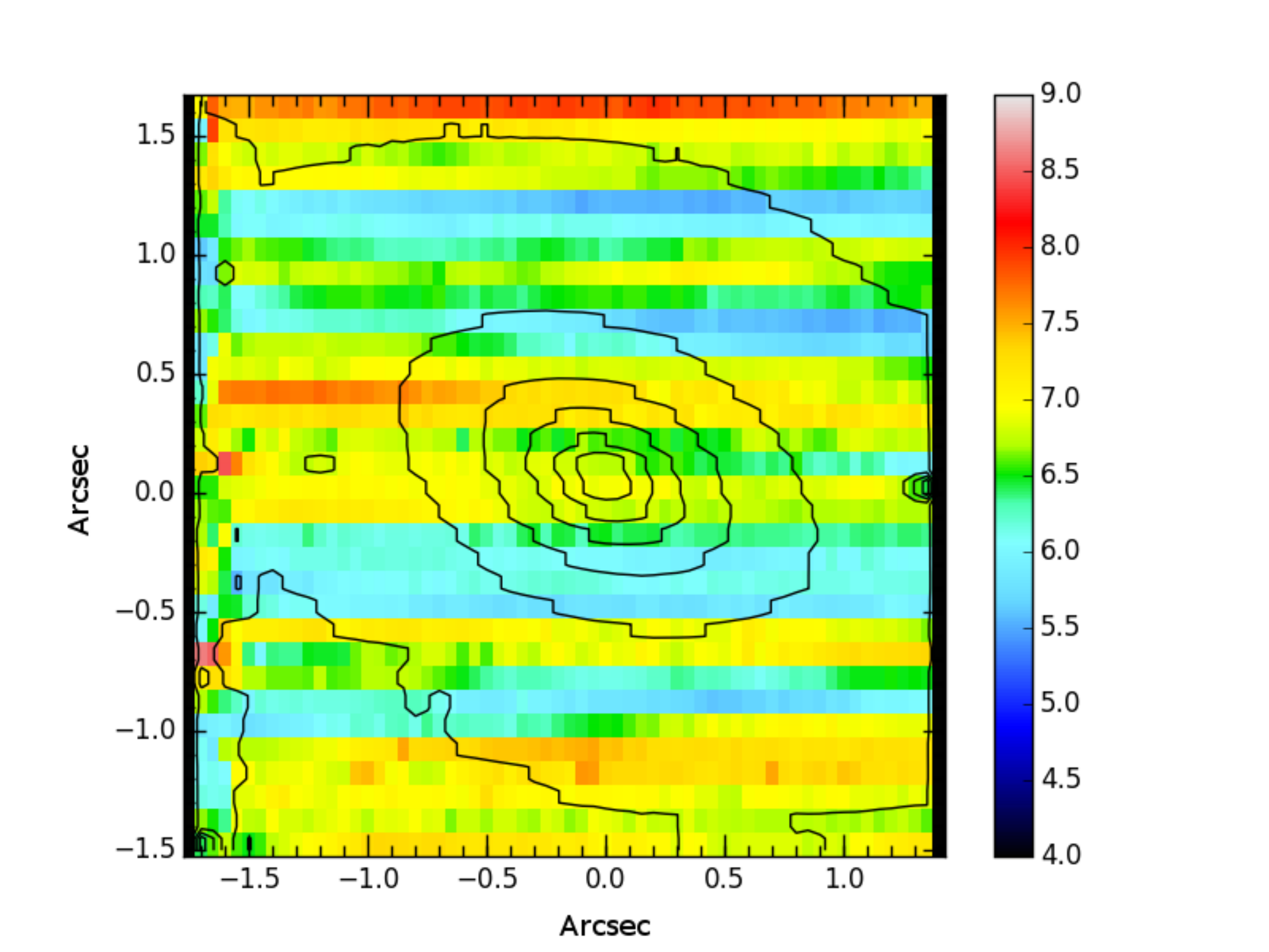}
      \caption{Example of the spectral resolution inhomogeneity across the SINFONI detector. The spectral resolution for each spaxel was derived from arc line observations of NGC 584. The spectral resolution varies significantly in the vertical direction of the detector with values ranging from  5.5~\AA\, to 7.7~\AA\, FWHM.}
      \label{ff:lfsgradient}
\end{figure}

In order to compare the spectra of the IFU with template spectra (which is needed for the extraction of the stellar kinematics, see Section~\ref{ss:ppxf}), it is necessary to quantify the intrinsic dispersion of the SINFONI instrument. Therefore, we determined the spectral resolution of the SINFONI data from  strong arc lines.
While attempting to determine the spectral resolution of the SINFONI data, we encountered a problem: The spectral resolution over the full $64\times 64$ spaxel FoV was very inhomogeneous (see Figure~\ref{ff:lfsgradient}) which was also recognized by \cite{Nguyen2018} and \cite{Voggel2018}. 
In order to better characterize the shape and inhomogeneity of the LSF for the merged cubes, we applied the same data reduction routines to the respective arc lamp 
(except for the sky subtraction). We then built an arc line data cube by combining the reduced arc lamp frames. We used the same dither pattern as for the science frames to ensure that the arc line cube would fully resemble the data cube of the science object. From the combined arc line cube we then measured the LSF using six isolated, strong arc lamp lines for each spaxel. This LSF cube was later used when fitting each spaxel with a stellar template (details in Section~\ref{ss:ppxf}).
The spectral resolution across the field-of-view has a median value of $6.8$~\AA\, FWHM ($\lambda/\Delta \lambda$ = 3820) with values ranging from 5.5~\AA\, to 7.7~\AA\, FWHM.

\subsubsection{Voronoi binning}
The last step before determining the stellar kinematics was to ensure a sufficient and spatially uniform signal-to-noise (S/N) by spatially binning the final SINFONI data cubes with the adaptive Voronoi binning method\footnote{See footnote 2} \citep{Cappellari2003}, Python version 3.1.0. In this method, based on the initial S/N estimate, spaxels are co-added while keeping the geometrical constraint of nearly round bins. An initially approximated noise estimate was obtained by median smoothing each spectrum with a kernel of 30 pixels width and calculating the standard deviation of the difference between the smoothed and the original spectrum. This initial estimate was then passed on as input S/N to the Voronoi binning script. The input S/N was systematically chosen between 50 and 70 balancing the desire to keep the central spaxels (if possible) unbinned to ensure a sufficiently high resolution in the center while increasing the quality of the outward spectra for the extraction of the kinematics. In our final binning scheme, we establish typical bin sizes of < 0.1$\arcsec$ in the center, while 0.3-0.4$\arcsec$ diameter for bins at a radius larger than 1$\arcsec$.

\subsubsection{SINFONI spatial resolution}
The quality of our black hole mass measurements is indicated by the spatial resolution of the AO-corrected SINFONI data. We determined the spatial resolution by convolving high-resolution HST images with a double Gaussian model PSF and compared it to the collapsed image of the SINFONI IFU data. A detailed description of how we derived the spatial resolution is given in Appendix~\ref{ss:hst_psf}. The resulting parameters are given in Table~\ref{t:spatialres}.

\begin{table}
\caption{SINFONI spatial resolution}
\centering
\begin{tabular}{lcccc}
\hline\hline
Galaxy  & FWHM$_{N}$ & FWHM$_{B}$ & f$_{N}$ & Strehl\\
        & (arcsec)    &  (arcsec) &         & \\
(1)   & (2) & (3) & (4) & (5)    \\
\hline
NGC 584 & 0.20 $\pm$ 0.02 & 0.74 & 0.54 & 13 \%\\
NGC 2784 & 0.21 $\pm$ 0.02 & 0.50 & 0.74 & 11 \%\\
NGC 3640 & 0.19 $\pm$ 0.02 & 0.56  &  0.41 & 14 \%\\
NGC 4281 & 0.22 $\pm$ 0.04 & 0.90 & 0.86 & 10 \%\\
NGC 4570 & 0.18 $\pm$ 0.02 & 0.58 & 0.47 & 15 \%\\
NGC 7049 & 0.20 $\pm$ 0.03  & 0.61 & 0.67 & 13 \%\\
\hline
\\
\end{tabular}
\\
\tablefoot{The SINFONI PSF of the data was parametrized by a double Gaussian with a narrow and broad component. The parameters are given in the following columns. Column 1: Galaxy name. Column 2: FWHM of the narrow Gaussian component. Column 3: FWHM of the broad Gaussian component. Column 4: Relative intensity of the narrow component. Column 5: Strehl ratio of the data. } 
\label{t:spatialres}
\end{table}

\subsection{Large-field data}
\subsubsection{MUSE IFU data}\label{ss:muse}
The VLT/MUSE \citep{Bacon2010} data of NGC 584 was taken on July 1st, 2016 under the science program 097.A-0366(B) (PI: Hamer). They obtained a total exposure time of 2700s divided into three $900$s on-source integrations each yielding a field of view of $60 \times 60$ \arcsec ($\approx$ two effective radii of NGC 584). In addition, there was an off-source exposure of a blank field which can be used to estimate the sky. The fields were oriented such as to map the galaxy along the major axis with a large overlap, as every frame contained the nucleus of the galaxy. We reduced the data using the MUSE data reduction pipeline \citep{Weilbacher2015}, version 1.6. The reduction followed the standard steps, first producing master calibration files (bias, flat and skyflat), the trace tables, the wavelength solution and the line-spread function for each slice. Each on-target observation was reduced using these calibration files and closest in time illumination flats to account for temperature variations. In addition, a separate sky field and a standard star were reduced in the same way. From these, we extracted a sky spectrum and its continuum, as well as the flux response curve and an estimate of the telluric correction. The sky spectrum was applied to all three on-target frames, where we let the pipeline model the sky lines based on the input sky spectrum and the line-spread function. As all on-source frames contained the nucleus, we recorded its relative positions between the frames and applied the offsets with respect to the first one, prior to merging with MUSE pipeline merging procedure. In the final cube each pixel has the size of 0.2\arcsec$\times0.2$\arcsec\, and a spectral sampling of 1.25 \AA\, per pixel. For our MBH determination we only needed the high S/N central $30\arcsec\times 30\arcsec$ of the MUSE data cube and cut this region out. We then Voronoi-binned the cutted central region to a target S/N of 60 resulting into bin diameter sizes of $0.5\arcsec$ in the center and $3\arcsec$ at radii larger than 12\arcsec.

\subsubsection{VIMOS IFU data}\label{ss:vimos}
The large-field data for NGC~2784 and NGC~7049 were obtained between October 2006 and August 2007 using the VIsible Multi Object Spectrograph \citep[VIMOS, ][]{LeFevre2003} mounted on UT3 Melipal under the science programs 078.B-0464(B) and 079.B-0402 (B) (PI: Cappellari). 
\\
The VIMOS data reduction was performed by \cite{Lagerholm2012} making use of the ESO pipeline\footnote{\url{http://www.eso.org/sci/software/pipelines/}} (version 2.3.3) and some IRAF tasks. It includes bias and sky subtraction, flatfield calibration, interpolation over bad pixels, cosmic-ray removal, spatial rectification, wavelength with HeArNe lamp exposures, flux calibration with standard stars and fringe-like pattern removal. As described in \cite{Lagerholm2012}, they also corrected the fringe-like pattern in the spectral and the intensity variations in the imaging domain which were dominating the raw data.
After the data reduction, they merged the individual science frames into final data cubes. In the same manner as for the SINFONI and MUSE data, we also Voronoi-binned the VIMOS data to a target S/N of 60, obtaining bin sizes of 0.5\arcsec in the galaxy center and 2-3\arcsec at radii larger than 7.5\arcsec.

\subsubsection{SAURON IFU data}
NGC 3640, NGC 4281 and NGC 4570 are part of the ATLAS$^{\textrm{3D}}$ galaxy survey \citep{Cappellari2011}. 
The observations were obtained with the Spectrographic Areal Unit for Research on optical Nebulae IFU \citep[SAURON,][]{bacon2001} at the 4.2-m William Herschel Telescope of the observatorio del Roque de los Muchachos on La Palma and reduced with the XSAURON software \citep{bacon2001} . The SAURON IFU has a FOV of $33\arcsec \times 41\arcsec$ with a sampling of $0.94\arcsec \times 0.94\arcsec$ square pixels, covering about 1-2 effective radii of our target galaxies.
A detailed description of the stellar kinematics extraction of the ATLAS$^{\textrm{3D}}$ sample is given in \cite{Cappellari2011}. 
The resulting velocity maps of NGC 3640, NGC 4281 and NGC 4570 were already presented in \cite{Krajnovic2011}, and we show the full kinematic set of these galaxies in Figure \ref{ff:atlas3d_kinematics}. In addition, NGC 4570 is part of the SAURON survey \citep{deZeeuw2002} and presented in \cite{Emsellem2004}. In this paper, we use the homogeneously reduced publicly available ATLAS$^{\textrm{3D}}$ data\footnote{\url{http://purl.org/atlas3d}} which was binned to a target S/N of 40.

\subsection{Imaging data}

For the high resolution central imaging of our target galaxies, we downloaded HST archival data. We obtained either Wide-Field Planetary Camera \citep[WFPC2, ][]{Holtzman1995} or Advanced Camera for Survey (ACS, Ford et al. 1998) data from the ESA Hubble Science Archive, which generates automatically reduced
and calibrated data. Cosmic rays were removed by taking the median of the aligned single CR-SPLIT images. Due to an unsuccessful sky subtraction in the archival data, 
the ACS image was reprocessed by applying the drizzlePac \footnote{\url{http://www.stsci.edu/hst/HST\_overview/drizzlepac}}
package of the Astroconda distribution. For the large field of view imaging of our targets of the southern hemisphere, NGC 584, NGC 2784, NGC 7049, we used images of the Carnegie Irvine Galaxy Survey Project \citep{Ho2011,Li2011,Huang2013}. For the other three targets we used SDSS DR7 r-band images \citep{Abazajian2009} which we received from the ATLAS$^{\textrm{3D}}$ collaboration \citep{Scott2013}.

\begin{table}
\caption{HST archival data}
\centering
\begin{tabular}{lccc}
\hline\hline
Galaxy &PID & Instrument & Filter\\
(1)  & (2) & (3) & (4)\\
\hline
NGC 584 & 6099 & WFPC2 & F555W\\
NGC 2784 & 8591 & WFPC2 & F547M\\
NGC 3640 & 6587 & WFPC2 & F555W \\
NGC 4281 & 5446 & WFPC2 & F606W \\
NGC 4570 & 6107 & WFPC2 & F555W \\
NGC 7049 & 9427 & ACS & F814W \\

\hline
\\
\end{tabular}
\\
\tablefoot{Column 1: Galaxy name. Column 2: programme identification number. Column 3 and 4: Camera on HST and the filters in which the data were taken.}
\label{t:hstbands}
\end{table}

\section{STELLAR KINEMATICS}
\begin{figure*}
  \centering
    \includegraphics[width=0.9\textwidth]{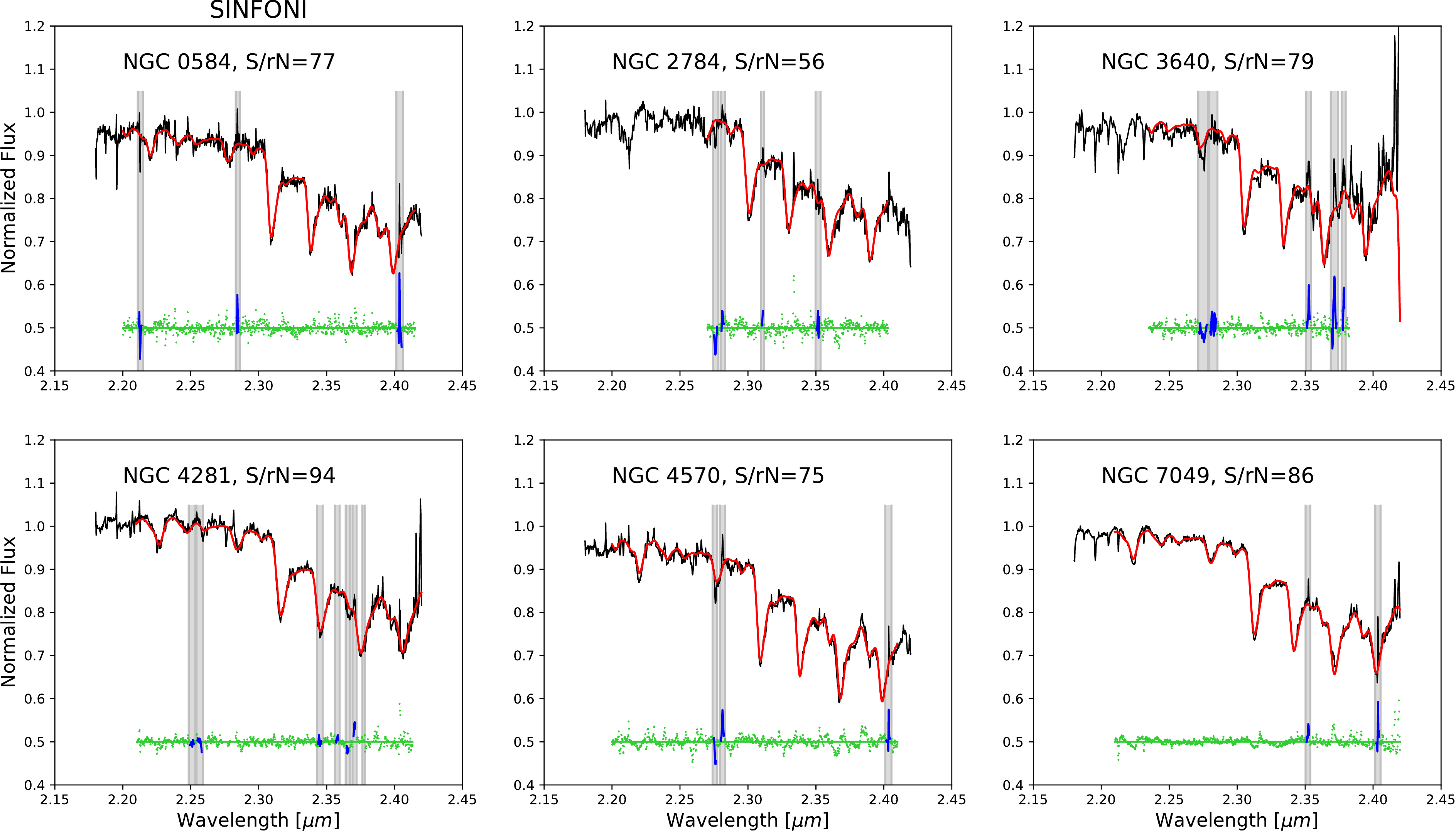}
    
    \vspace{0.5cm}
    
    \includegraphics[width=0.9\textwidth]{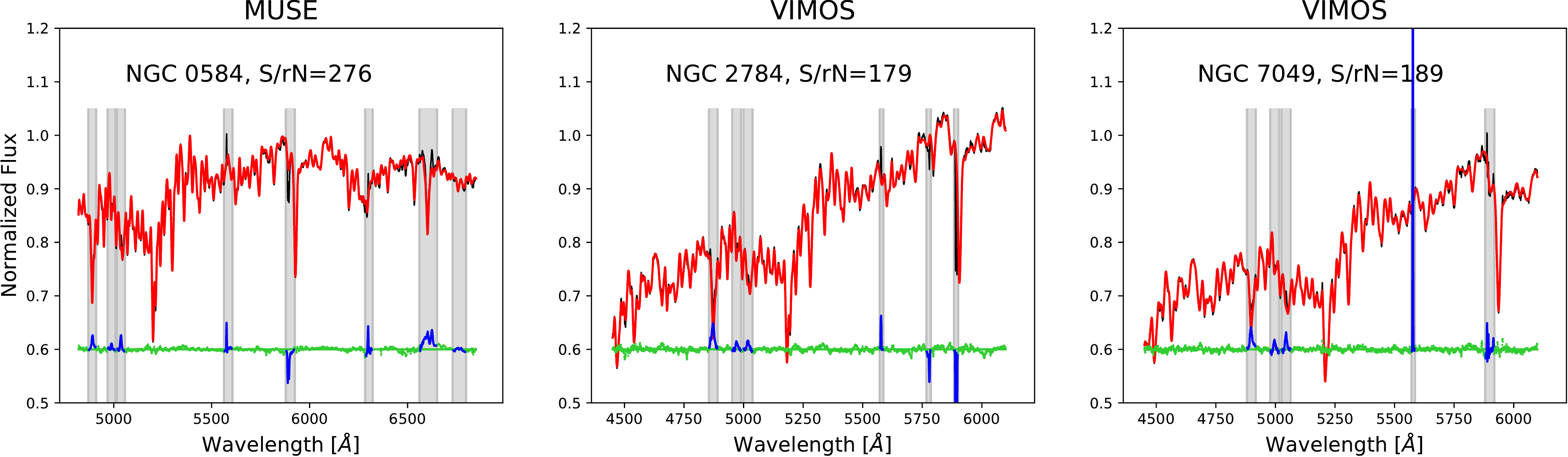}
      \caption{Integrated SINFONI, MUSE and VIMOS spectra and pPXF fits of our target galaxies. The integrated spectra (black solid lines) were obtained by summing up all spectra of the IFU data cubes and fitted using the pPXF routine (red lines) in order to derive an optimal template. The fitting residual between spectrum and best fitting model are shown as green dots and are shifted up by 0.5 (0.6 for the bottom panels). Regions which were masked in the fit (often due to emission lines or insufficient sky subtraction) are indicated as grey shaded regions and their residuals  are indicated in blue.}
      \label{ff:ppxf_overview}
\end{figure*}
\subsection{Method}\label{ss:ppxf}

\begin{figure*}
  \centering
    \includegraphics[width=0.88\textwidth]{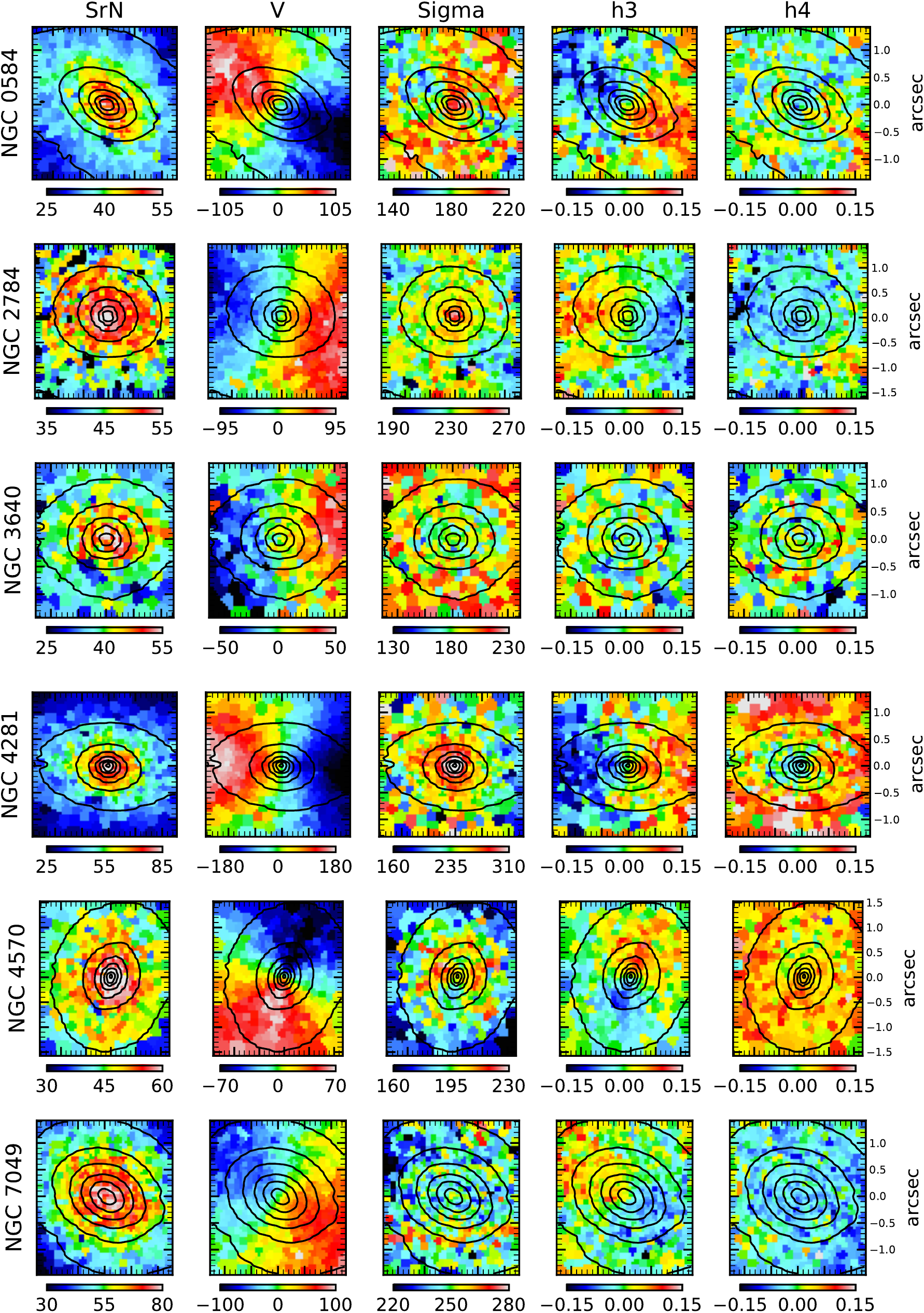}
      \caption{SINFONI stellar kinematics (derived from CO bandhead spectroscopy) of our target galaxies (from top to bottom) NGC 584, NGC 2784, NGC 3640, NGC 4281, NGC 4570 and NGC 7049. From left to right the panels show maps of signal-to-residual noise (S/N), mean velocity (V), velocity dispersion ($\sigma$) and the Gauss-Hermite moments $h_{3}$ and $h_4$. The black contours indicate the galaxy surface brightness from the collapsed data cube. North is up and east to the left.
      }
      \label{ff:sinfoni_kinematics}
\end{figure*}

\begin{figure}
  \centering
    \includegraphics[width=0.45\textwidth]{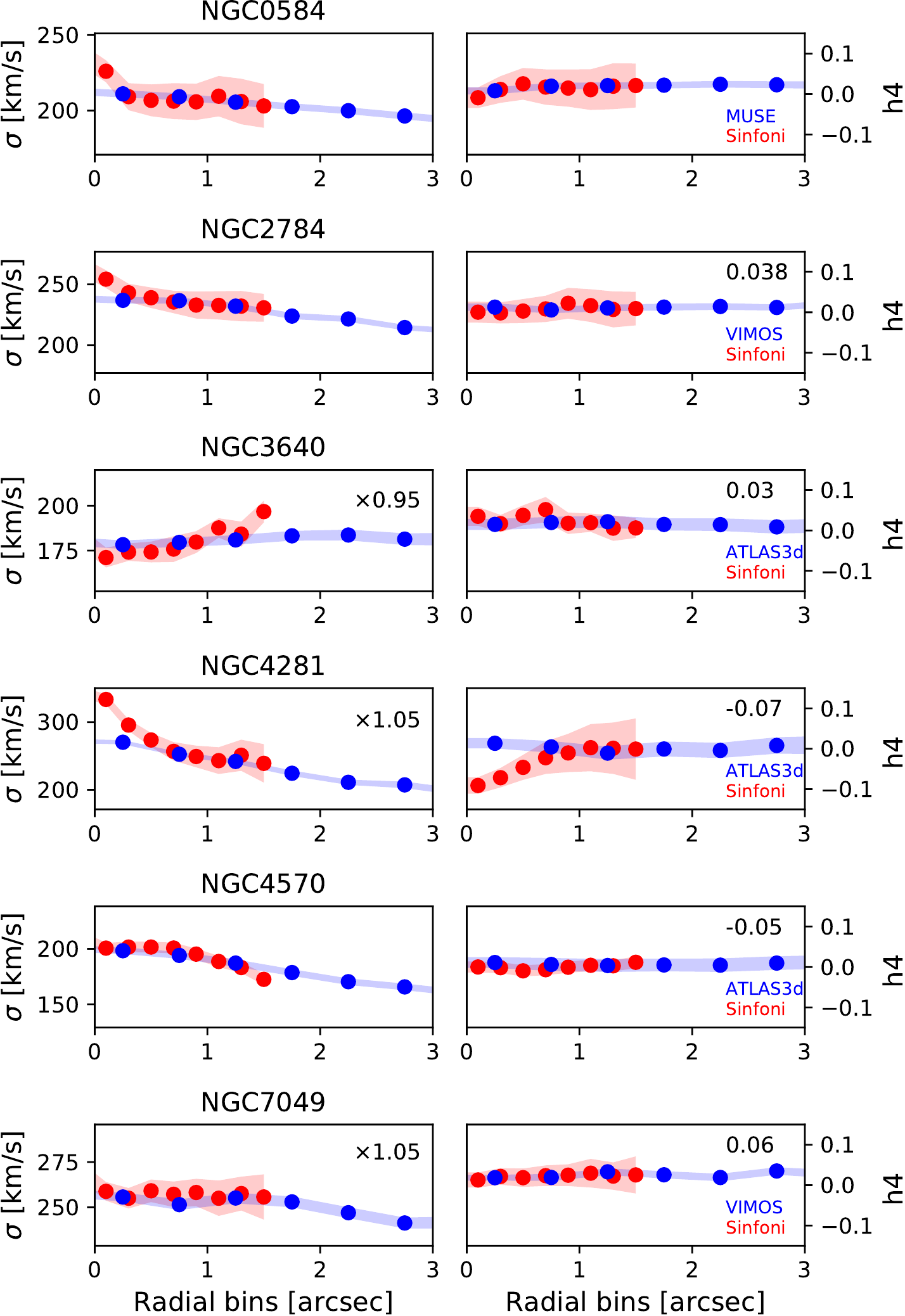}
      \caption{Comparison of velocity dispersion and $h_4$ profiles for the SINFONI (red) and the respective large scale data (blue). The values were averaged within circular annuli around the kinematic centre. The error range of the averaged values in the radial bins are calculated via error propagation and are shown as shaded regions. Applied shifts in the SINFONI maps are marked by the values in the upper right corner of each panel.}
      \label{ff:kin_comp}
\end{figure}

For each instrument, we independently measured the light-weighted stellar kinematics from the galaxy absorption line spectra using the Python implementation of the penalized Pixel Fitting method~\footnote{See footnote 2} \citep[pPXF, ][]{Cappellari2004,Cappellari2017}. pPXF fits the galaxy spectrum by convolving a stellar spectrum template with the corresponding stellar line-of-sight velocity distribution (LOSVD), which is parametrized by Gauss-Hermite polynomials \citep{Gerhard1993,vanderMarel1993}. In detail, the LOSVD is then specified by the mean velocity V, the velocity dispersion $\sigma$ and two additional quantities to describe asymmetric (h$_3$) and symmetric (h$_4$) deviations from a simple Gaussian. As the higher Gauss-Hermite polynomials are strongly coupled to the simple Gaussian moments, their relative weights are controlled by the so-called BIAS parameter which is dependent on the S/N of the data \citep{Cappellari2004,Emsellem2004}. For low S/N data, the BIAS parameter prevents spurious solutions by biasing the derived LOSVD towards a simple Gaussian. \\
We analogously derived a second set of kinematics for each of our sample galaxies where we parametrized the LOSVD with the first two moments (V, $\sigma$) only. In this case, the BIAS keyword is not used by the code. This set of kinematics was needed to construct the Jeans Anisotropic Models (Section \ref{ss:jam}) which only take into account the lower-order moments of the LOSVD.
\\
\\
The usage of pPXF is twofold in order to minimize statistical variations across the field and reduce the computational expense. The first step is the creation of an optimal template by running pPXF on the global galaxy spectrum. The optimal template is a non-negative linear combination of the stellar library and consisted of typically 2-5 stars for the SINFONI data and about 30 stars for the large-scale data. Depending on the spectral range of the observed data, we used either MILES \citep{Sanchez-Blazquez2006}, Indo-US \citep{Valdes2004} optical or Gemini Spectral Library of Near-IR Late-Type \citep{Winge2009} stellar template library spectra, which are further described in the following two subsections. The optimal template is then used to fit the spectra from each Voronoi bin using $\chi^2$ minimization. While running pPXF on our spectra, we also added an additive polynomial to account for the underlying continuum. Furthermore, emission lines and regions of bad sky subtraction were masked during the fit. We then compared the fitted spectrum with the original spectrum for each bin. The standard deviation of the residuals (i.e., shown as green points in Fig.~\ref{ff:ppxf_overview}) was used to derive a final signal-to-residual-noise (S/rN) which measures both the quality of the data and the precision of the fit.
\\ \\
The errors of the recovered kinematics were derived with Monte Carlo simulations, the standard approach for LOSVD extractions. The complete measurement process is repeated for a large number of data realizations (500) where each realization is the original spectrum perturbed by adding noise in the order of the standard deviation of the pPXF residuals. Applying pPXF (with BIAS parameter set to zero) on each realization (with the same optimal template) provides 500 measurements of the LOSVD. The final error of each bin is then the standard deviation of the LOSVD distributions of these 500 realizations. The kinematic errors are spatially anti-correlated with the S/rN distribution, low in the center and larger in the outer regions. Mean errors are shown in Figure~\ref{ff:kin_comp}, where we compare the large and small-scale kinematics with each other. It is at first glance visible that the large-scale kinematics have much smaller errors than the SINFONI data (velocity: $\approx 2.5-5$~km\,s$^{-1}$ versus $\approx 5-10$~km s$^{-1}$ and velocity dispersion: $2.5 - 6$~km s$^{-1}$ versus $6 - 12$~km\,s$^{-1}$).  

\subsection{SINFONI specifics}\label{ss:sinfonikinematics}
The SINFONI spectrograph in combination with adaptive optics provides spatially highly resolved spectra in the near-infrared regime yielding significant information about the motion of the stars surrounding the central black hole due its dust-transmissivity and its high resolution. A significant feature in the near-infrared is the CO absorption band head at about $2.3\,\mu$m which can be used to gain robust measurements of the LOSVD. We used the stellar template library by \cite{Winge2009} which consists of 23 late-type stars observed with the Gemini Near-Infrared Spectrograph (GNIRS) and 31 late-type stars observed with the Gemini Near-Infrared Integral Field Spectrometer (NIFS) to fit the SINFONI spectra in the range of 2.29 to $2.41\,\mu$m. Excluded from the fit were emission lines and incompletely reduced sky-lines which especially contaminated the third and fourth absorption line of the bandhead. Furthermore, to mitigate template mismatch effects in our kinematics extraction, we tested including both GNIRS and NIFS template stars as well as the restriction to only one instrument's template stars. While all three attempts gave generally consistent results, the NIFS template stars could not always reproduce the Calcium line (at $\sim$ 2.25 $\mu$m) very well. This slight template mismatch often led to systematically lower velocity dispersions (but within the statistical errors). During the fitting procedure, we carefully examined and compared all three template library combinations and always chose the one that gave the best fit to the SINFONI spectra. \\ \\
In order to recover reliable LOSVD measurements, we had to ensure that both the stellar templates and the SINFONI observations had a comparable spectral resolution before the fitting procedure. As the NIFS and GNIRS stellar spectra are provided at a better resolution than the SINFONI galaxy spectra, we had to degrade the template spectra to the same resolution as the SINFONI observations. Therefore, we convolved the template spectra ($\sigma_{\textrm{temp}} \approx 2.9 - 3.2$~\AA) with a Gaussian having the dispersion of the difference between the dispersion of the Gaussian assumed LSF of the data and the stellar template. Our final pPXF fits reproduce the observed galaxy spectra very well as illustrated in Figure~\ref{ff:ppxf_overview}. For NGC 2784, NGC 3640 and NGC 4281 we also excluded the region around the Na I atomic absorption line at $\sim 2.2\, \mu$m as none of our stellar templates could match the line strength fully. As \cite{Silva2008} point out this is an often seen discrepancy between pure old galaxies and Galactic open cluster stars. We extended the masked regions because the blue part of the spectrum is very noise-polluted and biases the kinematics to a more noisy solution. Including or excluding this region, the changes in the four moments stay within the derived kinematical errors.

\subsection{VIMOS \& MUSE specifics}\label{ss:vimoskinematics}
The kinematic extraction of the optical VIMOS and MUSE data was performed similarly to the ATLAS$^{\rm 3D}$ kinematic extraction. We re-extracted the kinematics for VIMOS data as the \cite{Lagerholm2012} extraction did not contain kinematic errors for each spaxel.\\
The optical IFU data matching stellar templates were taken from the medium-resolution Isaac Newton Telescope library of empirical spectra \citep[MILES, ][]{Sanchez-Blazquez2006,Falcon-Barroso2011} stellar library\footnote{\url{http://miles.iac.es/}} (version 9.1). We used the full sample consisting of 980 stars that span the wavelength range 4760-7400 \AA\, and fitted the wavelength range from 3800-6500 \AA\, in the galaxy spectra. As already mentioned, we also had to ensure that the instrumental resolution of the stellar templates and the optical data had comparable values. \cite{Beifiori2011} and \cite{Falcon-Barroso2011} report the instrumental dispersion of the MILES template library to be $\sigma_{\textrm{MILES}} = 2.51$~\AA. 
The MUSE spectral resolution was carefully measured by \cite{Guerou2017} based on sky emission lines, and the authors found a variation of the LSF with wavelength. In the for our analysis relevant wavelength range of 4800 to 6800~$\AA$ the spectral resolution changes from 2.5 to 2.9 $\AA$. In order to test the significance of this spectral resolution variation on our stellar kinematics measurement, we did an extraction using the extreme values of 2.5~$\AA$ and 2.9~$\AA$. The velocity dispersion changed in average by only 5 km/s which is within the kinematic error range, and we decided not to downgrade the MILES template library to the MUSE resolution. On the other hand, the VIMOS data have an instrumental dispersion of $\sigma_{\textrm{VIMOS}} = 2.1$~\AA\,\citep{Rawle2008}. Theoretically, a downgrading of the observed galaxy spectrum would be necessary here. 
As both instrumental spectral resolutions were relatively similar and we did not want to smooth relevant kinematic information, we did not convolve the VIMOS spectra to the lower resolution. All velocity dispersions are $\sigma>150$ km/s, so the effects of the slightly different resolutions on the derived kinematics are negligible. However, just like in \cite{Thater2017} we tested the effect of not downgrading the VIMOS spectra by using a well-defined sub-sample of the Indo-US stellar library \citep{Valdes2004} as stellar template in the pPXF fit. Our sub-sample of the Indo-US stellar library consists of 52 spectra and covers a wavelength range of 3460 to 9464 \AA\, at a spectral resolution of $\sigma_{\textrm{Indo-US}} = 1.35$ \AA\, \citep{Beifiori2011}. A comparison between the Indo-US kinematic extractions with the MILES extractions showed that the extracted kinematical maps displayed the same general features and trends. We could, however, discern a difference in the extracted values between the two stellar templates with the MILES velocity dispersions being systematically lower (10-20 km/s) and thus, more consistent with the stellar kinematics extraction from SINFONI. We furthermore recognized that the spectral fits were worse for the Indo-US fits such that we expected a template mismatch from the relatively small number of Indo-US template stars. Comparing our kinematic extraction with the extraction by \cite{Lagerholm2012} proved consistent results. We, therefore, decided to keep the MILES library for the rest of this work.

\subsection{Kinematic results}
In Figure~\ref{ff:sinfoni_kinematics}, we present the high-resolution SINFONI kinematic maps of the central $3 \times 3$~arcsecs of the galaxies resulting from the pPXF fits. 
The first column shows the S/rN map which we derived from the comparison between the pPXF fit and the input spectra (after applying the Voronoi binning). It visualizes well the quality of the pPXF fit and the quality of the data, as the S/rN is directly related to the errors of the kinematics, being large in the center and monotonically decreasing with radius. The S/rN maps show that our kinematics extraction works well (S/rN > 30) within 1 arcsec which is the region that we used for our dynamical modeling. The next four columns show the velocity, velocity dispersion, $h_3$ and $h_4$ maps for each of our galaxies.\\
As expected from our selection criteria, the derived kinematics show mainly regular features. For each of our six galaxies, a clear rotation pattern is visible with maximal relative velocities ranging from 50 to 180 km/s (after subtracting the systemic velocity). The velocity dispersions show various patterns for the different galaxies. NGC 2784 and NGC 4281 contain a clear sigma increase within the isophotal center. The sigma peak in NGC 2784 has a size of about 0.3\arcsec\, and goes up to 275 km/s, while we find a larger sigma peak in NGC 4281 ($\sigma \approx 310$ km/s). In NGC 3640 another clear feature is apparent: a slightly asymmetric velocity dispersion decrease in the center (down to 175 km/s) which spans the complete central region ($r<0.7~\arcsec$). This dip velocity is consistent with early work by \cite{Prugniel1988} and \cite{Davies1987}. \cite{Prugniel1988} also point out that this galaxy might be in an advanced merger state which would significantly affect our dynamical models. Large scale signatures of this merger (such as shells) are also visible in the MATLAS images from \cite{Duc2015}, also shown in \cite{Bonfini2018}. However, \cite{Krajnovic2011} analyzed the ATLAS$^{\mathrm{3D}}$ kinematics of NGC 3640 with Kinemetry \citep{Krajnovic2006} and found only very small residuals and a very regular shape within one R$_{\mathrm e}$, indicating that the center of NGC 3640 is relaxed now. We, therefore, believe that our MBH mass measurement is robust and not likely affected by the advanced merger state \citep{Prugniel1988}.
The velocity dispersion map of NGC 584 shows an hourglass-shape which can be attributed to a dynamically cold disc component (low velocity dispersions). Also, NGC 4570 shows signatures of a central disk. Its velocity dispersion goes up to 230 km/s, and we see maximal rotational velocities of 60 km/s which is fully consistent with the HST/Faint Object Spectrograph kinematics from \cite{vandenBosch1998a}. The velocity dispersion map of NGC 7049 is very unusual: it shows a very flat velocity dispersion profile without a clear sigma rise being visible in the kinematics of this galaxy. \\ 
The $h_{3}$ Gauss-Hermite moment maps show the typical anti-correlation to the velocity for each galaxy. The $h_{3}$ map of NGC 3640 may look chaotic at first glance, but the anti-correlation trend is also slightly visible here. 
\\ \\
The visual comparison of the near-infrared central kinematic maps with the optical large-scale maps (Appendix: Fig.~\ref{ff:atlas3d_kinematics},~\ref{ff:vimos_kinematics}) shows globally consistent results and similar trends even though we probe both very different scales and very different wavelength regions. The kinematic details of the SINFONI maps are generally not present on the large-scale kinematic maps, as the spatial resolution of the latter is comparable to the SINFONI FOV. In a second more quantitative comparison, we compared the Gauss-Hermite profiles from the four-moment pPXF fit of the two data sets. For the "point-symmetric" velocity dispersion and $h_4$ moment, we averaged the bins within concentric circular annuli around the kinematic center and repeated this process with growing radius. The bins were chosen such that the luminosity-weighted center was within the respective annulus. We present the comparison in Figure~\ref{ff:kin_comp}. For some cases, we had to slightly shift the velocity dispersion and $h_4$ values of the SINFONI data (values are shown in Figure~\ref{ff:kin_comp}) in order to perfectly match the large and small-scale data. The shifts are about 5 \% for three of our galaxies, no shifts for the remaining galaxies. Even before the shifts, all measured SINFONI velocity dispersions and most $h_4$ profiles were in very good agreement with the large-scale data. Some discrepancy can be seen in the $h_4$ profile of NGC 4281 which has a positive gradient for SINFONI and constant value for ATLAS$^{\textrm{3D}}$. We believe that this discrepancy arises from the ATLAS$^{\textrm{3D}}$ spatial resolution which flattens out the central features of the $h_4$ moment. \cite{Krajnovic2018} test the significance of the shifting with respect to the measured black hole mass and conclude that they add an uncertainty of about 80\% of the measured black hole mass by shifting the velocity dispersion by about 8\%. This means that we possibly add an uncertainty of 50\% in mass for NGC 3640, NGC 4281 and NGC 7049.\\

\section{DYNAMICAL MODELLING}\label{ss:dynamicalmodelling}
We derived the central black hole masses of our target galaxies using two different and independent dynamical modeling methods: Jeans Anisotropic Models \citep[JAM, ][]{Cappellari2008} for constraining the parameter space and three-integral \cite{Schwarzschild1979} orbital superposition models for deriving the final black hole masses. In the past, the Schwarzschild method has successfully been used to reproduce detailed models for spherical, axisymmetric and triaxial nearby galaxies. On the other hand, the JAM method is less general than orbit-based methods but far less computationally time-consuming. Furthermore, it provides a good description of galaxies based on two-dimensional stellar kinematics. Previous works, on almost forty galaxies, have shown that, although starting from different assumptions, both techniques provide generally consistent SMBH mass results \citep{Cappellari2010,Seth2014,Drehmer2015,Thater2017,Feldmeier-Krause2016a,Krajnovic2018,Ahn2018}, such that modeling the observed stellar kinematics with both independent methods provides a more robust measurement. Recently, \cite{Leung2018} compared the results from both Schwarzschild and JAM models against circular velocities derived from molecular gas for 54 galaxies with CALIFA \citep{Sanchez2012} integral-field stellar kinematics. They found that JAM and Schwarzschild recover consistent mass profiles (their Fig. D1). Moreover, JAM was found to recover more reliable circular velocities than the Schwarzschild models, especially at large radii where the gas velocities are more accurate (their fig. 8). Their study illustrates the fact that the reduced generality of the JAM method, with respect to the Schwarzschild method, is not necessarily a weakness and highlight the usefulness of comparing both methods as we do here.

\subsection{The mass model}\label{ss:mass_model}
\begin{figure*}
  \centering
    \includegraphics[width=0.95\textwidth]{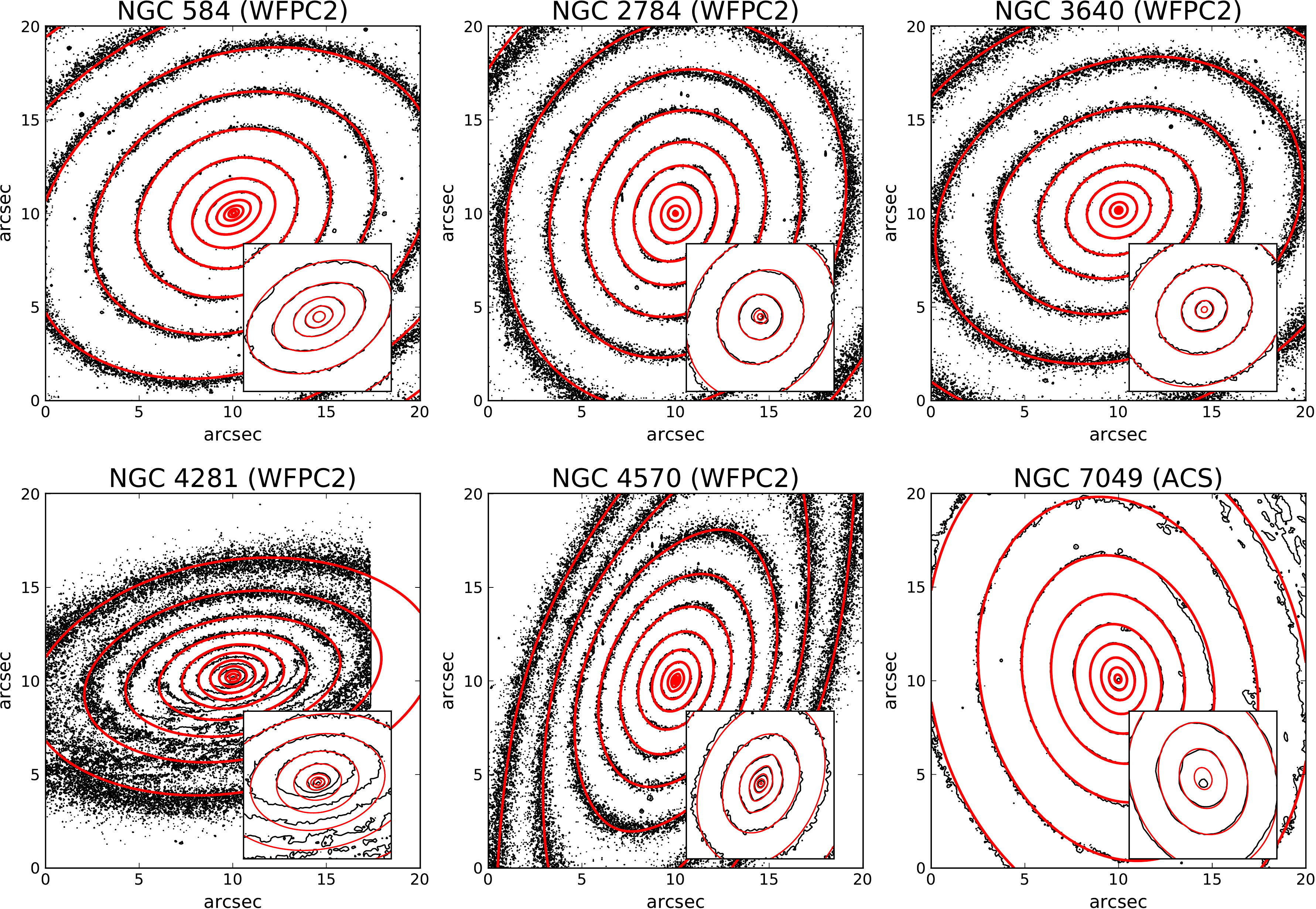}
      \caption{Isophotal maps of the WFPC2 and ACS images of our target galaxies within a FoV of $20 \times 20$ \arcsec. In the bottom right of each panel we show a  cutout of the central $3 \times 3$ \arcsec.  The contours of our best-fitting MGE model (red) are superimposed on the HST images. For the models, foreground stars and close galaxies were masked. For NGC~2784, NGC~4281 and NGC~7049 a dust-correction of the internal dust rings had to be applied before MGE modelling their surface brightness. The MGE models were build from the combined photometric information of HST (r < 10\arcsec) and wide-field of view data (r > 10\arcsec) from ATLAS$^{\textrm{3D}}$ and the CGS \citep{Ho2011} survey.}
      \label{ff:mge}
\end{figure*}
The gravitational potential of the galaxy is a composition of the potential of the stars, the potential from the central black hole which is assumed as a point mass, and the potential of dark matter. In order to find the mass of the central black hole, it is crucial to determine the stellar and dark matter contribution of the total galaxy mass as precisely as possible. The stellar mass density of the galaxy can be inferred from the galaxy luminosity density multiplied by the stellar M/L, which itself can be derived by modeling the stellar surface brightness of the galaxy. An efficient tool to provide an analytical description of the surface brightness of galaxies is the Multi-Gaussian Expansion (MGE) developed by \cite{Emsellem1994} and \cite{Cappellari2002} in which a sum of two-dimensional concentric Gaussians parametrizes the galaxy surface brightness.\\ \\
We performed the MGE modelling on both highly-resolved HST and deep wide-field ground-based SDSS (presented by \cite{Scott2013}) or CGS \citep{Ho2011} imaging data, simultaneously, using the MgeFit Python package\footnote{See footnote 2} Version 5.0 of \cite{Cappellari2002}. Except for NGC 7049, the SDSS images were in the r-band, while from HST we chose images taken with the WFPC2 camera in bands which matched the SDSS r-band best. NGC 7049 was only observed with the ACS camera in the F814W filter, and we matched it with I-band data from the CGS survey. We aligned the surface brightness profiles by re-scaling the large FOV imaging data to the central HST profiles and used the HST imaging for the photometric calibration. Foreground stars and nearby galaxies were carefully masked before applying this procedure. Furthermore, we had to apply a dust-correction to NGC 4281 and NGC 7049 to improve the modeling of the underlying galaxy surface brightness. Dust can have a significant effect on the stellar mass model as it alters the shape of the stellar surface brightness and dilutes the observed galaxy light. A careful dust correction is necessary to optimize the reproducibility from the model and the actual shape of the galaxy. We used the same method as in \cite{Cappellari2002b} and \cite{Scott2013} to dust-correct the SDSS and CGS images and the dust-masking method outlined in \citet{Thater2017} to mask dust-rings visible in the HST images, which had only a single image available (for details see Appendix~\ref{s:dustcorrection}). We also visually inspected the HST images of our galaxies for nuclear star clusters, but could not find any evidence. This is expected as galaxies with a mass of more than $10^{11} M_{\odot}$ usually do not harbor nuclear star clusters \citep{Ferrarese2006,Wehner2006,Seth2008, Graham2009}.\\
The final MGE fits converge for between 10-12 Gaussian components centered on the galaxy nucleus and with the major axis aligned with the galaxy photometric major axis. For most of our lenticular galaxies, we can see a clear trend of the axial-ratio change with radius. They show rather round isophotes in the central bulge region and flattened and discy isophotes for larger radii due to the outer disc.  

We converted the MGE parameters from pixel-space into physical units of L$_{\odot}$ pc$^{-2}$ following the guideline given by the MGE readme and \cite{Holtzman1995}. For the transformation we needed to account for the absolute Vega magnitude of the sun\footnote{\url{http://mips.as.arizona.edu/~cnaw/sun.html}} M$_{\rm F555W}= 4.85$, M$_{\rm F606W} = 4.66$ and M$_{\rm F814W}=4.15$. Furthermore, we corrected for the foreground Galactic Extinction applying the values found in the NASA/IPAC Extragalactic Database\footnote{\url{https://ned.ipac.caltech.edu/}} which were derived by \cite{Schlafly2011}. The final MGE parameters are presented in Table~\ref{t:mge}: for each galaxy, we list the index of the Gaussian component, the surface brightness in units of L$_{\odot}$ pc$^{-2}$, the Gaussian dispersion $\sigma_j$ in arcseconds and the axial ratios q$_j$. In Figure~\ref{ff:mge}, we show a comparison of our resulting best-fit MGE models and the observed HST WFPC2 and ACS images. Except for nuclear dust patterns (NGC 2784, NGC 4281, NGC 7049), the modeled MGE surface brightness are in good agreement with the surface brightness of each of the six galaxies. Especially for NGC 4281 a large dust mask had to be applied to correct the MGE model for the large amount of dust in this galaxy. \\ \\
The next step for determining the mass model is the de-projection of the surface brightness into a three-dimensional luminosity density. Therefore, it is necessary to impose an assumption on the structure of the potential. For each of our target galaxies we adopted the assumption of an axisymmetric potential, such that, assuming a given inclination ($i>0$), the luminosity density can directly be deprojected from the MGE model. 
We used the built-in MGE regularization in order to bias the axial ratio of the flattest Gaussian to be as large as possible to prevent strong variations in the mass density of the MGE model. The MGE deprojection assumption does not remove the intrinsic degeneracy of the deprojection \citep{Rybicki1987,Gerhard1996}, which, especially at low inclination, can lead to major uncertainties and constitutes a fundamental limitation to the accuracy of any dynamical model \citep[e.g.][]{Lablanche2012}.
In the center of the galaxies, which is probed by our data, stars mainly contribute to the mean potential of the galaxy. 
This means that the galaxy mass density $\rho$ can simply be described as the product of the galaxy luminosity density and a dynamical mass-to-light ratio M/L. The gravitational potential generated by this mass density can then be obtained with the Poisson equation, $\nabla^2  \Phi = 4\pi G \rho$, and is one of the ingredients for the dynamical models in the next two sections. For further details regarding the MGE de-projection, we refer to the original work by \cite{Emsellem1994} and \cite{Cappellari2002}.

\subsection{Jeans Anisotropic Models}\label{ss:jam}
\begin{figure}
  \centering
    \includegraphics[width=0.48\textwidth]{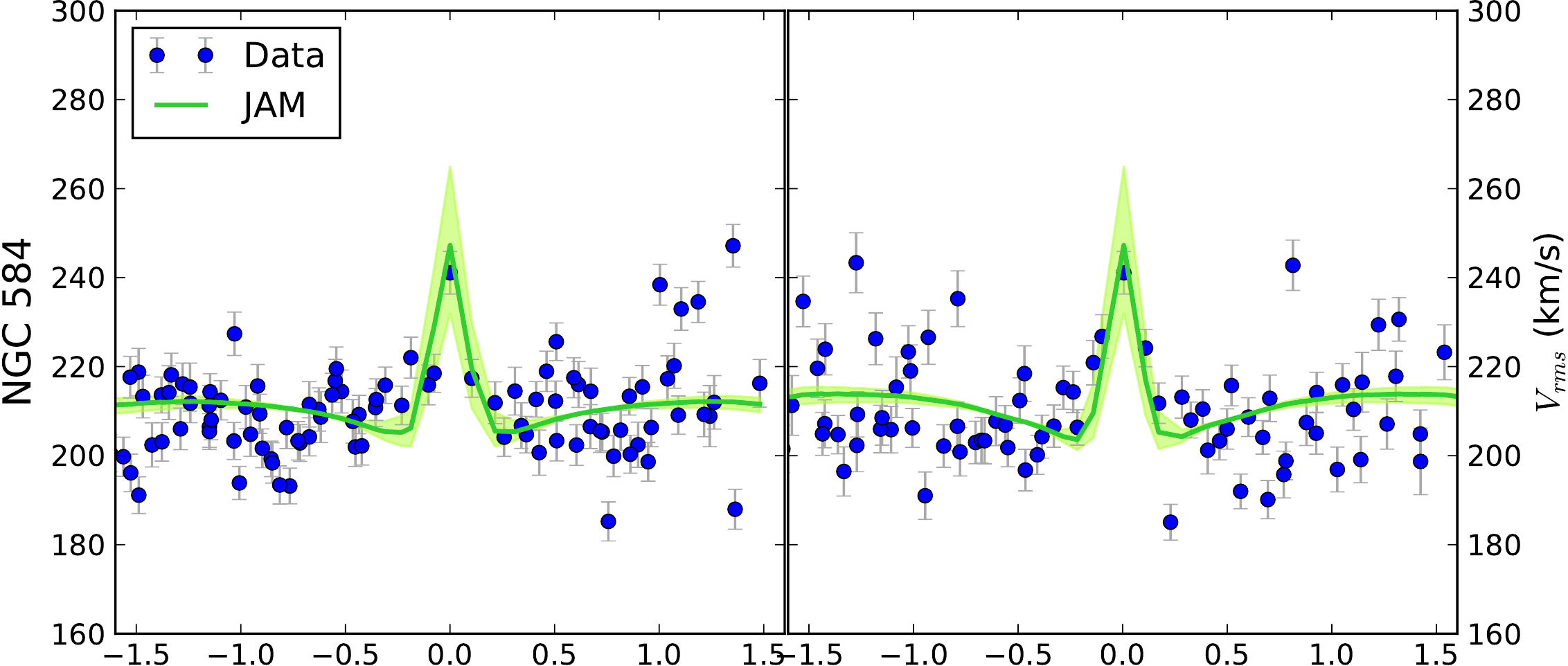}
    \includegraphics[width=0.48\textwidth]{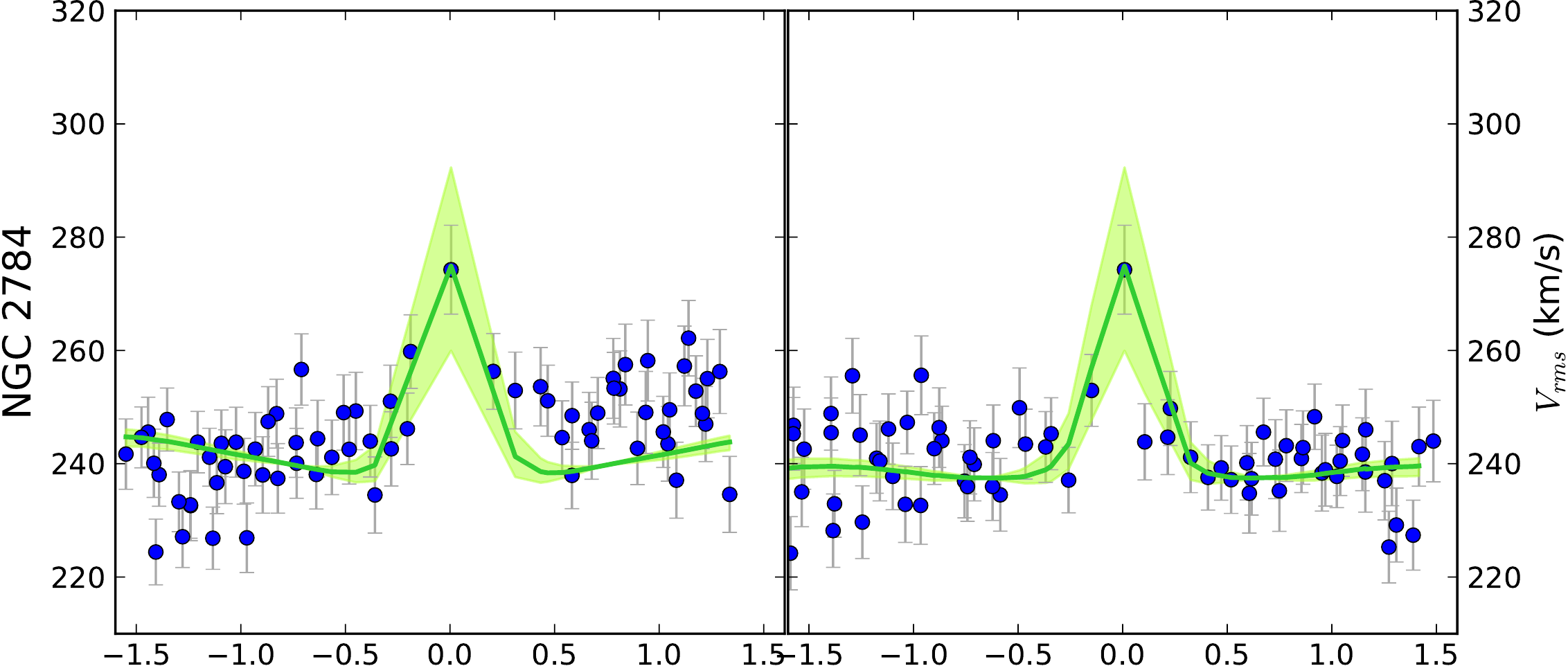}
        \includegraphics[width=0.48\textwidth]{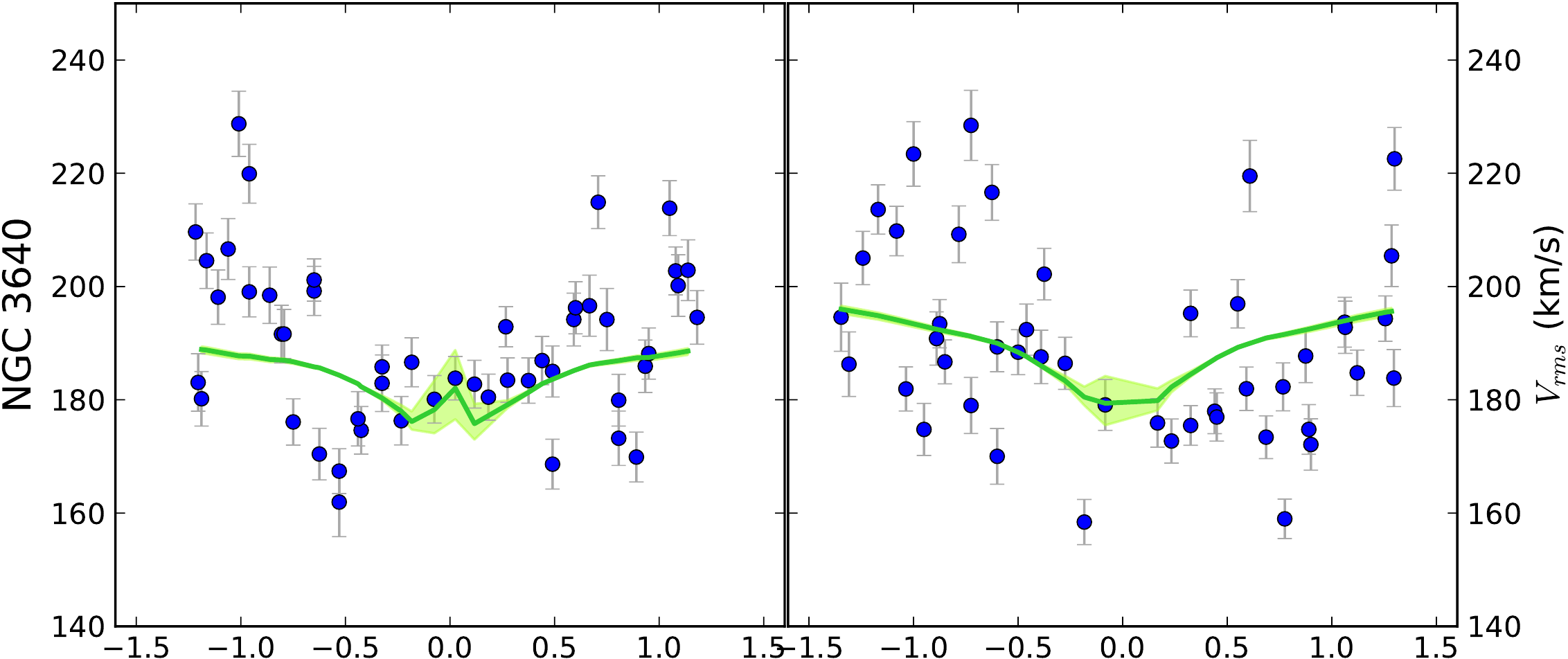}
    \includegraphics[width=0.48\textwidth]{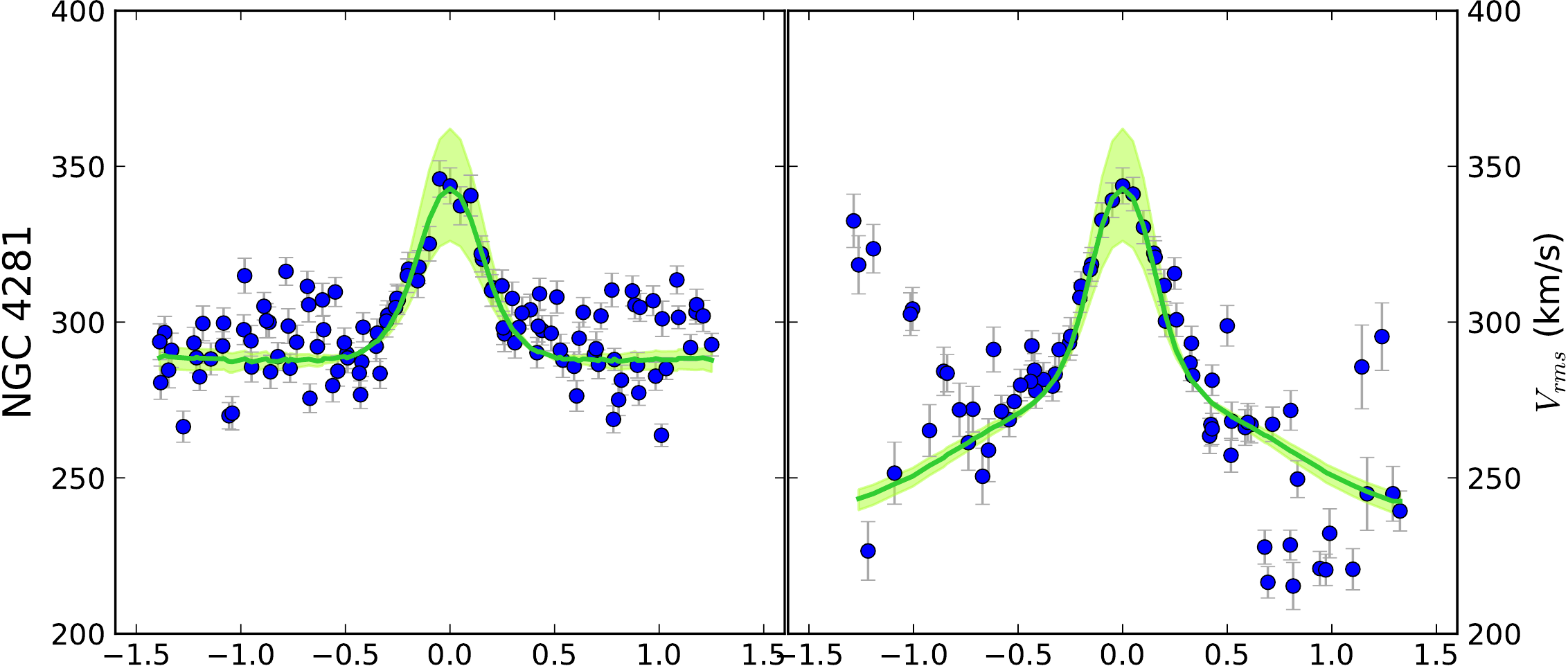}
        \includegraphics[width=0.48\textwidth]{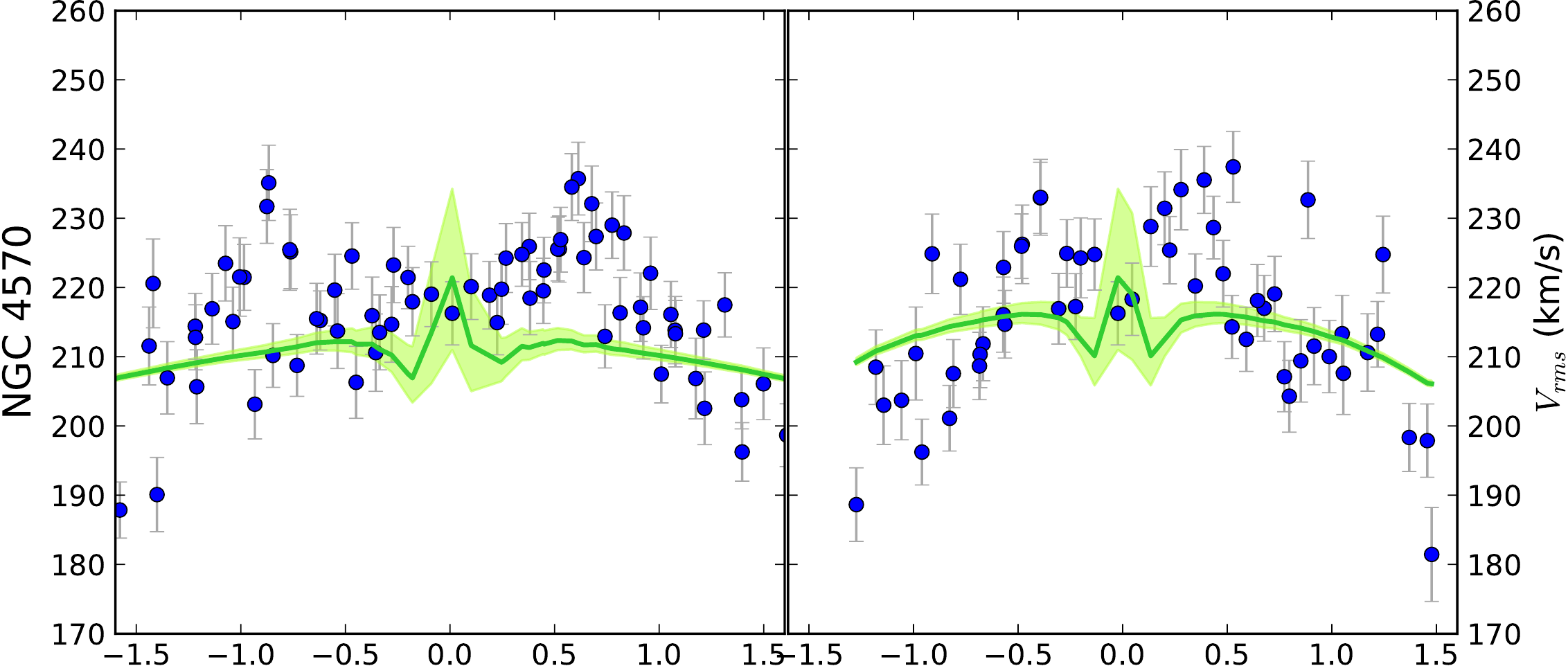}
    \includegraphics[width=0.48\textwidth]{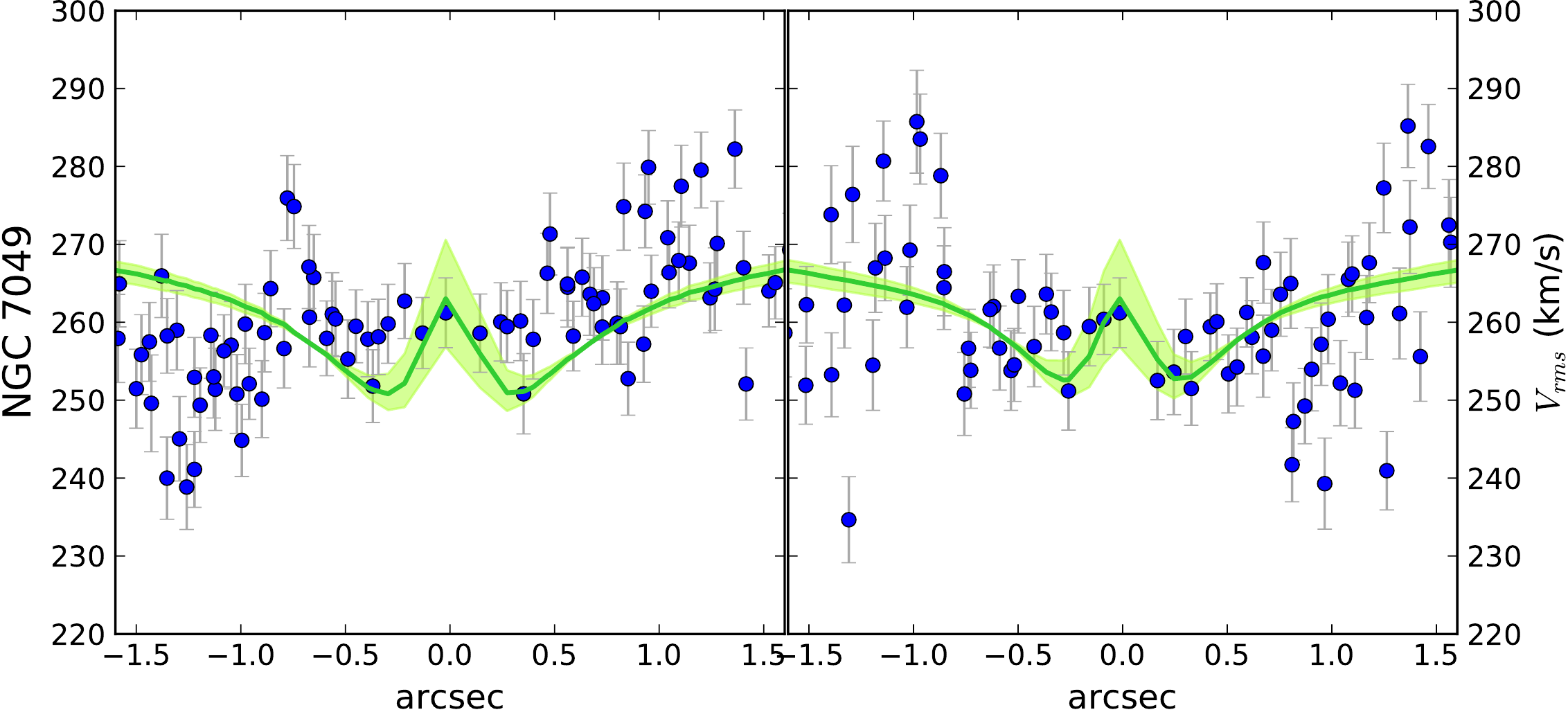}
      \caption{Comparison of the $V_{rms}$ profiles between the SINFONI data (blue) and best-fitting JAM models (green) along the major (left) and minor (right) axis. The green shaded region shows JAM models with varying black hole mass by a factor of 1.3 either larger or smaller than the best-fitting mass. }
      \label{ff:jam}
\end{figure}

The motion of a collection of stars in a gravitational field can be described by the \cite{Jeans1922} equations. They provide the basis for the JAM method \citep{Cappellari2008}, which predicts the second velocity moment by solving the Jeans and Poisson equations for the mass density derived from the MGE model. Projected along the line-of-sight of the model the second velocity moment is a function of four free parameters: the mass of the black hole $M_{BH}$, the anisotropy parameter $\beta_z$, the mass-to-light ratio M/L and the inclination angle i. The anisotropy parameter describes the orbital distribution by relating the velocity dispersion parallel to the rotation axis and in the radial direction: $\beta_z=1-\sigma_z^2/\sigma_R^2$ assuming that the velocity ellipsoid is aligned with cylindrical coordinates. We used the JAM method in order to model the second velocity moment in the potential defined by our MGE models, which is assumed to be axisymmetric. The modeled second velocity moment was then compared to the observed $V_{rms}=\sqrt{V^2+\sigma^2}$ with V being the mean velocity and $\sigma$ the velocity dispersion which was measured from the high-resolution SINFONI stellar kinematics (assuming a parametrization of the LOSVD of a simple Gaussian). Unlike the Schwarzschild models (Section~\ref{ss:schwarzschild}), we here only fit the innermost high-resolution SINFONI kinematics to be robust against possible gradients in the M/L or the anisotropy.\\ \\
We found the posterior distributions and the best-fitting values of the JAM parameters by applying a Bayesian framework in the form of Markov chain Monte Carlo (MCMC) inference method \citep{Hastings1970}. We used the emcee software package \cite{Foreman-Mackey2013} which is a python implementation of the \cite{Goodman2010} affine invariant Markov chain Monte Carlo ensemble sampler\footnote{http://dfm.io/emcee/current/}. JAM is generally fit to the data using Bayesian approaches and MCMC as this makes it easy to detect degeneracies between parameters and marginalize over uninteresting parameters \citep[e.g., ][]{Cappellari2012,Barnabe2012,Cappellari2013,Watkins2013,Cappellari2015,Mitzkus2016,Poci2016,Kalinova2017,Li2017,Bellstedt2018,Leung2018}, and in the context of massive black hole determination by \cite{Krajnovic2018} and \cite{Ahn2018}. For our dynamical JAM modeling, we followed a similar approach as \cite{Cappellari2013b}.  In the burn-in phase, a set of walkers explores the pre-defined parameter space, where each successive step is evaluated based on the likelihood of each walker.  We used 100 walkers and tracked them for 200 steps until the fit converged. After the exploration of the parameter space, we continued the MCMC for 500 steps (post-burn-phase) and used the final walker positions to generate posterior distributions and model properties.  \\
We built models with the four free parameters ($\log M_{BH}$, $\beta_z$,M/L,$i$) and compared them with the observed $V_{rms}$ using a $\chi^2$ statistic. The logarithmic likelihood probability of our data is defined as \\
\begin{equation}
\log P\,(V_{rms}\, \vert \, i,\,\mathrm{M/L},\, \beta_z,\, \log M_{BH}) \propto -\frac{1}{2} \sum_n \underbrace{\left(\frac{V_{rms}-\langle v^2_{los} \rangle^{1/2}}{\delta V_{rms}}  \right)^2}_{\chi^2}
\end{equation}
which is a sum over all good spaxels and where $\delta V_{rms}$ are the errors derived by the Monte Carlo simulations of the kinematic data and error propagation. 
In order to ensure that the fitting converges, we set reasonable priors on the parameters. We used uninformative priors (assumption of maximal ignorance) for the different parameters, which are uniform within the bounds of the likelihood function: $\log M_{BH} \in [4.8,9.8]$, $\beta_z \in [-1,+1]$, M/L $\in [0.1,20]$ and the inclination was allowed to vary over the full physical range (only limited by the flattening parameter $q_{\rm min}$ of the flattest Gaussian of the MGE model $\cos^2 i = q^2$). We made sure that the MCMC chain converges by visually checking our burn-in plots and running several Markov chains. \\ \\

\begin{figure*}[!htb]
  \centering
    \includegraphics[width=1\textwidth]{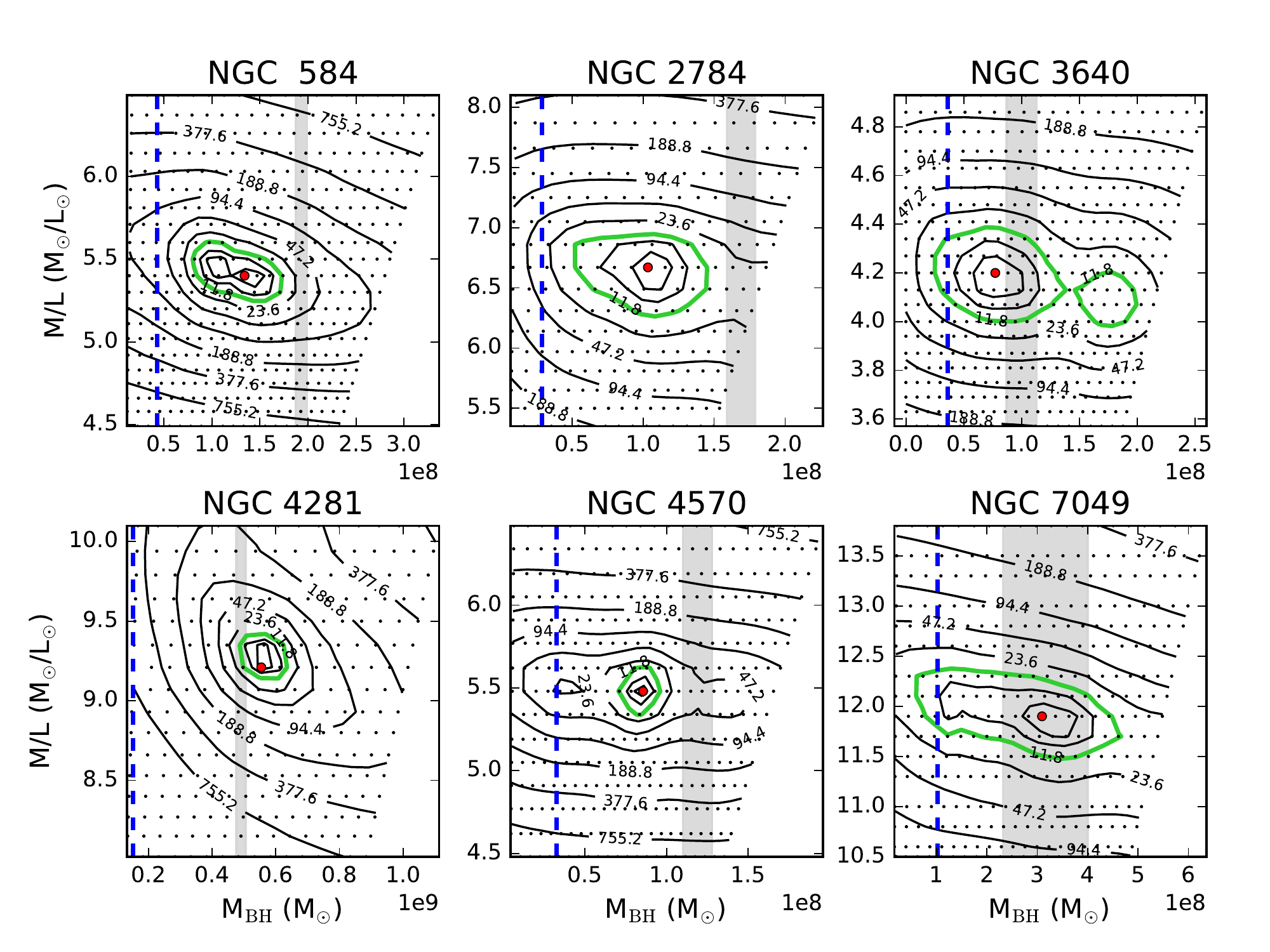}
      \caption{Grids of Schwarzschild models (indicated by the black dots) over different mass-to-light ratios M/L and black hole masses $M_{BH}$. The best-fitting model, derived as the minimum of $\chi^2$, is indicated by a large red circle. Contours are the $\Delta \chi^2 = \chi^2 - \chi^2_{min}$ levels where the thick green contour shows the $3\sigma$ level of the two-dimensional distribution. In addition, we have added the 3-sigma limits on the best-fitting black hole masses of the JAM models (grey shaded regions). The dashed blue line indicates the mass of the black hole which has the radius of the sphere of influence of half the resolution of our LGS AO data (inferred from the narrow component of the AO PSF), which is approximately the lowest black hole measurement that we expect to be detectable based on our data.}
      \label{ff:schwarzschild_grid}
\end{figure*}

\begin{figure*}[!htb]
  \centering
    \includegraphics[width=0.86\textwidth]{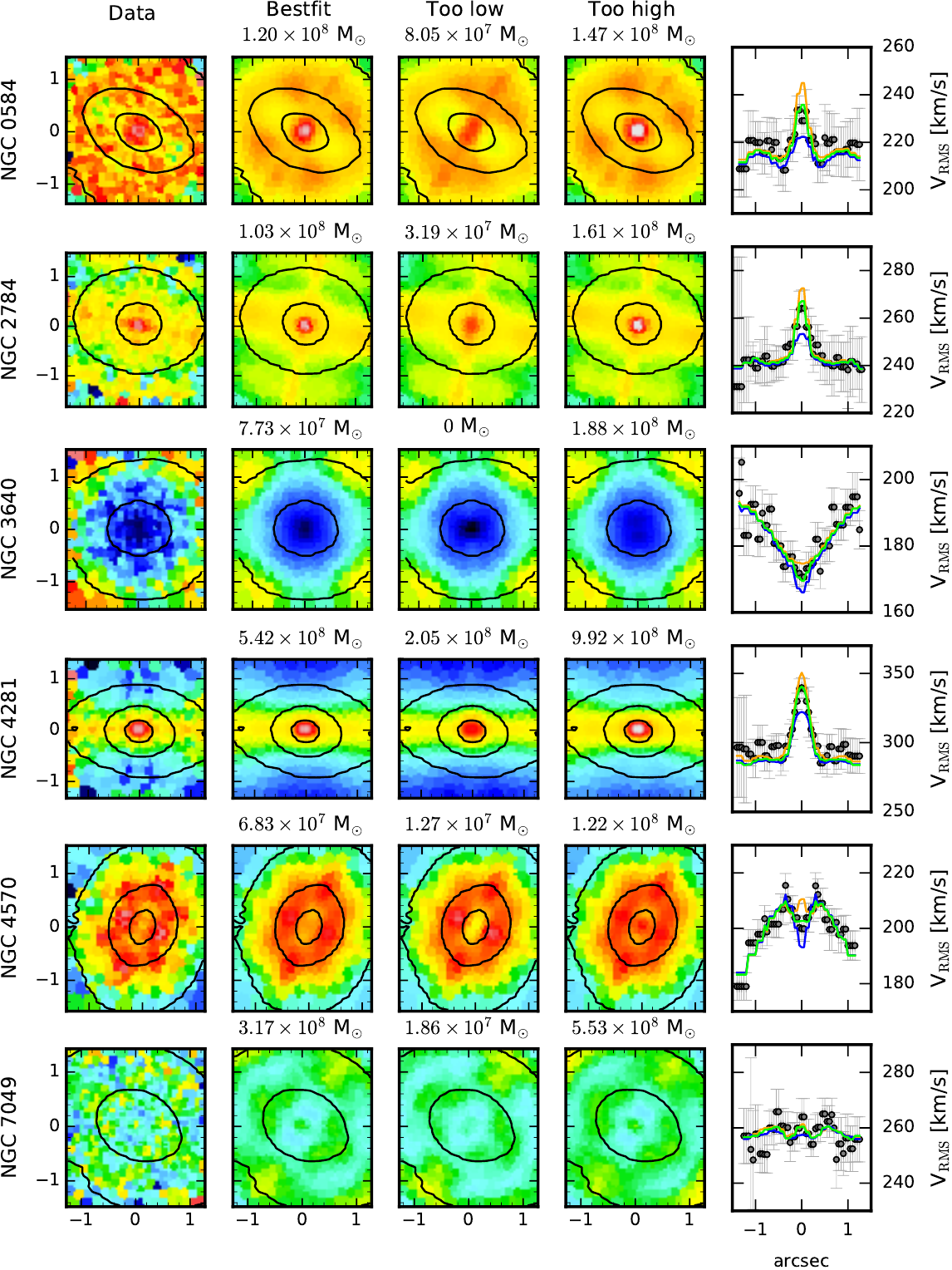}
      \caption{Comparison of the $V_{\rm RMS}=\sqrt{V^2+\sigma^2}$ maps from the SINFONI data and the Schwarzschild models. Each row shows the maps of one galaxy, respectively. From left to right we present the observed symmetrized $V_{\rm RMS}$ from the SINFONI data, and the $V_{\rm RMS}$ maps of the Schwarzschild models from the best-fitting, a too low and a too high $M_{BH}$ as well as the profiles along the x=0 axis. The too low (blue) and too high (orange) black hole masses are chosen to be just outside of the $3\sigma \chi^2$ contours. All models are shown at the respective best-fitting M/L. The high and low mass models are clearly ruled out for all galaxies.}
      \label{ff:schwarzschild_sigma_comparison}
\end{figure*}

In Appendix E (Figure~\ref{ff:mcmc_corerplot}), we present the MCMC posterior probability distributions of the different JAM model parameters for each galaxy. The contour plots show the projected two-dimensional distributions for each parameter combination and the histograms show the one-dimensional distributions for each parameter. As clearly indicated by the contour plots, our $M_{BH}$ and $\beta_z$ parameters are not degenerate for NGC 584, NGC 2784, NGC 3640 and NGC 4570, which shows that these measurements are robust. The $\beta_z$ parameters and the inclinations are naturally correlated but do not affect the black hole mass measurement. Generally, the derived inclinations are not well constrained and tend to be larger than expected from the literature. This is expected behavior as we only fit the central kinematics of our galaxies. That is why we decided to use the literature inclinations for the Schwarzschild modeling analysis. Furthermore, NGC 4281 and NGC 7049 show a degeneracy but still clearly constrain the black hole mass. We used the posterior probability distributions to calculate the best-fit value and their corresponding 3$\sigma$ uncertainties. The median values of the posterior distribution are given in Table~\ref{t:results}. A visual comparison between the observed $V_{rms}$ profiles and the best-fit jam models of the MCMC routine are presented in Figure~\ref{ff:jam}. In all cases the derived models reproduce the central peak of the observed $V_{rms}$ well, while the outer kinematics often suffer from scatter. Our derived best-fitting JAM MBH masses and M/L were used as an initial guess to constrain the Schwarzschild models.

\subsection{Schwarzschild models}\label{ss:schwarzschild}

In our second dynamical modeling approach, we used the axisymmetric Schwarzschild code which was optimized for two-dimensional IFU data and described in \cite{Cappellari2006}. The method is based on the numerical orbit-superposition method originally invented by \cite{Schwarzschild1979} and further developed to fit stellar kinematics \citep{Richstone1988,Rix1997,vanderMarel1998,Cretton1999a}. The basic idea of the Schwarzschild method is that the mass distribution of the galaxy is well described by the sum of time-averaged orbits in a stationary galaxy potential. The method consists of basically two steps which are repeated for each modeled black hole mass, respectively.
First, assuming a stationary galaxy potential, a representative orbital-library is constructed from the galaxy potential which itself is derived from the mass density from Section~\ref{ss:mass_model}.
Regular orbits in axisymmetric potentials are characterized by three integrals of motion: the binding energy E, the vertical component of the angular momentum $L_z$ and a nonclassical third integral $I3$ introduced by \cite{Ollongren1962}, see also \cite{Richstone1982}
, which are equally sampled by the orbit library. We typically trace each orbit for 200 oscillations through the system to have a representative characteristic within the entire equilibrium phase of the galaxy.\\
In a second step, each orbit is projected into the plane of the observables and the complete set of orbits is combined to match the light distribution and the LOSVD of the galaxy by assigning a weight in a non-negative least-squared (NNLS) fit \citep{Lawson1974}. Compared to the JAM models, where we approximated the velocity second moments as the dispersion of a Gaussian, the Schwarzschild modeling method fits the full LOSVD. 
\\ \\
We constructed the Schwarzschild models along a grid of radially constant dynamical mass-to-light ratio (M/L) and the mass of the black hole $M_{BH}$.
We began the modeling procedure by running coarse parameter grids centered on the best-fitting parameters ($M_{BH}$, M/L) derived from the JAM models in Section~\ref{ss:jam}. These models were improved iteratively by running finer and finer grids centered on the respective $\chi^2$ minimum of the coarse grid. Our final grids were then built with 21 $M_{BH}$ and 21 M/L equally spaced values for each galaxy. We only had to compute the orbit libraries for the different black hole masses as the orbits depend on the shape of the galaxy potential. The different M/L values only scale the potential and thus the orbit libraries can be re-scaled to match the different M/L a posteriori. Each orbit library consists of $21\times8\times7\times2$ orbit bundles, which are composed of $6^3$ dithers, making in total 508 032 orbits per black hole mass. These orbit libraries were then fitted to the symmetrized stellar kinematics and to the photometric model in a NNLS fit and $\chi^2$ values were calculated by fitting our Schwarzschild models to both small and large-scale kinematics. We excluded the large-scale kinematics in the central $0.8 \arcsec$ such that in the central regions only the more reliable high-resolution data was fitted. For the nnls fitting, we applied a regularization of $\Delta=10$ (analogous to \cite{Krajnovic2009a,vanderMarel1998}) to impose an additional smoothing on the distribution function of the orbit weights. We present our final grids of Schwarzschild models for each of our six galaxies in Figure~\ref{ff:schwarzschild_grid}. Plotted on the grid is the $\chi^2$ distribution as a function of $M_{BH}$ and dynamical M/L from which we deduced the best-fitting parameters within $3\sigma$ significance ($\Delta \chi =11.8$). 
In order to smooth the topology of the $\chi^2$ contours, we applied the local regression smoothing algorithm LOESS \cite{Cleveland1979}, adapted for two dimensions \citep{Cleveland1988}  as implemented by Cappellari et al. (2013a, see footnote 3). 

\begin{table*}
\caption{Summary of dynamical modelling results}
\centering
\begin{tabular}{c|lcccc|lcc|l}
\hline\hline
 & JAM &  &  & &   & Schwarzschild & & & \\
Galaxy & M$_{\rm BH}$ & M/L & $\beta$ & $i$ & $\chi^2/DOF$  &   M$_{\rm BH}$ & M/L  & $\chi^2/DOF$ &  r$_{\rm SoI}/ \sigma_{\mathrm{PSF}}$ \\
 & $(\times 10^8~$M$_{\sun})$ & (M$_{\sun}$/L$_{\sun})$ & & ($^{\circ}$) &   & $(\times 10^8~$M$_{\sun})$ &  (M$_{\sun}$/L$_{\sun})$ &  &    \\
(1) & (2) & (3) & (4) & (5)  & (6) & (7) & (8) & (9) & (10)\\
\hline
 & &  & &   & & & &\\
NGC 584 &  $1.93^{\pm 0.06}$ & $5.4^{\pm 0.1}$ & $0.05^{\pm 0.01}$  & $89^{\pm 3}$ &  4.35 & $1.34^{\pm 0.49}$ & 5.4$^{\pm 0.2}$ &  0.99  & 1.6\\
 & &  &   & & & & &\\
NGC 2784 & $ 1.69^{\pm 0.1} $ & $7.7^{\pm 0.3}$ & $0.04^{\pm 0.04}$ & 77$^{\pm 13}$ &  1.86  & $1.03^{\pm 0.54}$ & $6.7^{\pm 0.7}$ & 1.08 & 1.8\\
 & &  &   & & & & &\\
NGC 3640 & $ 0.99^{\pm 0.13} $ & $4.2^{\pm 0.1}$ & $0.14^{\pm 0.04}$  & 85$^{\pm 9}$ & 13.90& $0.77^{\pm 0.51}$ &   $4.2^{\pm 0.2}$ & 1.65 &  1.1\\
 & &  &   & & & & & \\
NGC 4281 & $4.91^{\pm 0.15} $ & $12.5^{\pm 0.2}$ & $-0.03^{\pm 0.02}$ & 75$^{\pm 9}$ &  9.93 & $5.42^{\pm 0.80}$ & $9.3^{\pm 0.3}$  & 3.98 & 2.1\\
 & &  & &  &  & & &\\
NGC 4570 &  $1.19^{\pm 0.09}$ & $6.6^{\pm 0.1}$ & $0.22^{\pm 0.02}$  & 74$^{\pm 1}$ &  3.98 & $ 0.68^{\pm 0.20}$&   5.5$^{\pm 0.1}$  & 1.87 & 1.1\\
 & &  & &  &  & & &\\
NGC 7049 &  $3.16^{\pm 0.84} $ & $11.4^{\pm 0.4}$ & $0.04^{\pm 0.04}$& 44$^{\pm 10}$ &  3.18& $3.17^{\pm 0.84}$ & $11.9^{\pm 0.3}$ & 1.38 & 1.6\\
 & &  &   & & & & &\\
\hline

\end{tabular}
\\
\tablefoot{Column 1: Galaxy name. Column 2-6: Parameters of the JAM models (black hole mass, mass-to-light ratio, velocity anisotropy parameter and stellar mass of the galaxy). Column 7-9: Parameters of the Schwarzschild models (black hole mass and mass-to-light ratio in the HST band specified in Table~\ref{t:hstbands}). Column 10: Comparison of the black hole sphere-of-influence (calculated with the central velocity dispersion $\sigma_0$) and the spatial resolution of the observations (measured by the narrow component of the AO PSF).}
\label{t:results}
\end{table*}
For each galaxy, we can constrain the upper and lower limit of the black hole masses. The best-fit values are presented in Table~\ref{t:results}. Figure~\ref{ff:schwarzschild_grid} also includes our JAM black hole mass measurements (MBH values within 99.7\% intervals from posterior, namely 3$\sigma$) as grey shaded regions and the lowest possible black hole measurement based on the data resolution in combination with the sphere-of-influence argument (blue dashed line). NGC 3640, NGC 4281 and NGC 7049 have a clear overlap between the $3\sigma$ uncertainties of the JAM and Schwarzschild models meaning they are fully consistent with each other. For the remaining galaxies we measure slightly smaller black hole masses than with the JAM method. We note that the presented uncertainties on our black hole mass measurements are predominantly formal random errors from the dynamical modeling and as such they under-estimate the fuller systematic uncertainties which we discuss in Section
\ref{ss:errorbudget}. In Figure~\ref{ff:schwarzschild_sigma_comparison}, we compare the $V_{RMS}$ maps between the SINFONI data and the Schwarzschild models for the best-fitting, a lower and higher MBH mass (just outside the $3\sigma$ contours) as well as the profiles along the x-axis. The different models are even visually very different, such that we can clearly constrain the upper and lower limit of the black hole mass.
A full comparison between our observed (symmetrized) kinematic maps and the best-fitting Schwarzschild models with all our LOSVD parameters for both the SINFONI and large-scale data is shown in Appendix~\ref{ss:comparison_schwarz}. The models can reproduce all of the kinematic features very well, both on the high-resolution SINFONI data and the large-scale data. 

NGC 4281 and NGC 7049 have an unusually large M/L, but roughly comparing the derived value of NGC 4281 (F606W-band) with the value from \cite{Cappellari2013b} who derived a value of 9.1 for the r-band by applying dynamical JAM models on the ATLAS$^{\textrm{3D}}$ data only, our value is fully consistent.

\section{DISCUSSION}
The results that we recovered from our dynamical models are only robust when the assumptions on the models are valid. Therefore, we further investigated a number of systematic error sources that could have affected our results. In that respect, the choice of distance D does not influence our conclusions but sets the scale of our models in physical units. Specifically, lengths and masses are proportional to D, while M/L scales as D$^{-1}$. 

\subsection{Systematic uncertainties}\label{ss:errorbudget}

\subsubsection{Variations in stellar populations}\label{ss:ml_variation}
\begin{figure}
  \centering
    \includegraphics[width=0.45\textwidth]{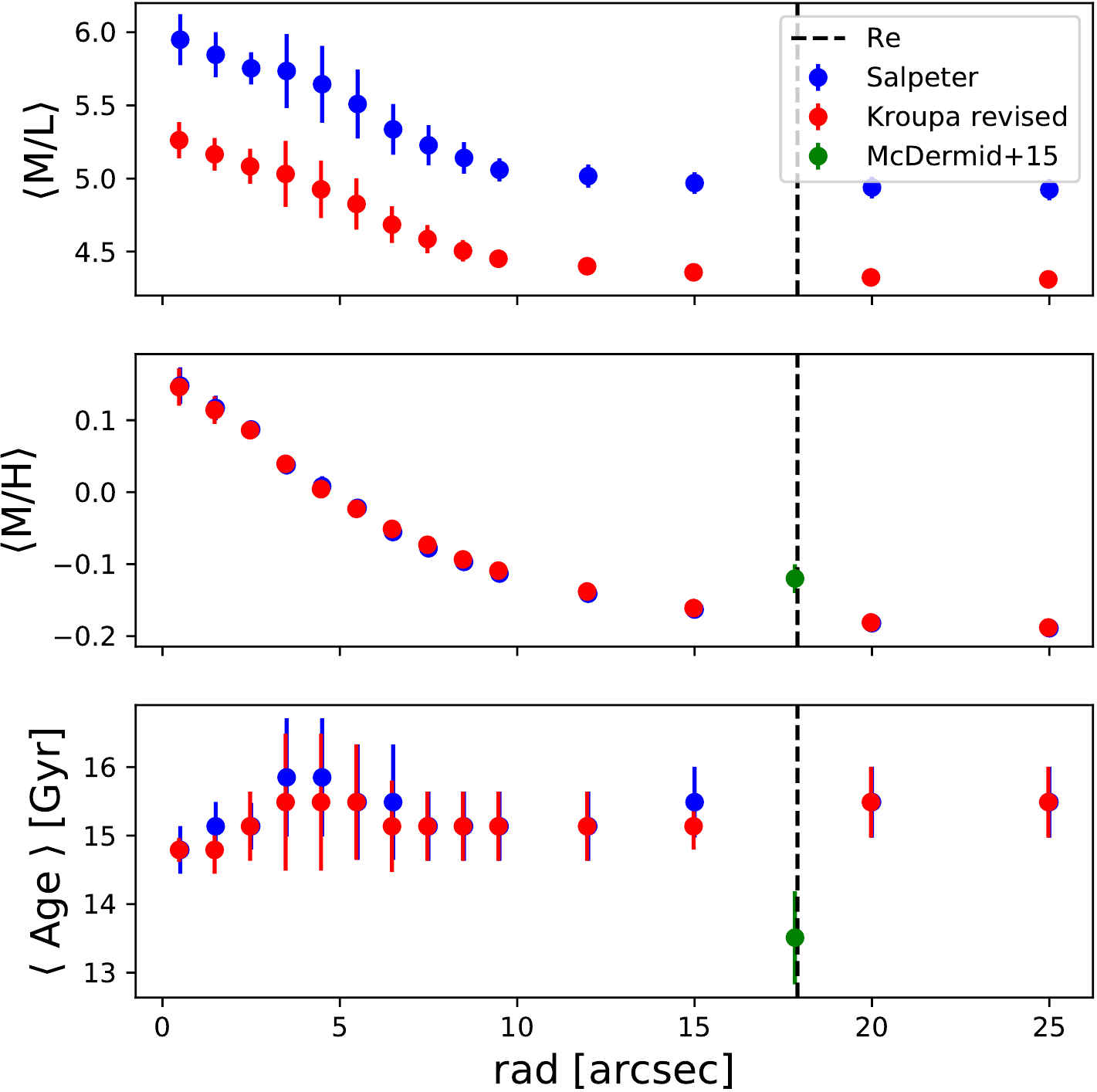}
      \caption{M/L profiles derived from stellar population analysis of the SAURON spectra combined over different aperture sizes between 0.5 and 25\arcsec. The two different colours specify whether the MILES stellar templates were created based on a Salpeter IMF ($\alpha$=1.3) (blue) or a revised Kroupa IMF (red). The dashed line denotes the effective radius of the galaxy where we compare our measurements with measurements from McDermid 2015 as a consistency check. The discrepancy between the measurements (lower age and larger metallicity) arises from them using a different stellar library as they capped their MILES library at 14 Gyrs.}
      \label{ff:mlgradient}
\end{figure}

Various radial gradients have been found for different stellar population properties in early-type galaxies. For instance, early-type galaxies typically show color gradients, the central regions being redder than the galaxy outskirts \citep{Peletier1990,Wu2005}. Metallicities often follow a negative trend with radius (i.e., the metallicity decreases when the radius increases), while the age gradient is moderately flat \citep{Kuntschner2010,Li2017}. The mentioned gradients imprint their signature on the stellar M/L which is thus expected to increase towards the center. Furthermore, variations in the stellar IMF corresponding to a larger fraction of low-mass stars can have an additional effect on the M/L variation. Negative stellar M/L gradients were observationally confirmed for local early-type galaxies \citep[e.g., recently in][]{Boardman2017, Sarzi2018, Vaughan2018}. Possibly problematic, in the previous section, we assumed the M/L to be constant for simplicity. However, ignoring the stellar M/L gradients can lead to overestimating the dynamical M/L 
and therefore also the central black hole mass \citep{McConnell2013b,Krajnovic2018b}. 
On the other hand, the stellar M/L usually runs contrary to the dark matter content which is low in the center but increases towards the outskirts of the galaxy. Therefore, including a nearly constant dynamical M/L must not always be a bad assumption in dynamical modeling \citep[e.g.][]{Thater2017} in particular, when modeling the stellar kinematics observed over a wide range of radial scales. \\ \\
In our first dynamical modeling attempt, we assumed a constant dynamical M/L for both Jeans and Schwarzschild dynamical models. Comparing the dynamical M/L derived from the JAM models (where we only used the central kinematics <1.5\arcsec) and the Schwarzschild models, we noticed a significant (>10\%) difference for half of our sample: NGC 2784, NGC 4281 and NGC 4570. We considered that the dynamical M/L difference could be caused by stellar population variations. In order to study the effect of spatial variations in the stellar populations, we followed the same method as in \cite{McDermid2015} and \cite{Thater2017} and applied a mass-weighted stellar population synthesis for NGC 4570. We chose NGC 4570 for this test as its data had the best S/N and it did not suffer from dust contamination.\\ \\
The ATLAS$^{\textrm{3D}}$ IFU spectra of NGC 4570 were co-added in growth curves with increasing circular aperture sizes having radii between 0.5 and 25 arcsec and then fitted with a linear combination of MILES simple stellar population (SSP) model spectra \citep{Vazdekis2010} using the pPXF routine. We used two different sets of template model spectra assuming either a unimodal initial mass function (IMF) of slope 1.30 (which equals a \cite{Salpeter1955} IMF) or a \cite{Kroupa2001} revised IMF. For each IMF choice, we used 350 SSP template spectra spanning a grid of 50 ages logarithmically spaced between 0.06 to 17.78 Gyr and 7 metallicities [Z/H] = [-2.32, -1.71, -1.31, -0.71, -0.40, 0.00, 0.22]. In addition, we also kept track of the stellar and stellar remnant mass $M_*$ and the r-band luminosity $L_r$ of each stellar model of the template library. Each of the template SSP spectra is assigned weights in the pPXF fit, which are smoothed out for models having similar ages and metallicities to ensure a smooth star formation history solution and suppress the noise in the final weights distribution. The smoothing is applied by adding a linear regularization to the pPXF fit which is chosen such that the difference in $\chi^2$ between regularised and non-regularised fit equals  $\sqrt[]{2N}$, where N is the number of good pixels in the spectrum. 
We then calculated the mass-weighted stellar M/L for each radial bin using the tracked stellar mass and r-band luminosity from the SSP models and using equation (5) from \cite{Thater2017}.\\ \\
The derived M/L profiles of NGC 4570 for the two different IMFs, the metallicity and age profiles are shown in Figure~\ref{ff:mlgradient}. Within the effective radius, a negative M/L gradient in the order of 10-20 \% of the central M/L is clearly visible, which has to be accounted for in the dynamical models. The gradient is very strong between 3\arcsec and 10\arcsec and flattens out for larger distances. Furthermore, while the shape of the M/L profile does not depend on the choice of the IMF, we note that their values differ by about $0.66\,M_{\odot}/L_{\odot}$ due to the ratio of high-mass to low-mass stars within the different IMFs. Recent papers suggest a trend in IMF with $\sigma_e$, in such a way that low-$\sigma_e$ (<250 km/s) typically follow Kroupa-like IMFs, while galaxies with large $\sigma_e$ follow Salpeter-like or heavier IMFs \citep[e.g.][]{Cappellari2012,Cappellari2013b,Posacki2014,Li2017}. Having velocity dispersions between 170 km/s and 245 km/s our sample galaxies thus would likely follow a Kroupa-like IMF but are located in the transition zone. In addition, IMF gradients have been found to follow the radial trend of the stellar metallicity \citep[e.g. ][]{MartinNavarro2015,Sarzi2018} which gives even more reasons to consider also bottom-heavy IMF forms. That is why we decided to derive the M/L for both IMFs and test them in our dynamical models.  \\
The derived M/L values were then multiplied with the luminosity model MGE at the respective distance from the galaxy center (assuming an aperture size of the order of MGE $\sigma_j$ from Section~\ref{ss:mass_model}) and included in the dynamical Schwarzschild models as mass density. We emphasize that we only included the M/L gradients in the Schwarzschild models as the JAM models only trace the galaxy potential within 1.5\arcsec where the stellar M/L is approximately constant. However, when constructing the Schwarzschild models, we also include stellar orbits from greater distances which could feel the effect of the observed M/L gradient. In order to account for the two possible IMFs, we ran the Schwarzschild grid for the two M/L profiles independently. We present the final black hole masses derived from the Schwarzschild models in combination with a variable M/L in Table~\ref{t:variableML}. We find two main results from this analysis: 1) both IMFs give very consistent results which was expected as the shape of their M/L gradient is very similar, 2) including M/L variations in the Schwarzschild models reduces the derived black hole mass by about a factor of 1.5 (30\%). The mass of the SMBH is decreased as more mass is included in the stellar component, and the impact on the black hole mass may have been even more important if we could have accounted for stellar population gradients down to the resolution of our SINFONI data. Our test agrees with \cite{McConnell2013b} who have noticed that the MBH mass decreases by about 20-30\% by taking M/L gradients into account. On the other hand, \cite{Cappellari2002b} only found negligible variations when allowing for M/L gradients which were within the statistical uncertainties. While this test provides an interesting implication on the SMBH scaling relations, we will postpone a more detailed discussion for a future paper in the series when we can apply a detailed test to all 18 galaxies of the sample. This test will be crucial for the three galaxies of our sample that contain nuclear disks which are often accreted and thus likely have a different stellar populations and varying M/L gradients. Furthermore, it will be interesting to test if our dusty galaxies follow positive M/L gradients due to ongoing star formation and how much these gradients will affect the derived black hole measurements. A solution to the uncertainties introduced by possible unknown population gradients consists of allowing the total mass profile to differ from the distribution of the tracer population producing the kinematics \cite[as done, e.g. in][]{Mitzkus2016,Poci2016,Li2017}. In black hole studies, this is generally done by allowing for a dark matter profile in addition to a stellar component (see Sec. 6.1.2), but the very same approach will account for gradients in the stellar M/L as well.

\begin{table}
\caption{Schwarzschild dynamical modeling results with variable M/L}
\centering
\begin{tabular}{lccc}
\hline\hline
Galaxy &   M$_{\rm BH}$ [M$_{\sun}]$ & M/L$_{\mathrm{dyn}}$/M/L$_{*}$  & IMF\\
&   $(M_{\sun})$ &  & \\
(1)  & (2) & (3) & (4)\\
\hline
NGC 4570 & $4.04_{-1.2}^{+0.9}\times10^7$ & $0.98_{-0.02}^{+0.02}$  & Salpeter \\
& & & \\
 & $4.2_{-0.4}^{+0.6}\times 10^7$ & $1.14_{-0.03}^{+0.03}$ & Kroupa rev\\
\hline
\\
\end{tabular}
\\
\tablefoot{Column 1: Galaxy name. Column 2: Derived black hole mass. Column 3: Derived ratio between stellar and dynamical M/L and 5: Assumed IMF for deriving the stellar M/L.}
\label{t:variableML}
\end{table}

\subsubsection{Dark matter}
Our dynamical models only work under the assumption of self-consistence (mass follows light). Breaking this assumption by having significant amounts of dark matter in the center can lead to systematic changes in the black hole mass. \citep{Gebhardt2009,Schulze2011a,Rusli2013}. We tested the significance of the dark matter in the central regions of our galaxies using the radial acceleration relation \citep{McGaugh2016,Lelli2017}. As long as the galaxies stay in the linear regime of the radial acceleration relation (g$_{\mathrm{\mathrm{dyn}}}$ > g$_{\mathrm{\mathrm{crit}}} = 1.2 \times 10^{-10}$ m s$^{-2}$) it is expected that the dark matter does not contribute extensively to the galactic potential. The total acceleration can be derived from the gravitational potential by g$_{\mathrm{dyn}} (R)=- \nabla \Phi_{\mathrm{tot}}(R) =V_c^2/R$ where $V_c$ is the circular velocity. 
We used the mass density (derived in Section~\ref{ss:mass_model}) assuming the dynamical M/L of the best-fitting Schwarzschild model (Table~\ref{t:results}) to calculate a model circular velocity at a radius equal to the edge of our large-scale kinematical data for each of our target galaxies. 
Our analysis yielded total accelerations between $8.8 \times 10^{-9}$ and $9 \times 10^{-10}$ m s$^{-2}$ with the smallest acceleration found in NGC 3640. All values lie well above the critical acceleration, and we conclude that our galaxies have likely a negligible contribution of dark matter in the central region which will not affect the dynamical modeling significantly. This is consistent with more direct estimates of the DM content of our galaxies from \cite{Cappellari2013b} and \cite{Poci2016}. Furthermore, the total accelerations determined in our galaxies are consistent with the accelerations of other ATLAS$^{\rm 3D}$ early-type galaxies analyzed by \cite{Lelli2017}.

\subsection{Black hole - host galaxy scaling relations} \label{ss:scalingrelations}
\begin{figure}
  \centering
    \includegraphics[width=0.48\textwidth]{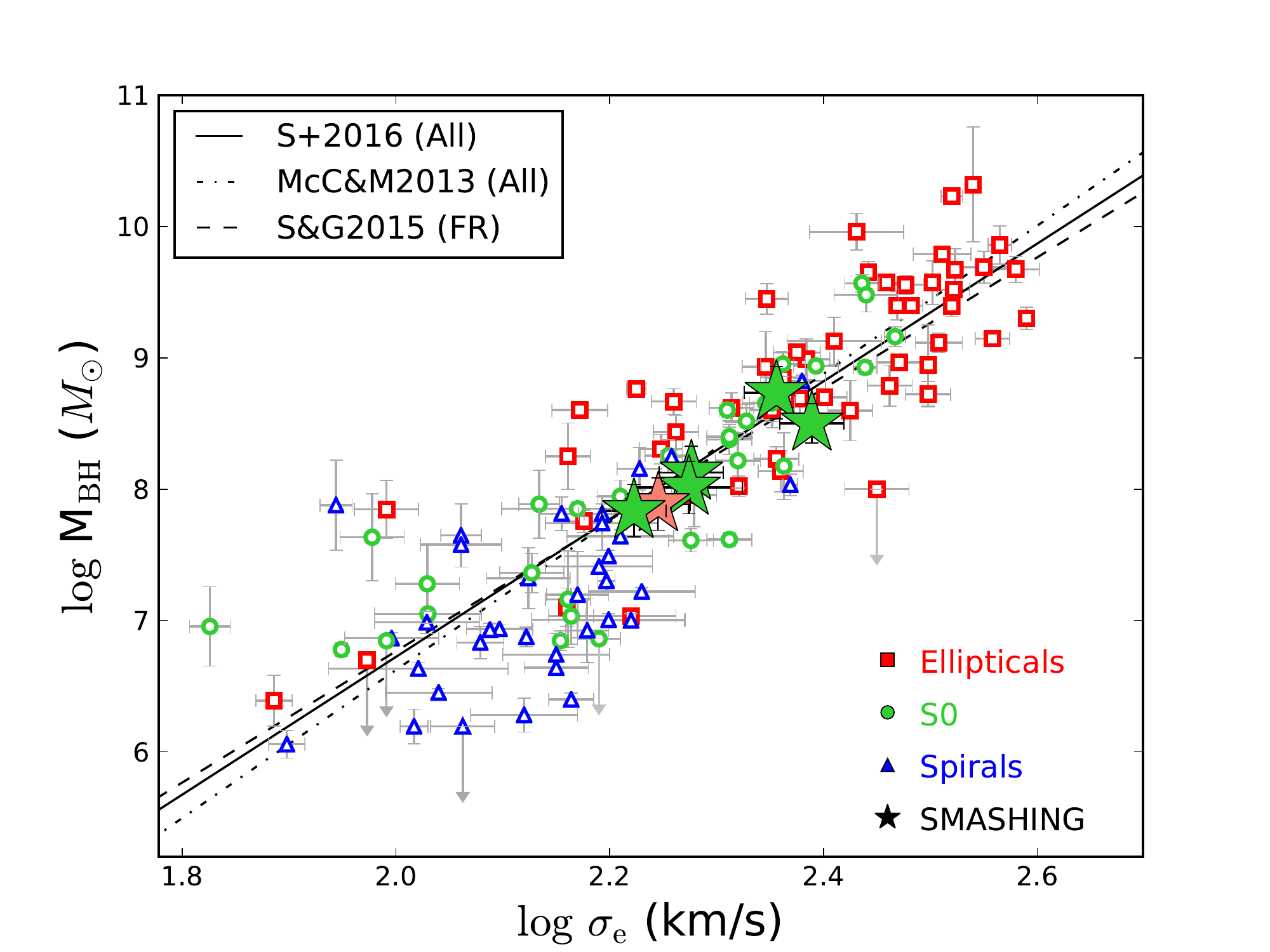}
      \caption{Supermassive black hole mass- effective velocity dispersion relation based on the compilation of \cite{Saglia2016}. The colour scheme indicates the morphological type of the galaxies: elliptical (red), lenticular (green) and spiral (blue). In order to visualize the general trend we have added the global scaling relations by \cite{Saglia2016} and \cite{McConnell2013} for all galaxy types (solid and dashed-dotted line) as well as \cite{Savorgnan2015} for fast-rotator galaxies (dashed-dashed line). Our measurements (highlighted as stars) lie very well on the two scaling relations. }
      \label{ff:scalingrelation}
\end{figure}

We populated the M$_{\mathrm{BH}}-\sigma_{\mathrm{e}}$ diagram with the compilation of dynamical black hole masses from \cite{Saglia2016} and \cite{Krajnovic2018}. We then added our derived Schwarzschild MBH measurements in combination with the bulge effective velocity dispersions from \cite{Cappellari2013b}(see Table~\ref{t:properties}). The diagram is shown in Figure~\ref{ff:scalingrelation}. Our measurements are located in the intermediate mass regime for early-type galaxies where the scatter is very tight. In Figure~\ref{ff:scalingrelation}, we furthermore show the scaling relations derived in \cite{McConnell2013}, \cite{Saglia2016} and \cite{Savorgnan2015}. All of our black hole mass measurements follow the black hole scaling relation closely. Except for NGC 4281, our measurements are slightly below the scaling relation (but within the $1\sigma$ scatter of the relation), NGC 7049 deviating slightly more from the scaling relation. 
\\
The massive black holes in NGC 584 and NGC 3640 have already been measured indirectly in the literature. In their study \cite{Dullo2014} recognized that NGC 584 and NGC 3640 show signatures for a partially depleted core which can be translated into a black hole mass of $M_{\mathrm{BH, dep}}=(1.95\pm 1.1) \times 10^8 M_{\odot}$ for NGC 584 and $M_{\mathrm{BH, dep}}=(9.77\pm 1.1) \times 10^7 M_{\odot}$ for NGC 3640. These values are consistent with our dynamical mass measurements of $M_{\mathrm{BH, dyn}}=(1.34 \pm 0.49) \times 10^8  M_{\odot}$ for NGC 584 and of $M_{\mathrm{BH, dyn}}=(7.73 \pm 0.51) \times 10^  M_{\odot}$ for NGC 3640. We also note that our dynamical mass measurements derived from the Jeans modeling match perfectly with the MBH masses derived from the depleted cores. 
\\\\
While our measurements follow the general trend of previous mass measurements, a systematic offset seems to emerge between MBHs in early-type and late-type host galaxies, the latter being significantly lower. \cite{Graham2013} note that this offset is also seen for barred versus non-barred galaxies (barred galaxies having larger velocity dispersion). The authors note that most of the late-type galaxies on the M$_{\mathrm{BH}}-\sigma_{\mathrm{e}}$ relation are actually barred, and it is not clear at this moment if the departure of the late-type galaxies from the scaling relation for the early-type galaxies is driven by bars or is typical for all late-type galaxies. Furthermore, black hole masses measured via $\rm H_{2}O$ megamasers \citep[e.g.,][]{Greene2010} in possibly barred galaxies also seem to be systematically lower than dynamical black hole mass measurements \citep[which is nicely visualized in Figure 1 of][ see also \citealt{Davis2018a,Davis2018b}]{Bosch2016}. In Section~\ref{ss:ml_variation} we have recognized that by taking into account a variable M/L the dynamical mass measurements could shift down by a factor of about 1.5. The radial M/L variation might even be more important for late-type galaxies, except for cases where the MBH is estimated by directly observing tracers within the MBH sphere of influence, such as H$_2$O megamasers.
We will investigate this implication in a future paper of this series. Independent from the M/L variation, together with the recently published dynamical mass measurements by \cite{Krajnovic2018} our measurements strengthen the idea of early-type galaxies having more massive black holes than (barred) late-type galaxies and thus following different M$_{\mathrm{BH}}-\sigma_{\mathrm{e}}$ relations \citep[e.g., ][]{Terrazas2017,Davis2018a,Martin-Navarro2018}.

\section{SUMMARY AND CONCLUSION}
In this work, we have presented the black hole mass measurement of six nearby early-type galaxies (NGC 584, NGC 2784, NGC 3640, NGC 4281, NGC 4570 and NGC 7049). Our measurements are based on AO-assisted K-band SINFONI IFU observations complemented by ground-based IFU data from MUSE, VIMOS, and SAURON from the ATLAS$^{\textrm{3D}}$ survey. All of our target galaxies show regular rotation and except for NGC 3640 and NGC 7049 a velocity dispersion increase towards their center. NGC 3640 contains a velocity dispersion dip of about 30 km/s, while NGC 7049 seems to have a flat velocity dispersion profile. This finding is consistent with the kinematic features of the large-scale SAURON data from the ATLAS$^{\textrm{3D}}$ survey and our VIMOS and MUSE data. We combined our kinematic results with photometric mass-models based on the composition of HST, SDSS and CGS survey data to build dynamical models to measure the mass of the central black holes. We constrained the parameter space of possible masses and mass-to-light ratios using Jeans Anisotropic Modelling on our SINFONI data and then created axisymmetric orbit-superposition Schwarzschild modeling based on both central and large-scale IFU data to derive robust results. We derive black hole masses of $(1.3\pm 0.5) \times 10^8 M_{\sun}$ for NGC 584, $(1.0\pm 0.6) \times 10^8 M_{\sun}$ for NGC 2784, $(7.7\pm 5) \times 10^7 M_{\sun}$ for NGC 3640, $(5.4 \pm 0.8) \times 10^8 M_{\sun}$ for NGC 4281, $(6.8\pm 2.0) \times 10^7 M_{\sun}$ for NGC 4570 and $(3.2\pm 0.8) \times 10^8 M_{\sun}$ for NGC 7049 which fit well with the recent black hole - $\sigma_{\mathrm{e}}$ scaling relations.
\\ \\
For three galaxies we find a slight discrepancy in the derived dynamical M/L of the two different methods, the central values being larger than the M/L derived from the combination of small-scale and large-scale data. Dynamical models typically assume a constant M/L for simplicity reasons which are usually not the case in observed galaxies. To test this assumption, we derive the stellar M/L profile from stellar population modeling for the test case of NGC 4570, which does not suffer from dust contamination and has the best quality data. The stellar population modeling shows a negative gradient of about 20\% within the effective radius of the galaxy which is based on variations in stellar age and metallicity. This negative gradient is then included in the dynamical Schwarzschild models, and we derive a black hole mass which is lower by almost 30\%, irrespective of further possible stellar M/L re-scaling due to radially constant IMF variations. We conclude that the inclusion of M/L variations has an effect of the order of the general uncertainty of the measurement, but it should be included in dynamical models to lower the systematic uncertainties which are still very large in dynamical modeling. As was already suspected in different works, this has an interesting implication on the black hole scaling relations as different dynamical methods still suffer from partially inconsistent results. We caution that careful additional analysis of the effect of M/L variations on dynamical models is urgently needed in future studies.    

\section*{Acknowledgements}
The authors want to thank Aaron Barth of the Department of  Physics and Astronomy, University of California, Irvine, for sharing the large scale imaging data of NGC 584, NGC 2784 and NGC 7049 from the CGS survey with us. MC acknowledges support from a Royal Society University Research Fellowship.\\
\\
Based  on  observations  collected  at  the  European  Organisation  for  Astronomical  Research  in  the Southern Hemisphere under ESO programme 075.B-0495(A), 078.B-0464(B), 079.B-0402(B), 080.B-0015(A), 085.B-0221(A), 091.B-0129(A), 097.A-0366(B), 291.B-5019(A). 
Based on observations made with the NASA/ESA Hubble Space Telescope, obtained from the Data Archive at the Space Telescope Science Institute, which is operated by the Association of Universities for Research in Astronomy, Inc., under NASA contract NAS 5-26555. These observations are associated with program 
This research has made use of the NASA/IPAC Extragalactic Database which is operated by the Jet Propulsion Laboratory, California Institute of Technology, under contract with NASA.


\bibliography{papers}
\bibliographystyle{aa}

\newpage
\begin{appendix}

\section{Dust correction and masking}\label{s:dustcorrection}
\begin{figure}
  \centering
    \includegraphics[width=0.45\textwidth]{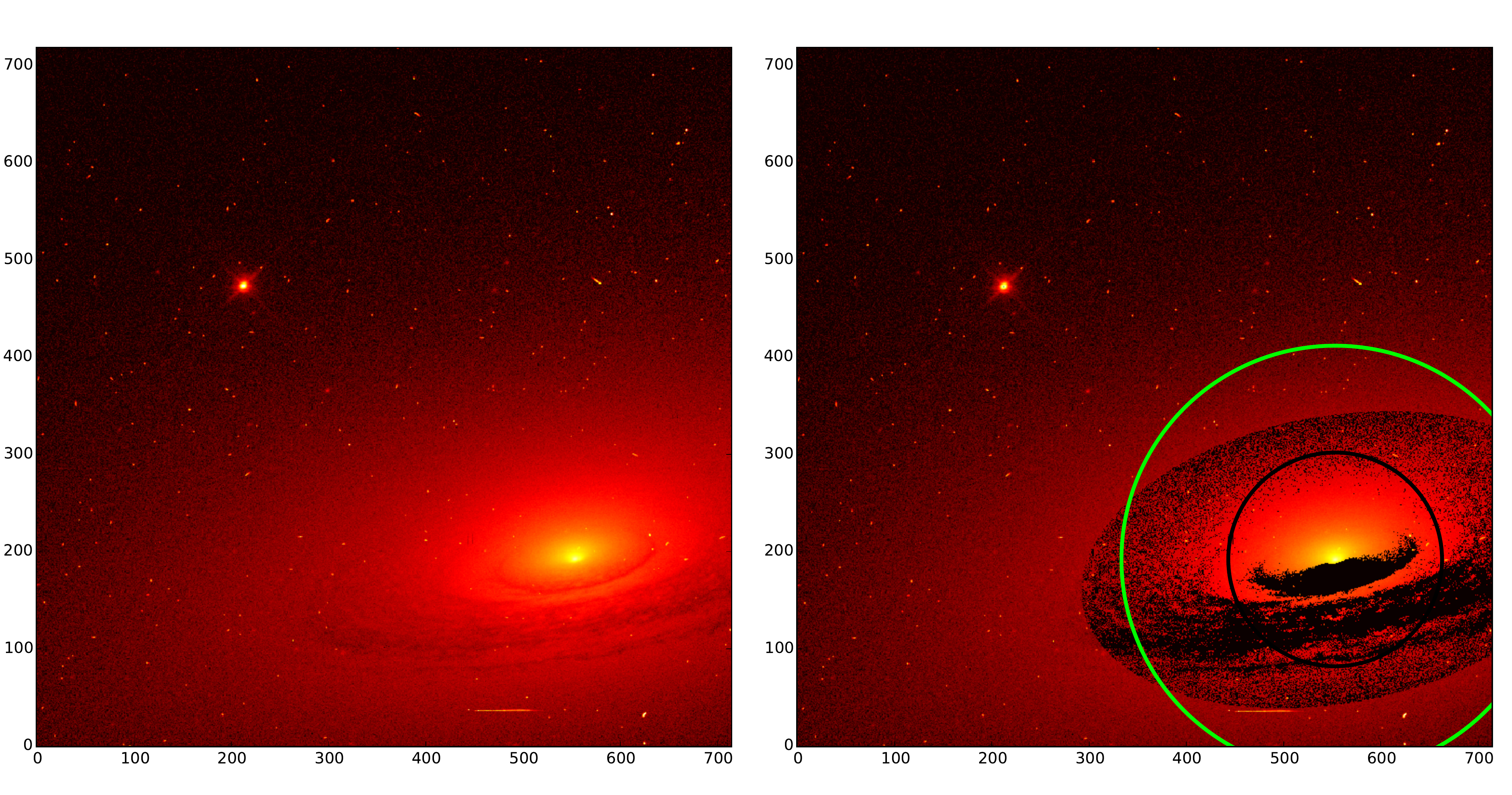}
    \includegraphics[width=0.45\textwidth]{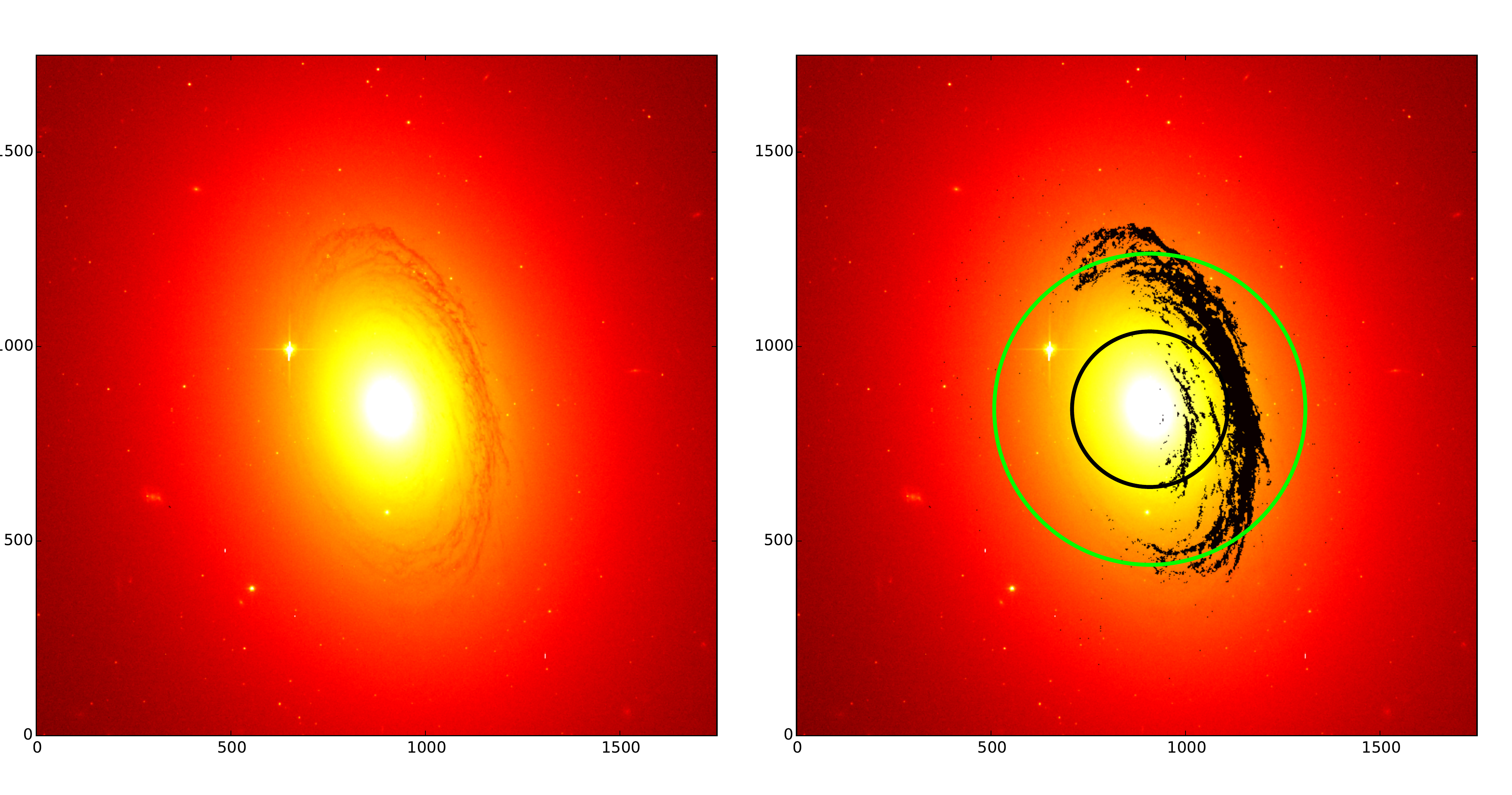}
      \caption{Dust-masked region of the HST WFPC2 and ACS images of NGC 4281 (upper panels) and NGC 7049 (lower panels), respectively. The left panels show the original HST images, the right panels show the same image overplotted with the dust mask (black) and circular regions with $r<5''$ (black) and $r<10''$ (green) which encompass the region being used for the MGE modeling. }
      \label{ff:dustcorrection1}
\end{figure}

As the presence of dust can have a crucial effect on the galaxy mass modeling, we had to correct and mask the dust polluted image pixels before constructing the mass models. NGC 4281 and NGC 7049 contain extended nuclear dust rings which are well visible in the HST images (see Figure~\ref{ff:dustcorrection1}). Furthermore, we found a small dust ring in the HST image of NGC 2784. As carefully tested in \cite{Thater2017} we applied a dust mask for the HST small scale images for all three galaxies and a dust correction for the SDSS large scale image of NGC 4281.

\subsection{HST images}
For NGC 2784, NGC 4281 and NGC7049 the nuclear dust ring reaches into the very central regions of the galaxies. It was therefore necessary to also correct for dust in the HST small scale images. The careful dust correction in the HST images is crucial as we probe the direct vicinity of the black hole with them. However, due to the lack of HST images in different bands, we applied the dust masking method which was developed in \cite{Thater2017}. The method is based on the assumptions 1) that the presence of dust attenuates the light emitted within the galaxy and 2) the major part of the galaxy surface brightness is not significantly affected by dust. In order to derive the dust affected images we fitted the lower envelope of the characteristic surface brightness profile with an appropriate function (four parameter logistic function). Masked were all pixels which had a surface brightness below this fit. This method corrects patchy dust structures which can be clearly distinguished from the unattenuated regions, but misses thin dust structures and dust layers (see Figure~\ref{ff:dustcorrection1}).

\subsection{Large scale images}
\begin{figure}
  \centering
    \includegraphics[width=0.5\textwidth]{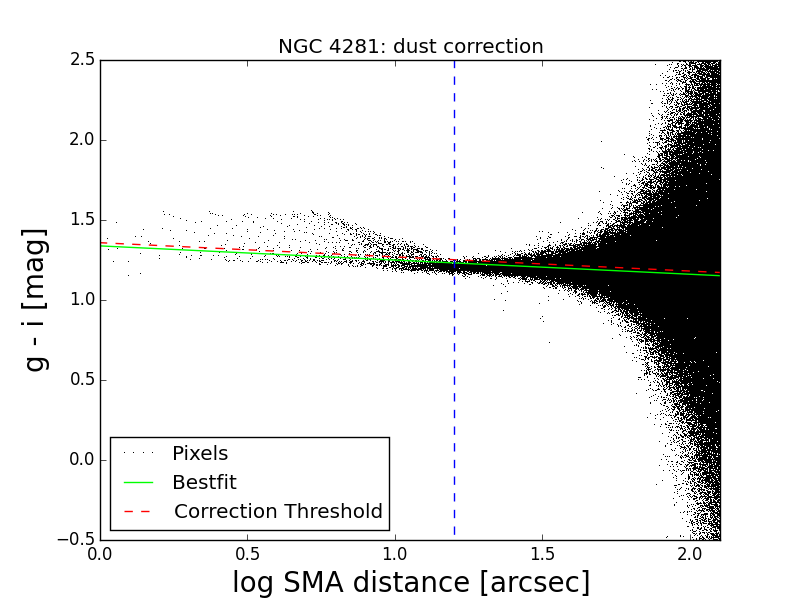}
    \includegraphics[width=0.5\textwidth]{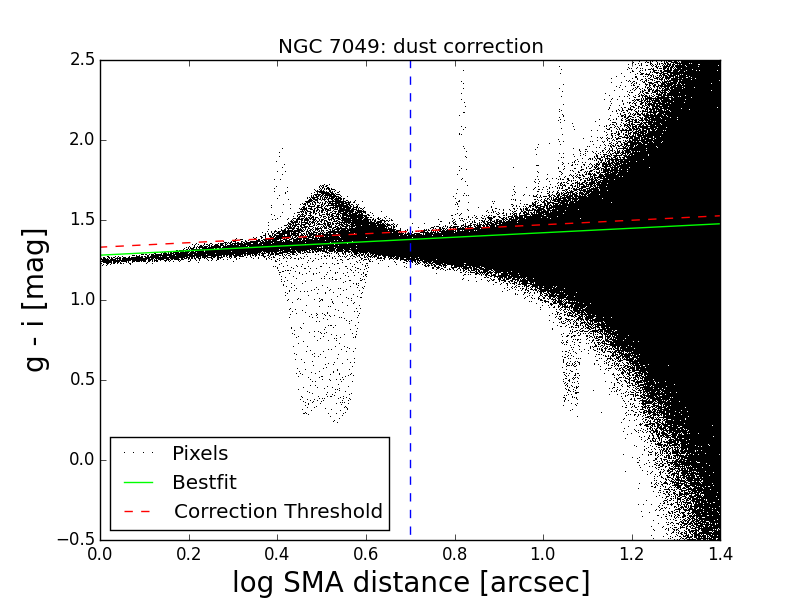}
      \caption{Colour profile of the NGC 4281 and NGC 7049 SDSS (g-i) images used for the dust correction. In the upper left quadrant one can clearly identify the central dust rings. The best-fitting linear function obtained from a robust fit is shown in green, the slightly shifted red line is the correction threshold. For NGC 4281, all pixels above the red line and for $\log(d)<1.2 (\approx 16)$ [arcsec] were corrected for dust extinction.}
      \label{ff:dustcorrection}
\end{figure}
\begin{figure}
  \centering
    \includegraphics[width=0.5\textwidth]{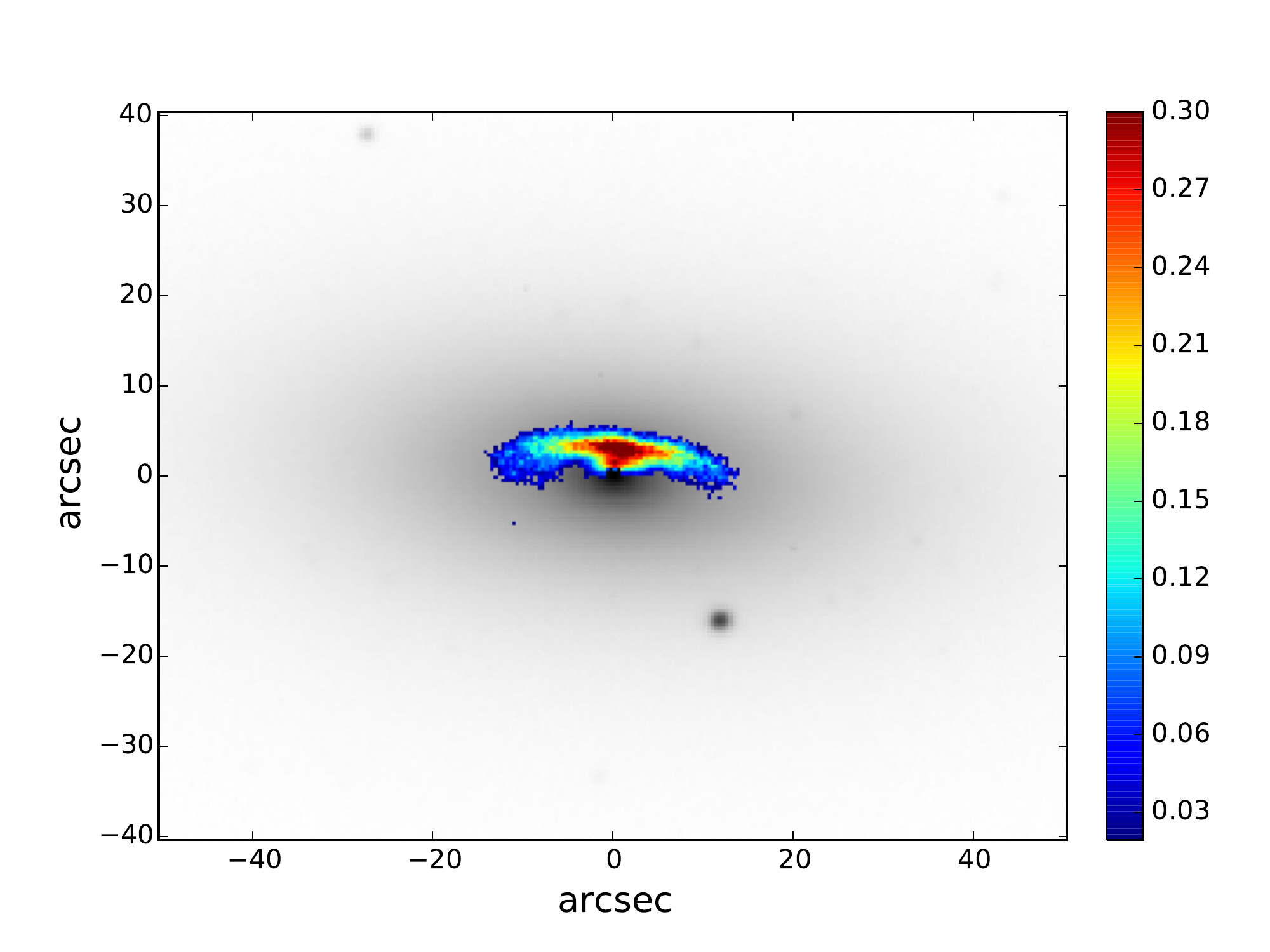}
      \caption{Dust-corrected central region of the SDSS r-band
image of NGC~4281. The correction was only applied within a major axis distance of 16$''$
(see caption of Fig.\ref{ff:dustcorrection}). The over-plotted colour coding indicates the degree of the dust correction where 0.1 means that the observed flux increased by 10 \%. }
      \label{ff:dustcorrection2}
\end{figure}

In order to correct the dust attenuation in the SDSS and CGS images we applied a method that was developed by \cite{Carollo1997a} and applied and further advanced by \cite{Cappellari2002b}, \cite{Krajnovic2005} and \cite{Scott2013}. The main assumption is that dust between the observer and the stellar emission can be folded into a screen which dims the observed light wavelength-dependent. Due to the extinction, the observer sees a change in colour in the dust-affected regions of the galaxy which are assumed to have the same intrinsic colour as their adjacent areas. Using the Galactic extinction law \citep{Schlegel1998} we derived the r-band extinction for NGC~4281 from the colour excess between the g- and i-band images $A_r=1.15\,E(g-i)$ 
. In order to derive the colour excess we created the respective colour profile as function of the
logarithm of the semi-major axis distance along ellipses with fixed position angle and ellipticity. The colour profile of NGC 4281 is presented in Figure \ref{ff:dustcorrection} showing a very slight colour gradient with radius and the signatures of the nuclear dust ring in the reddened pixels above. We fitted a robust linear function to the colour gradient to reduce the influence of the dust-affected pixels. The colour excess $E\,(g/r -i)$ of each pixel is now computed by subtracting its assumed intrinsic colour value (approximated by the linear fit) from its measured value. All pixels above an arbitrary threshold value, here chosen to be $E\,(g - i) > 0.02$ mag for NGC 4281, are assumed to have a significant dust extinction and were corrected using the Galactic extinction law. Figure \ref{ff:dustcorrection2} also shows quantitatively how much of the measured
flux was corrected (where 0.1 means 10\% (blue) and 0.25 means 25\% (orange)). The largest correction was approximately 35\% of the measured flux. Figure~\ref{ff:dustcorrection} also shows the colour profile of NGC 7049. Based on this plot and an additional visual check of the image we realized that the dust content in this galaxy is mostly concentrated in the centre (within the HST PC FoV)  and we decided to only apply the dust-masking of the HST images.

\section{Determination of the HST, NIFS and VIMOS PSF}

\subsection{HST spatial resolution}\label{ss:hst_psf}
We generated a PSF image for the HST WFPC2/PC and ACS images using the online available Tiny Tim HST Modelling tool\footnote{http://www.stsci.edu/hst/observatory/focus/TinyTim} \citep{Krist2001} taking into account the imaging filter, the central position of the galaxy on the CCD chip and assuming the spectrum of a K giant star.
While we could use the HST WFPC2 images as they were provided by Tiny Tim, the procedure was slightly more complicated for the ACS images. The ACS camera has the disadvantage to be mounted away from the optical axis of HST which distorts the observed image significantly giving it the shape of a sheared rectangle (TinyTim manual\footnote{http://tinytim.stsci.edu/static/tinytim.pdf}). Distortion is not corrected by the internal optics, but must be accounted for during the data-reduction, both for the science image and the PSF image. In order to do so, we used a method applied by \cite{Rusli2013}. We substituted the distorted PSF image into the flat-fielded galaxy ACS image, replacing the centre of the galaxy. We then corrected the full substituted ACS image using the drizzelpac task in Astroconda with the same input parameters as during the science image data reduction. We then cut out the PSF image from the resulting distortion-corrected ACS mosaic and applied the same following procedures as for the WFPC2 images.  
\\
The resulting PSF image was parametrized by a sum of normalized concentric circular Gaussians using the MGE method \citet{Cappellari2002}. All MGE parameters and relative weights of each Gaussian and for each galaxy are given in Table~\ref{t:hst_psf}.

\begin{table*}\label{t:hst_psf}
\caption{MGE parametrisation of the HST PSF}
\centering
\begin{tabular}{lc|lc|lc|lc|lc|ll}
\hline\hline
N584 &  &N2784&  & N3640 &   & N4281 &   & N4570 & & N7049 & \\
WFPC2 & F555W & WFPC2 & F547M & WFPC2  & F555W & WFPC2 & F606W & WFPC2 & F555W & ACS & F814W\\
\hline
norm & $\sigma$ & norm & $\sigma$ & norm & $\sigma$ & norm & $\sigma$ & norm & $\sigma$ & norm & $\sigma$  \\
 & (arcsec) &  & (arcsec) &  & (arcsec) &  & (arcsec) &  & (arcsec) &  & (arcsec)  \\
(1) & (2) & (1) & (2) &(1) & (2) &(1) & (2) &(1) & (2) &(1) & (2) \\
\hline \hline
0.1972 & 0.0173 & 0.2430 & 0.0173 & 0.1951 & 0.0173 & 0.1995 & 0.0173 & 0.1956 & 0.0173 & 0.1517& 0.0284\\
0.5832 & 0.0484 & 0.5591 & 0.0463 & 0.5847 & 0.0475 & 0.5631 & 0.0489 & 0.5863 & 0.0479 & 0.6483& 0.0649\\
0.0923 & 0.1251 & 0.0902 & 0.1186 & 0.0960 & 0.1200 & 0.0498 & 0.1158 & 0.0927 & 0.1258 & 0.0983& 0.1513\\
0.0676 & 0.3116 & 0.0685 & 0.3012 & 0.0687 & 0.3075 & 0.0680 & 0.1477 & 0.0668 & 0.3080 & 0.0620& 0.4047\\
0.0186 & 0.4724 & 0.0391 & 0.8523 & 0.0144 & 0.5062 & 0.0782 & 0.3278 & 0.0172 & 0.4619 & 0.0166& 0.8361\\
0.0409 & 0.8753 &  -- &  -- & 0.0411 & 0.8630 & 0.0414 & 0.8395 & 0.0414 & 0.8506 & 0.0231& 1.6155\\
\hline \hline
\end{tabular}
\\
\tablefoot{Specifics of the single Gaussians from the MGE parametrisation of the HST image PSF for each galaxy. The first columns show the normalised relative weights of each Gaussian and the second columns show the dispersion $\sigma$ of each Gaussian (converted into arcsec), respectively. }
\label{ff:hst_psf}
\end{table*}

\subsection{SINFONI and VIMOS spatial resolution}
\begin{figure*}
  \centering
  \vspace{0.1cm}
    \includegraphics[width=0.9\textwidth]{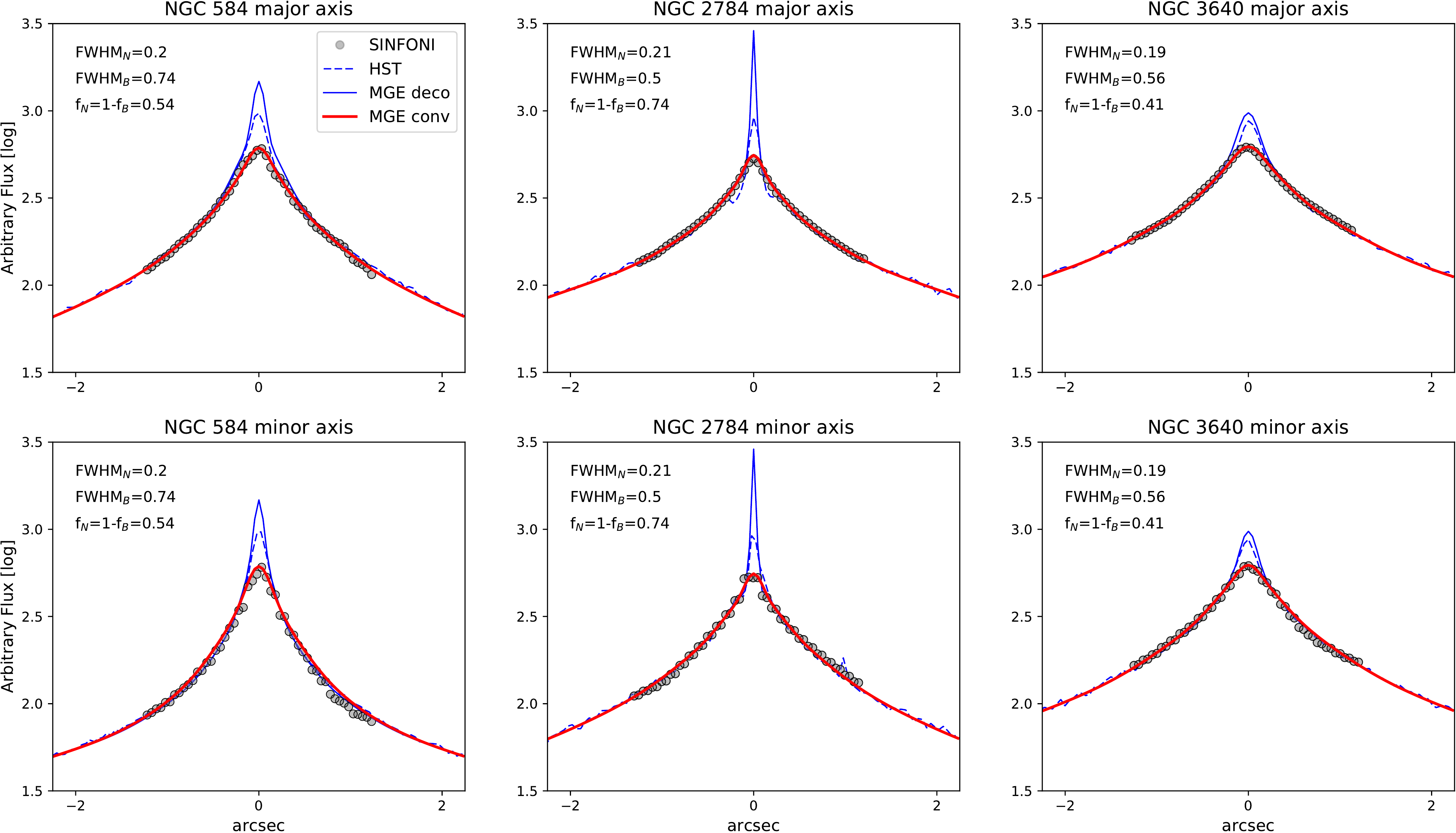}
    
    \vspace{1cm}
    
    \includegraphics[width=0.9\textwidth]{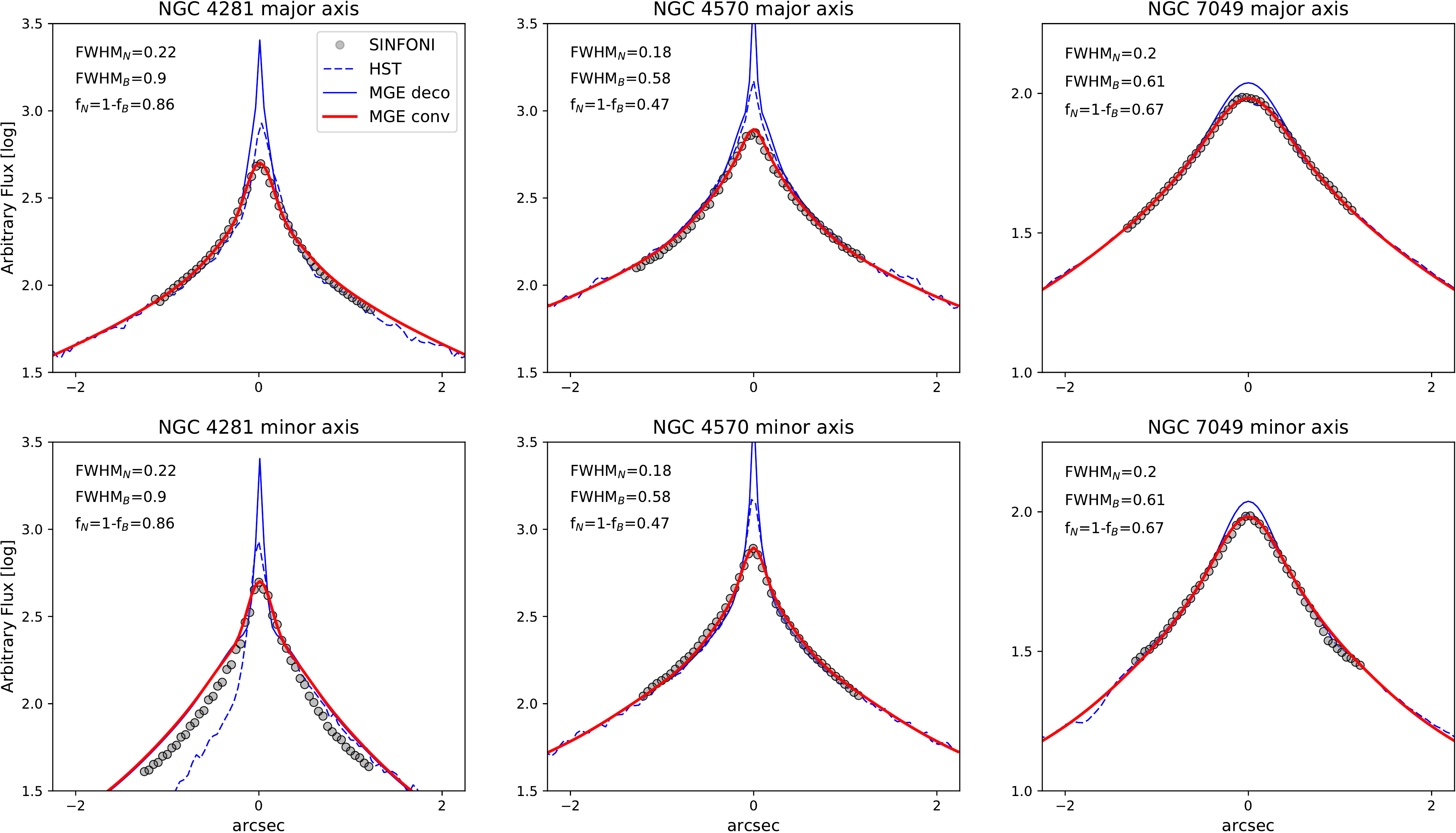}
   \vspace{0.1cm}
      \caption{Determination of the SINFONI AO spatial resolution by comparing the surface brightness from the SINFONI reconstructed images with the respective convolved MGE models. Shown are the surface brightness profiles along the galaxy semi major (top panel) and minor axis (bottom panel) of the SINFONI IFU image(red circles), the deconvolved MGE model (green dashed line), the convolved MGE model (black solid line) and the HST image (blue dashed line) used to create the MGE model. The light profiles of NGC 2784, NGC 4281 and NGC 7049 show clear signatures of nuclear dust. Before comparing the profiles, all images were rotated such that the major and minor axis would match the vertical and horizontal image axis. The parameters of the double Gaussians used to describe the SINFONI PSFs are given in the upper left corner.
      }
      \label{ff:psf_determination}
\end{figure*}

The spatial resolution of the IFU data sets a limit on the scales that can be probed with the central dynamics of our target galaxies. As such the resolution provides a quality argument of our black hole mass measurements which has to be carefully evaluated. When no point source is present in the FoV, the reconstructed IFU images need to be compared with reference images of significantly higher resolution. We used HST imaging data (used for the MGE models) in order to determine the PSF of the SINFONI observations (e.g. \cite{Shapiro2006,Davies2008,Krajnovic2009a}. Deconvolved HST images (a product from the MGE modelling) are convolved with a PSF (and such degraded) until they match the collapsed IFU image. A simple parametrisation of the SINFONI PSF can be obtained using a circular and concentric double Gaussian composed of a narrow and broad component of different relative weights. In order to compare the different images, they were aligned by rotating the images such that the major and minor axes would coincide with the vertical and horizontal image axes. Furthermore, after the convolution the MGE model was re-binned to the same pixel scale as the respective IFU observation. The best fitting parameters were found by minimising the residual between the convolved MGE model and the reconstructed SINFONI image using the python routine mpfit.    
\\
We furthermore performed a number of different tests with the PSF determination routines which involved changing the flux scales between the HST and collapsed IFU image, varying the rotation angle, the size of the fitted image and the kernel size. This led to additional systematic uncertainties of about 0.05 arcsec for the narrow Gaussian component. A comparison of the main axes between the SINFONI light profiles and convolved MGE models is presented in Fig.~\ref{ff:psf_determination}. Even though we probe very different wavelength regions, the fits agree remarkably well. The best-fitting parameters of the PSF fits are also given in Fig.~\ref{ff:psf_determination} and summarized in Table~\ref{t:spatialres}.\\

\subsection{Strehl ratio}\label{ss:strehl}
The  Strehl  ratio  is an indication of the  effect  of  wavefront  aberrations on the optical quality of the observations. It can be determined by calculating the ratio between the peak intensity of the measured PSF and the peak intensity of the ideal diffraction limited PSF assuming an ideal working LGS AO. We obtained the FWHM of the narrow Gaussian component of the SINFONI observation from the PSF determination (see Table~\ref{t:spatialres}). The diffraction limit of the telescope can be calculated as $\theta=1.22\times \lambda/D $ where $\lambda$ specifies the wavelength of the observed light and $D$ is the diameter of the primary mirror. At 2.3 microns, the diffraction limit of the VLT telescope (D= 8.2 m) is approximately 0.07~\arcsec. We created normalized 2D Gaussians with the resolution of the SINFONI observation and the diffraction limit. Dividing the peak intensities resulted into Strehl ratios of around 10 \% for our SINFONI observations (see Table~\ref{t:spatialres}).

\section{MGE parametrisation of target galaxies}
\begin{table*}
\caption{MGE parameters}
\centering
\begin{tabular}{l|cccc|cccc|cccc}
\hline\hline
&N584  &(F555W)  & & &N2784  &(F547M) & & &N3640   &(F555W) & &\\
\hline
j & log $M_j$ & log $I_j$ & $\sigma_j$ & $q_j$& log $M_j$ & log $I_j$ & $\sigma_j$ & $q_j$ & log $M_j$ & log $I_j$ & $\sigma_j$  & $q_j$\\
 & ($M_{\sun}$) & ($L_{\sun} pc^{-2}$) & (arcsec) & & ($M_{\sun}$) & ($L_{\sun} pc^{-2}$) & (arcsec) & & ($M_{\sun}$) & ($L_{\sun} pc^{-2}$) & (arcsec) & \\
(1) & (2) & (3) & (4) & (5) & (2) & (3) & (4) & (5) & (2) & (3) & (4) & (5) \\
\hline \hline
1 & 7.746 & 4.891 & 0.055 & 0.80 & 7.421 & 4.981 & 0.061 & 0.80 & 7.719 & 4.239 & 0.094 & 0.80 \\
2 & 8.371 & 4.496 & 0.201 & 0.64 & 8.100 & 4.313 & 0.287 & 0.80 & 8.403 & 3.999 & 0.259 & 0.90 \\
3 & 8.805 & 4.168 & 0.492 & 0.61 & 8.802 & 4.273 & 0.674 & 0.80 & 9.074 & 3.850 & 0.677 & 0.86 \\
4 & 9.137 & 3.872 & 1.014 & 0.61 & 9.169 & 3.998 & 1.413 & 0.80 & 9.542 & 3.620 & 1.553 & 0.82 \\
5 & 9.225 & 3.468 & 1.544 & 0.82 & 8.094 & 2.901 & 1.930 & 0.45 & 9.698 & 3.238 & 2.969 & 0.78 \\
6 & 9.625 & 3.399 & 2.667 & 0.81 & 9.403 & 3.690 & 2.635 & 0.80 & 10.074 & 3.052 & 5.736 & 0.76 \\
7 & 9.944 & 3.137 & 5.089 & 0.84 & 9.795 & 3.602 & 4.581 & 0.80 & 10.237 & 2.683 & 10.588 & 0.76 \\
8 & 9.759 & 2.666 & 8.320 & 0.61 & 9.993 & 3.043 & 10.944 & 0.80 & 10.222 & 2.408 & 13.376 & 0.86 \\
9 & 10.231 & 2.764 & 12.257 & 0.66 & 9.293 & 1.830 & 19.749 & 0.80 & 10.274 & 2.074 & 22.256 & 0.76 \\
10 & 10.304 & 2.210 & 26.342 & 0.61 & 10.042 & 2.785 & 20.772 & 0.45 & 10.345 & 1.807 & 29.307 & 0.95 \\
11 & 10.366 & 1.748 & 48.130 & 0.61 & 10.376 & 2.215 & 58.832 & 0.45 & 10.360 & 1.295 & 60.314 & 0.76 \\
12 & 10.274 & 1.046 & 77.768 & 0.95 & 9.182 & 0.355 & 94.984 & 0.80 & 10.676 & 1.286 & 78.209 & 0.95 \\
13 &--& -- & -- & -- & 10.204 & 1.614 & 96.422 & 0.45 & -- & -- & -- & -- \\
\hline \hline
\end{tabular}

\vspace{0.3cm}
\begin{tabular}{lcccc|cccc|cccc}
\hline\hline
& N4281 &(F606W) & & &N4570 &(F555W) & & &N7049  &(F814W) &\\
\hline
j & log $M_j$ & log $I_j$ & $\sigma_j$ & $q_j$& log $M_j$ &log $I_j$ & $\sigma_j$ & $q_j$ & log $M_j$ &log $I_j$ & $\sigma_j$ & $q_j$\\
 & ($M_{\sun}$) & ($L_{\sun} pc^{-2}$) & (arcsec) & & ($M_{\sun}$) & ($L_{\sun} pc^{-2}$) & (arcsec) & & ($M_{\sun}$) & ($L_{\sun} pc^{-2}$) & (arcsec) & \\
 (1) & (2) & (3) & (4) & (5) &(2) & (3) & (4) & (5) & (2) & (3) & (4) & (5) \\
\hline \hline
1 & 7.583 & 5.364 & 0.017 & 0.67 & 7.276 & 6.043 & 0.017 & 0.30 & 9.014 & 3.825 & 0.363 & 0.75 \\
2 & 8.472 & 4.907 & 0.086 & 0.60 & 7.738 & 4.647 & 0.096 & 0.70 & 9.744 & 3.680 & 0.981 & 0.77 \\
3 & 9.043 & 4.159 & 0.353 & 0.75 & 5.773 & 2.611 & 0.104 & 0.70 & 10.201 & 3.315 & 2.517 & 0.77 \\
4 & 9.513 & 3.904 & 0.812 & 0.75 & 7.063 & 3.709 & 0.199 & 0.30 & 10.595 & 2.918 & 6.396 & 0.74 \\
5 & 9.762 & 3.616 & 1.663 & 0.62 & 8.184 & 4.248 & 0.254 & 0.70 & 10.583 & 2.254 & 13.402 & 0.76 \\
6 & 9.815 & 3.127 & 3.560 & 0.47 & 7.713 & 4.071 & 0.277 & 0.30 & 10.861 & 2.110 & 21.652 & 0.77 \\
7 & 10.098 & 2.947 & 5.139 & 0.65 & 8.683 & 4.055 & 0.565 & 0.70 & 10.812 & 1.483 & 42.830 & 0.74 \\
8 & 10.338 & 2.552 & 12.854 & 0.45 & 9.167 & 3.862 & 1.230 & 0.70 & 10.979 & 0.914 & 94.544 & 0.83 \\
9 & 10.565 & 2.187 & 25.399 & 0.45 & 9.676 & 3.623 & 2.911 & 0.70 & 9.895 & 0.679 & 169.678 & 0.83 \\
10 & 10.238 & 1.371 & 44.588 & 0.45 & 8.024 & 1.938 & 4.617 & 0.30 & -- &-- & -- & --  \\
11 & 10.432 & 1.219 & 53.51 & 0.69 & 9.575 & 2.893 & 6.005 & 0.70 & -- &-- & -- & --  \\
12 & -- &-- & -- & -- & 9.670 & 2.720 & 12.491 & 0.30 & -- &-- & -- & --  \\
13 & -- &-- & -- & -- & 9.618 & 2.436 & 16.321 & 0.30 & -- &-- & -- & -- \\
14 & -- &-- & -- & -- & 10.059 & 2.431 & 27.262 & 0.30 & -- &-- & -- & -- \\
15 & -- &-- & -- & -- &  9.899 & 1.700 & 52.604 & 0.30 & -- &-- & -- & --  \\
16 & -- &-- & -- & -- &  9.573 & 0.865 & 61.819 & 0.70 & -- &-- & -- & --  \\
\hline \hline
\end{tabular}
\\
\tablefoot{Details of the MGE parametrisation for each galaxy. We show the number of the Gaussian component (1), the total mass (2), the surface brightness in the specified band (3), the velocity dispersion (4) and the axial ratio (5) for each Gaussian component. The dynamical M/L from the Schwarzschild models (Table~\ref{t:results}) was used to determine the mass of each Gaussian component.}
\label{t:mge}
\end{table*}
\section{Large scale kinematics}

\begin{figure*}
  \centering
    \includegraphics[width=1.\textwidth]{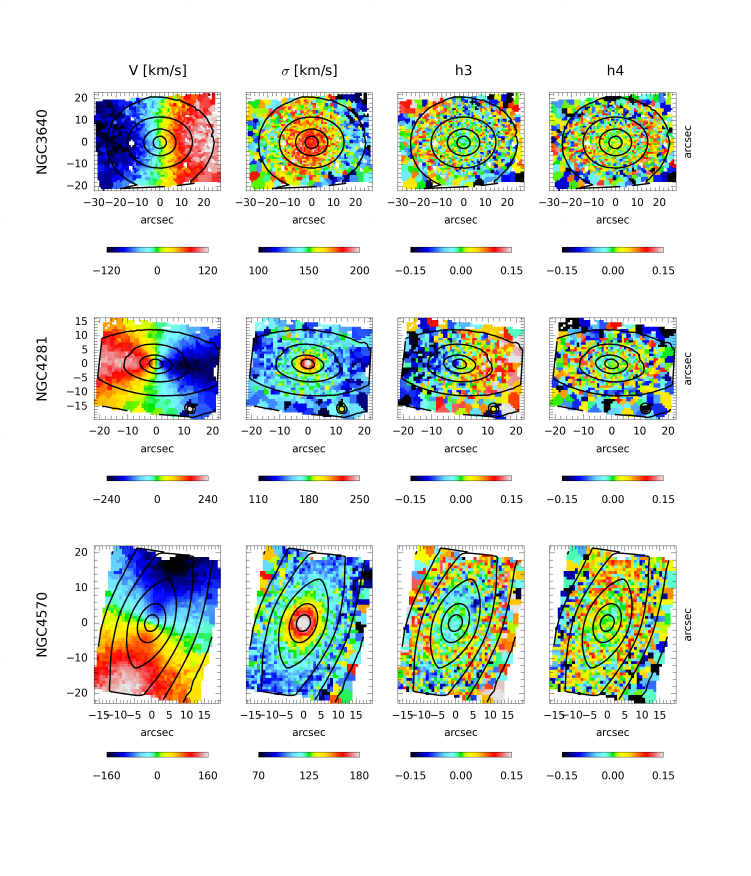}

      \caption{Large scale SAURON stellar kinematics of NGC 3640, NGC 4281 and NGC 4570. Shown are the mean velocity V, velocity dispersion $\sigma$ and the $h_{3}$ and $h_4$ Hermite polynomials extracted by using pPXF. The galaxies are part of the ATLAS$^{\textrm{3D}}$ project and detailed described in \citet{Cappellari2011}. The image orientation is such that north is up and east is left.}
      \label{ff:atlas3d_kinematics}
\end{figure*}

\begin{figure*}
  \centering
    \includegraphics[width=1.\textwidth]{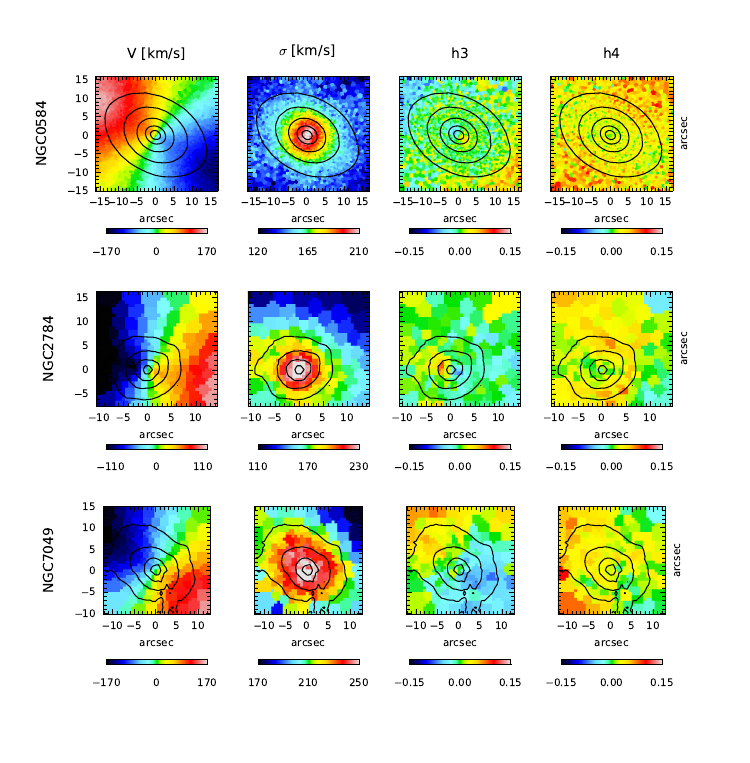}

      \caption{Large scale stellar kinematics of NGC 584 (MUSE), NGC 2784 (VIMOS) and NGC 7049 (VIMOS). Shown are the mean velocity V, velocity dispersion $\sigma$ and the $h_{3}$ and $h_4$ Hermite polynomials extracted by using pPXF. The data reduction and the extraction of the kinematics are detailed described in Section~\ref{ss:vimoskinematics}. The image orientation is such that north is up and east is left.}
      \label{ff:vimos_kinematics}
\end{figure*}

\section{JAM models}\label{ss:jam_models}

\begin{figure*}[!htb]
  \centering
        \includegraphics[width=0.42\textwidth]{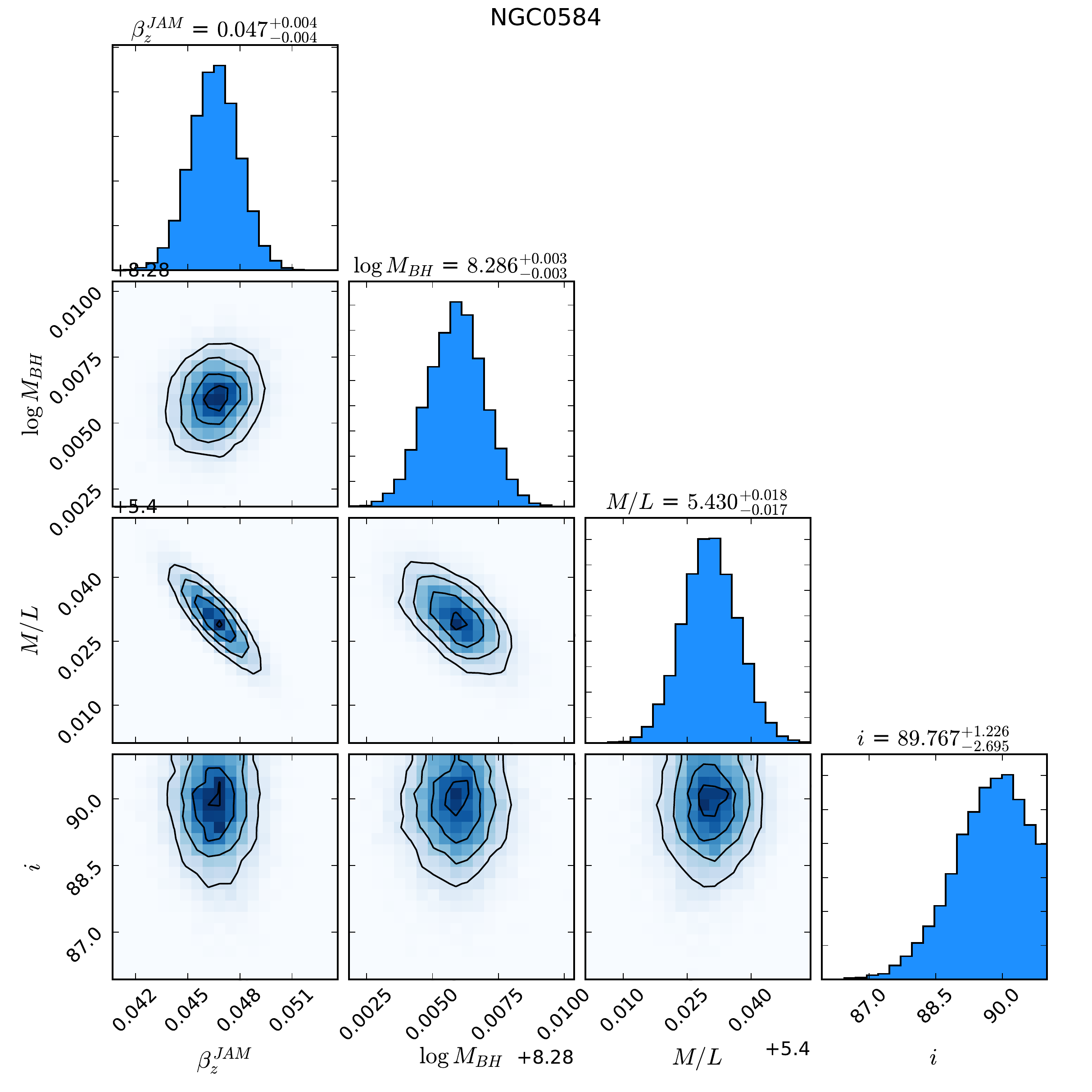}
        \includegraphics[width=0.42\textwidth]{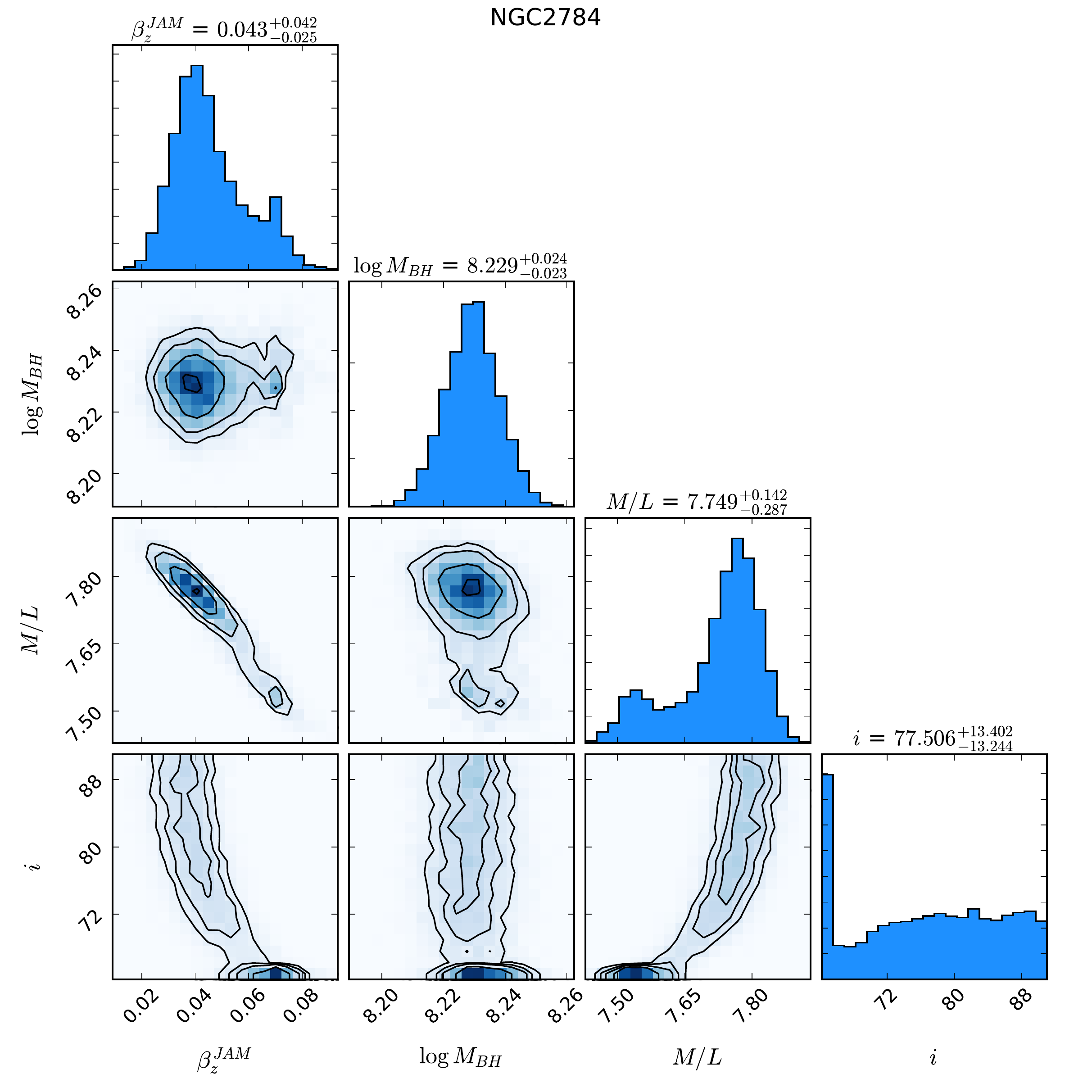} 
        \includegraphics[width=0.42\textwidth]{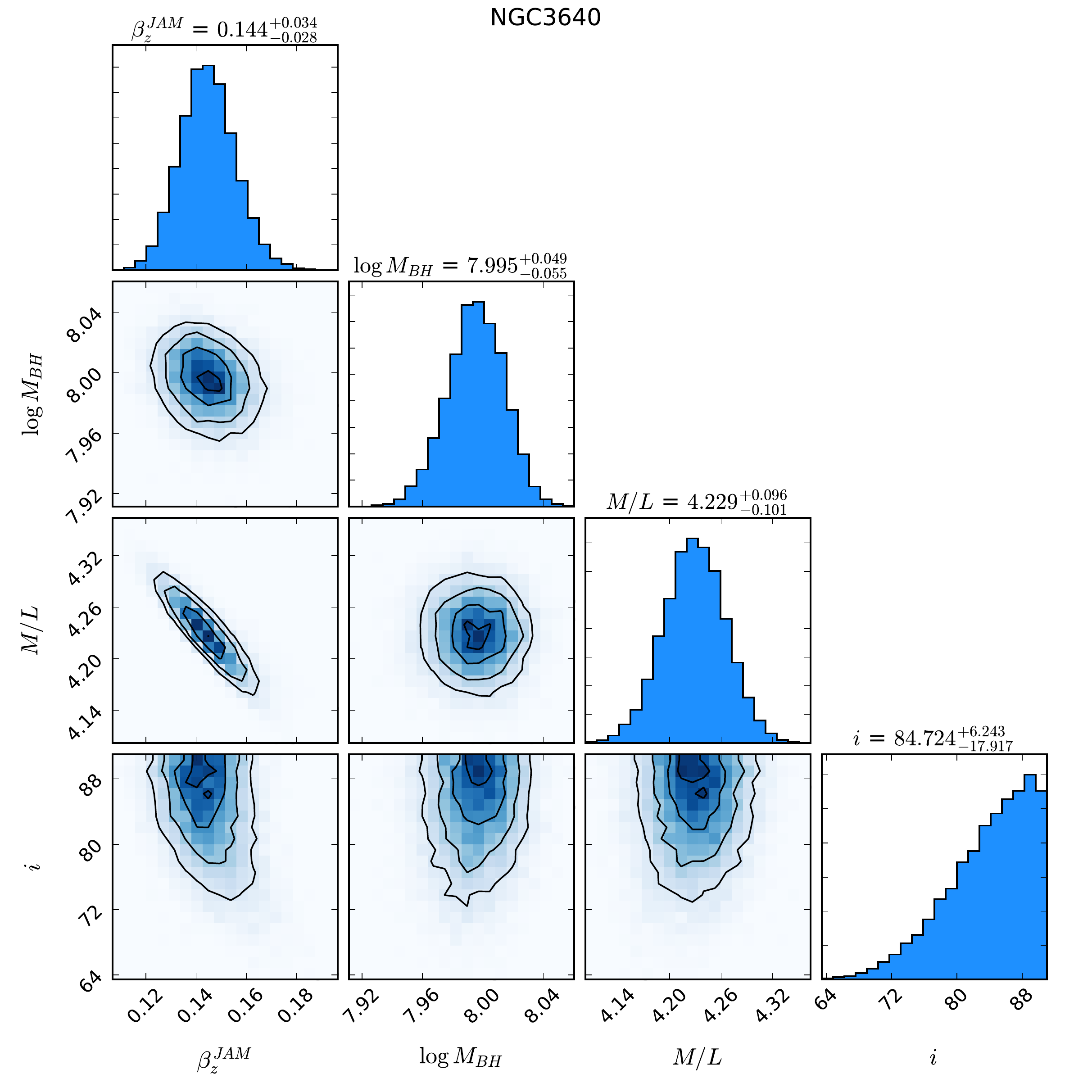} 
        \includegraphics[width=0.42\textwidth]{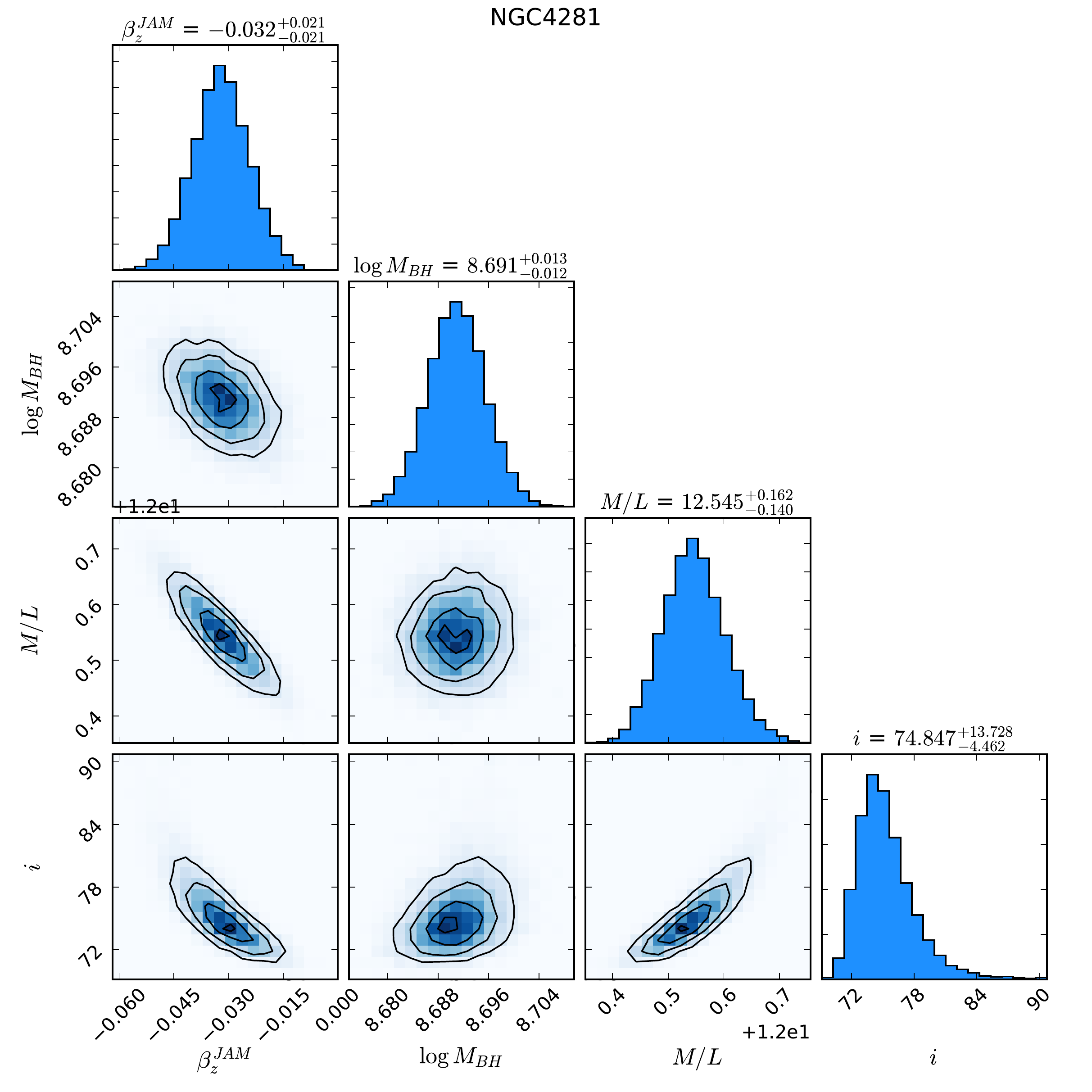}  
        \includegraphics[width=0.42\textwidth]{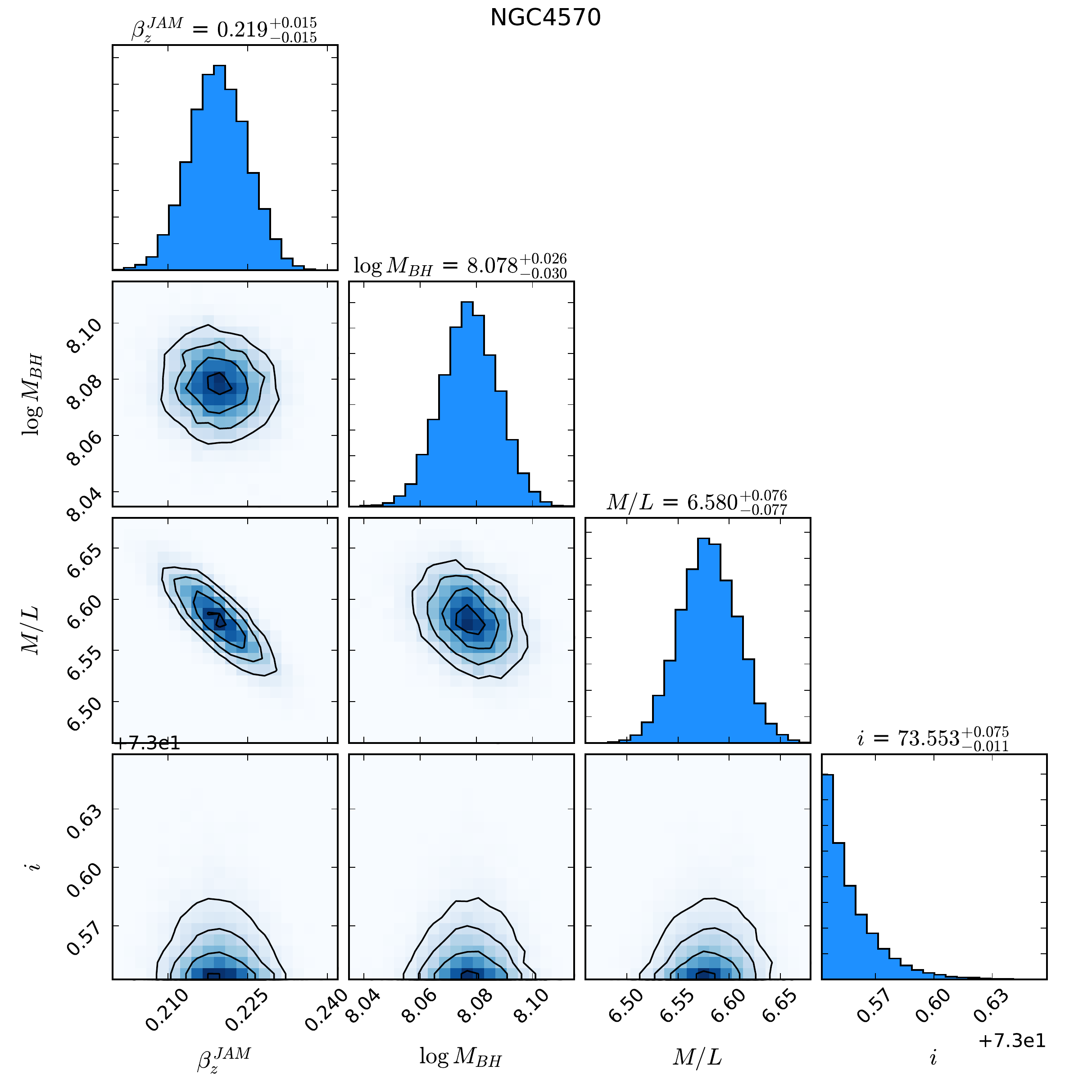}  
        \includegraphics[width=0.42\textwidth]{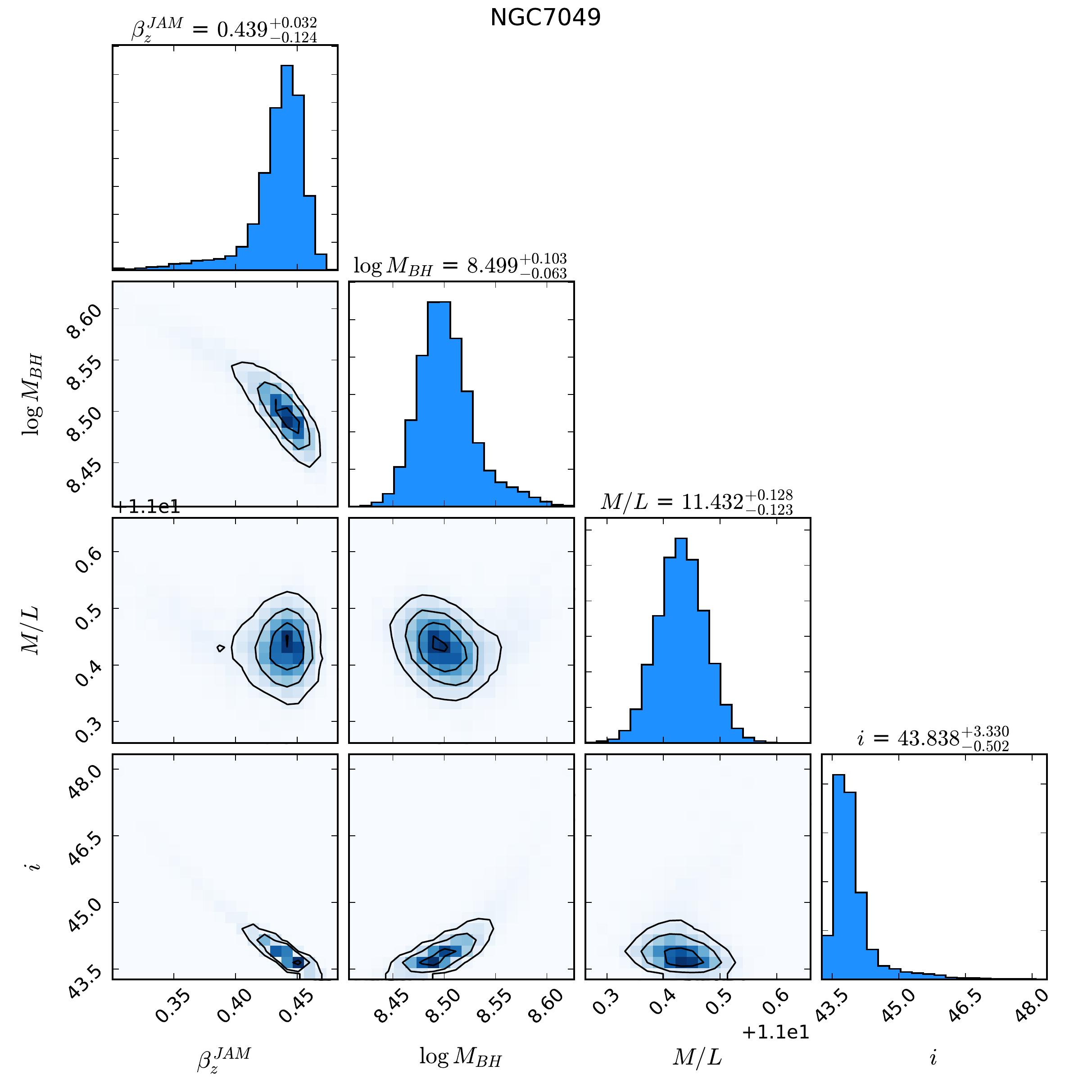}  

      \caption{MCMC posterior probability distribution of the JAM model parameters (M$_{\mathrm{BH}}$, $\beta$ and inclination) for each galaxy. The contour plots show the two-dimensional distributions for each parameter combination, the histograms show the projected one-dimensional distributions.                                            }
      \label{ff:mcmc_corerplot}
\end{figure*}

\section{Comparison of the Schwarzschild dynamical models with the symmetrised data}\label{ss:comparison_schwarz}

\begin{figure*}
  \centering
    \includegraphics[width=0.45\textwidth]{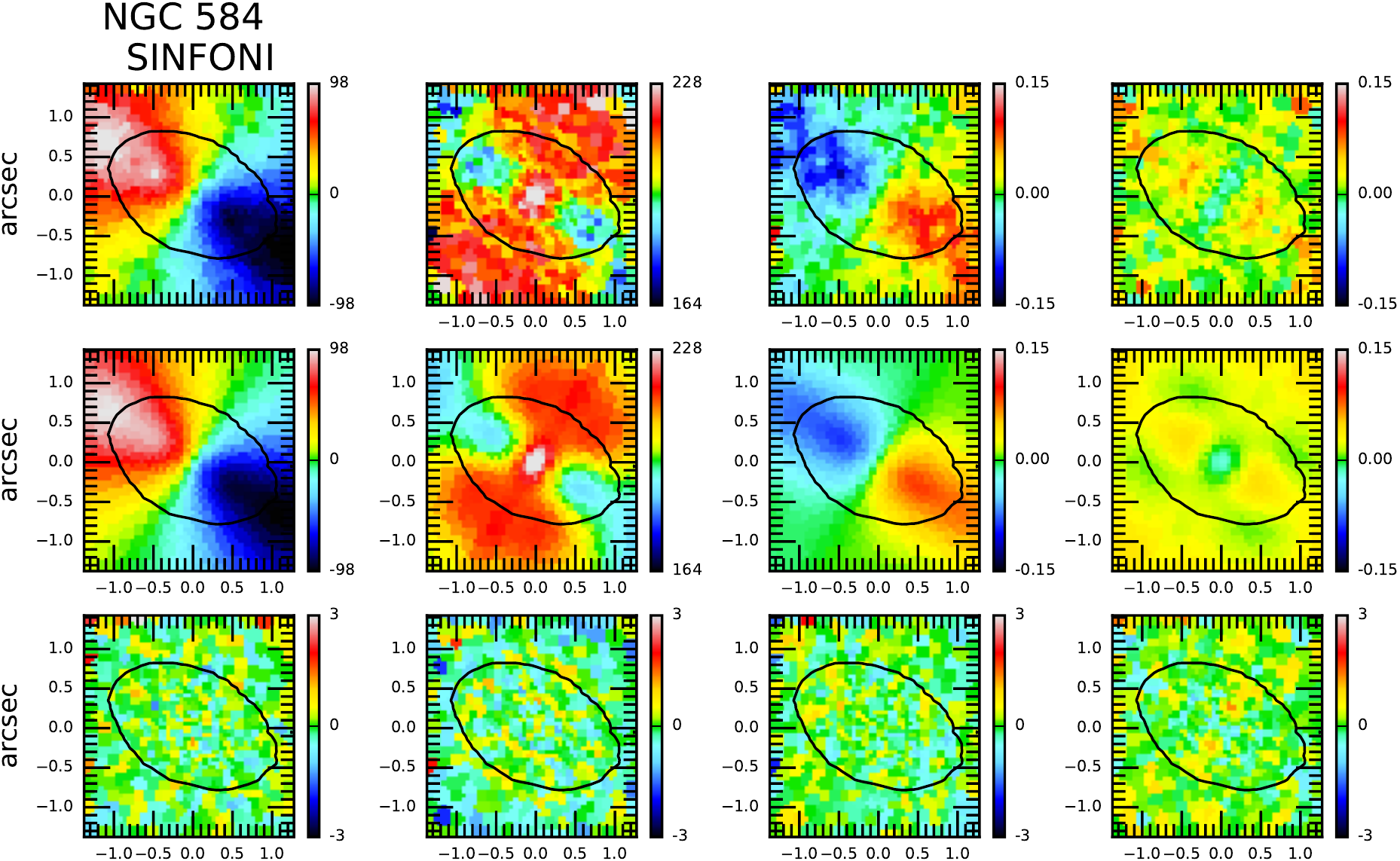}
    \includegraphics[width=0.45\textwidth]{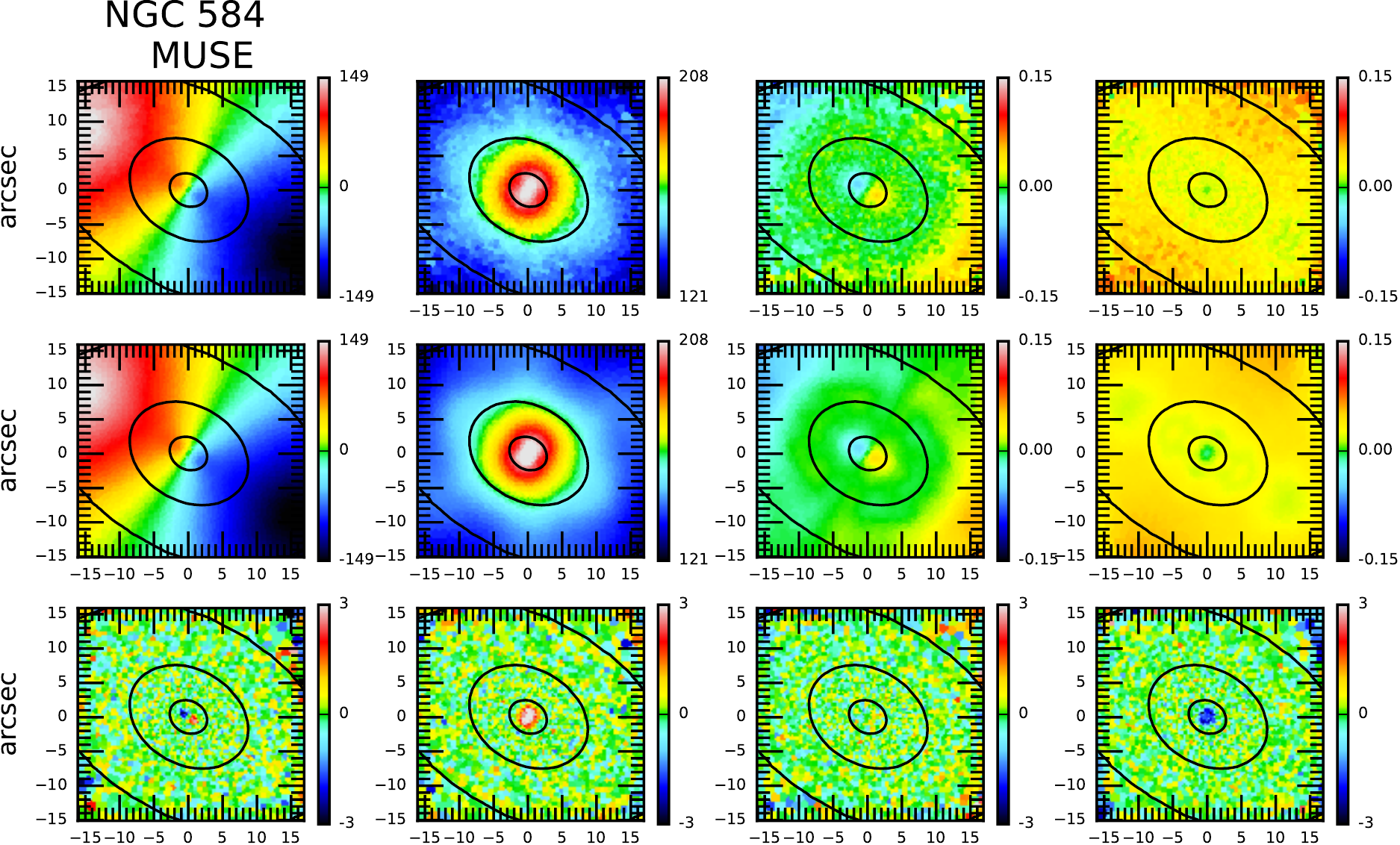}
    \includegraphics[width=0.45\textwidth]{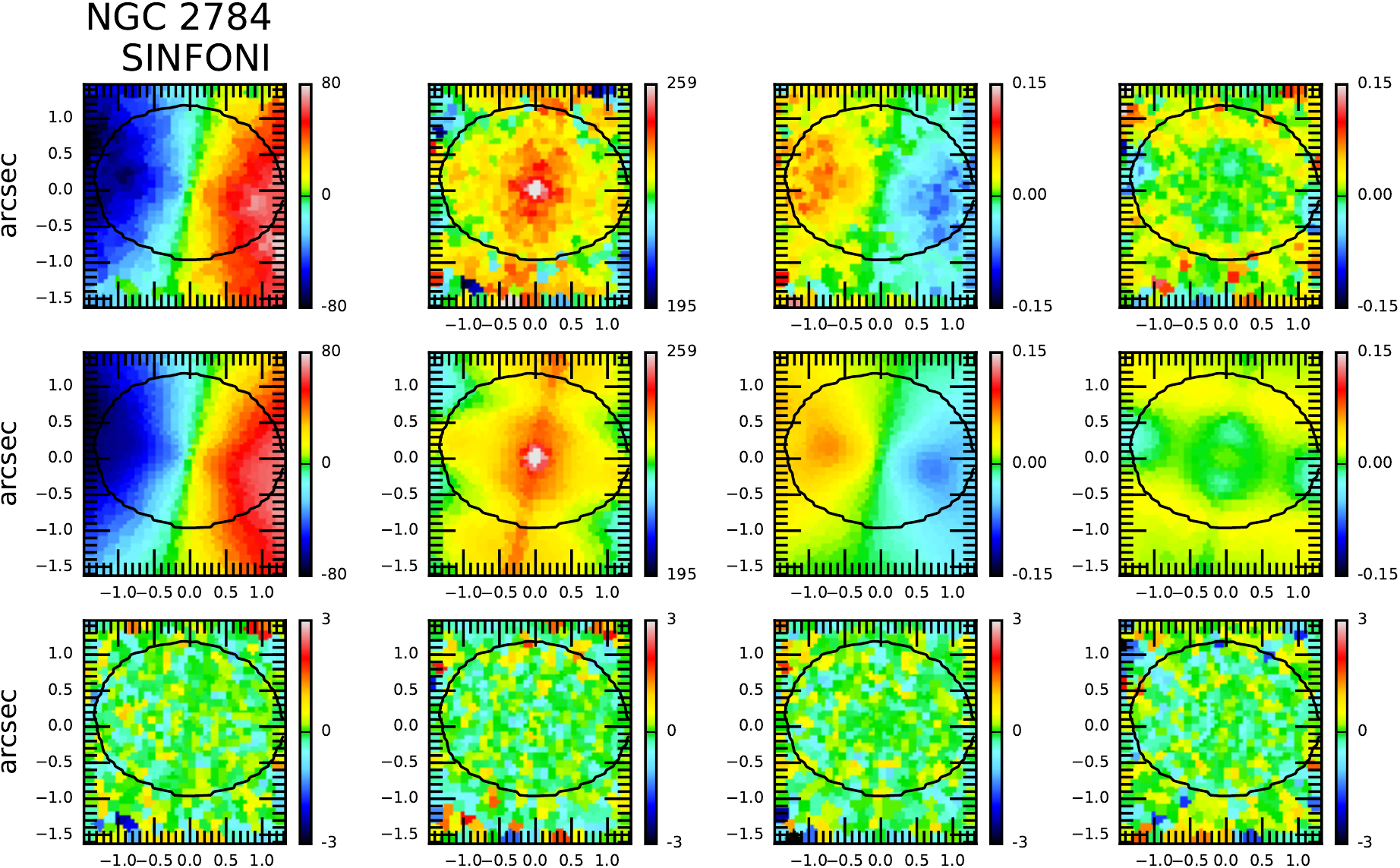}
    \includegraphics[width=0.45\textwidth]{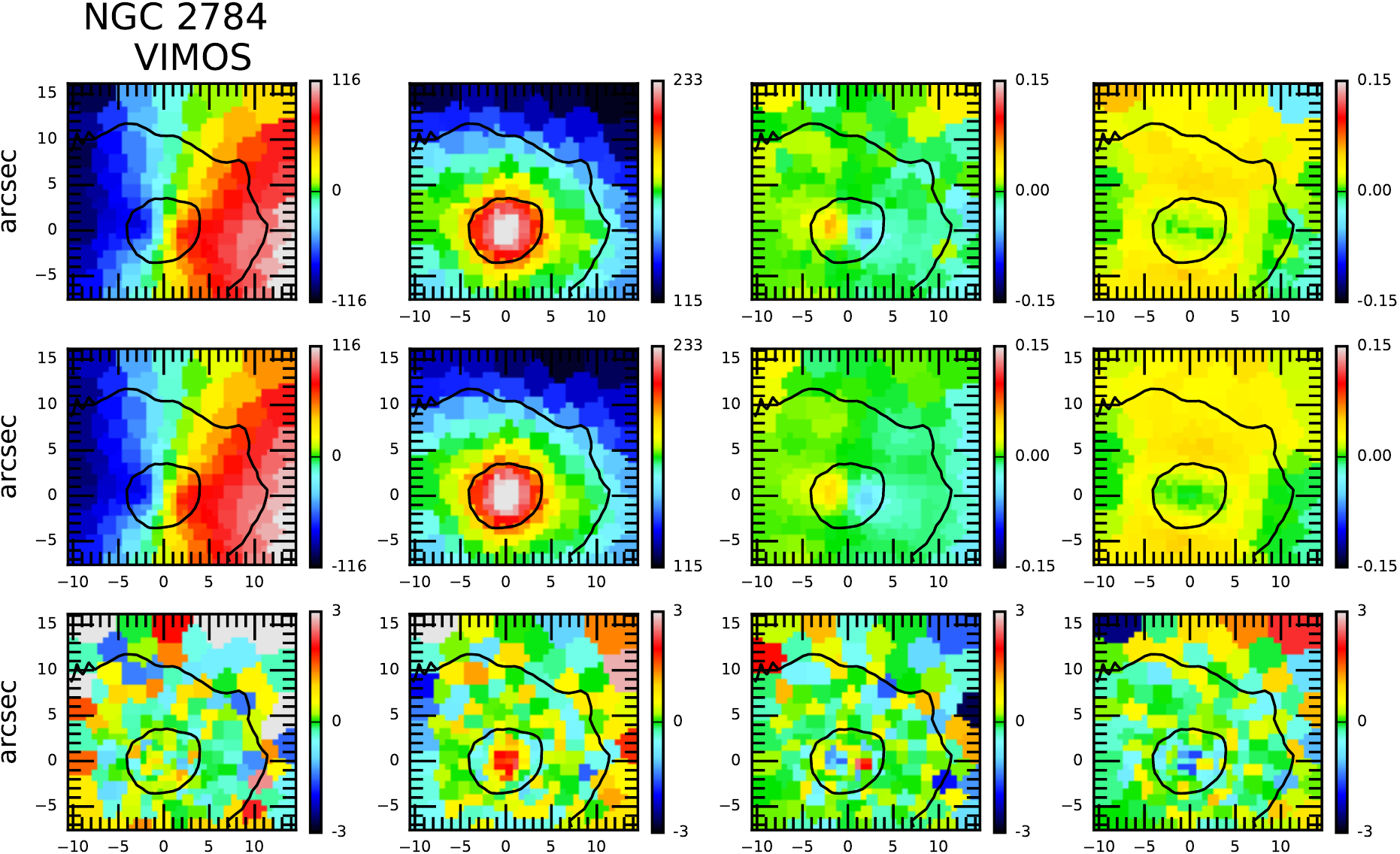}
    \includegraphics[width=0.45\textwidth]{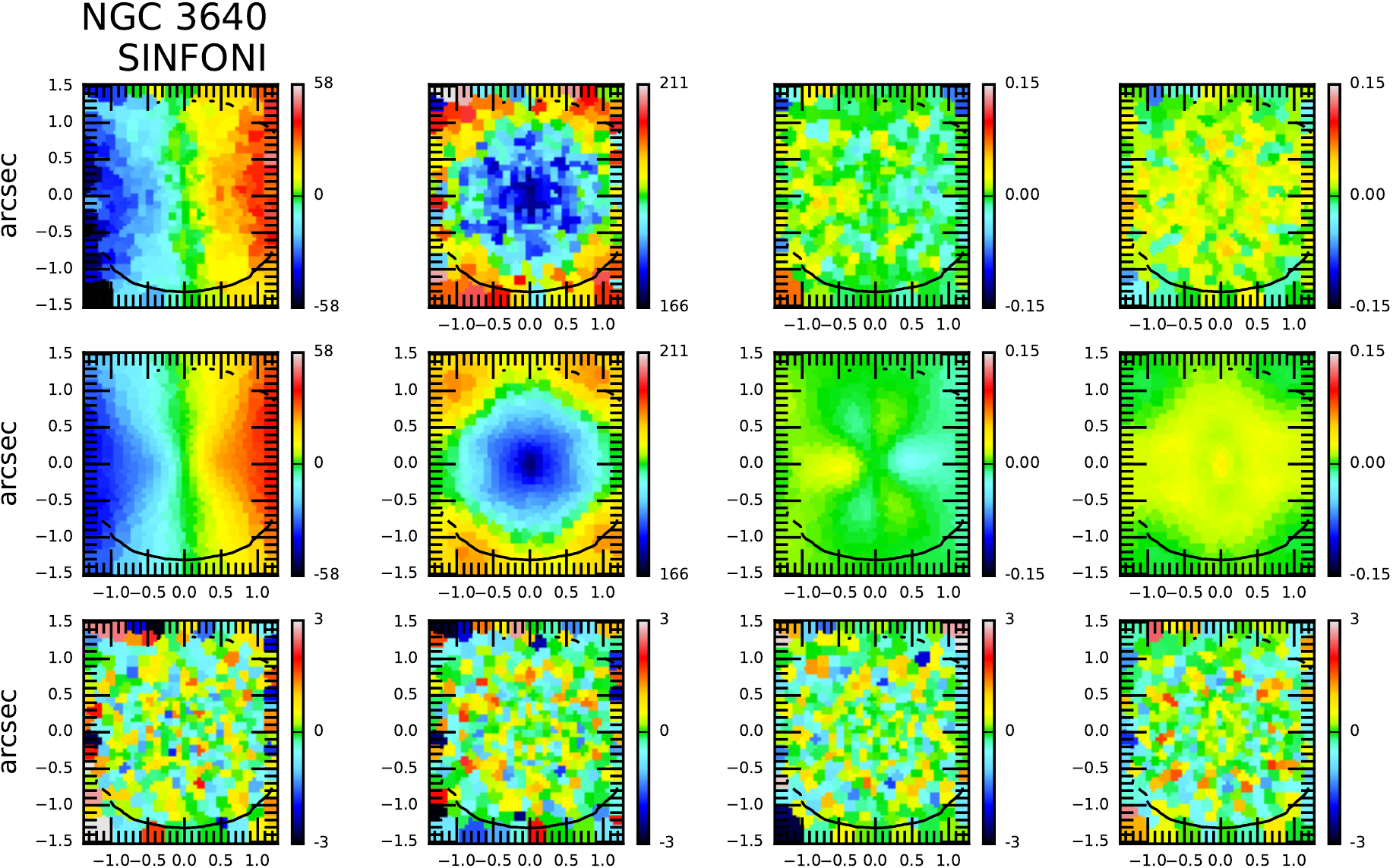}
    \includegraphics[width=0.45\textwidth]{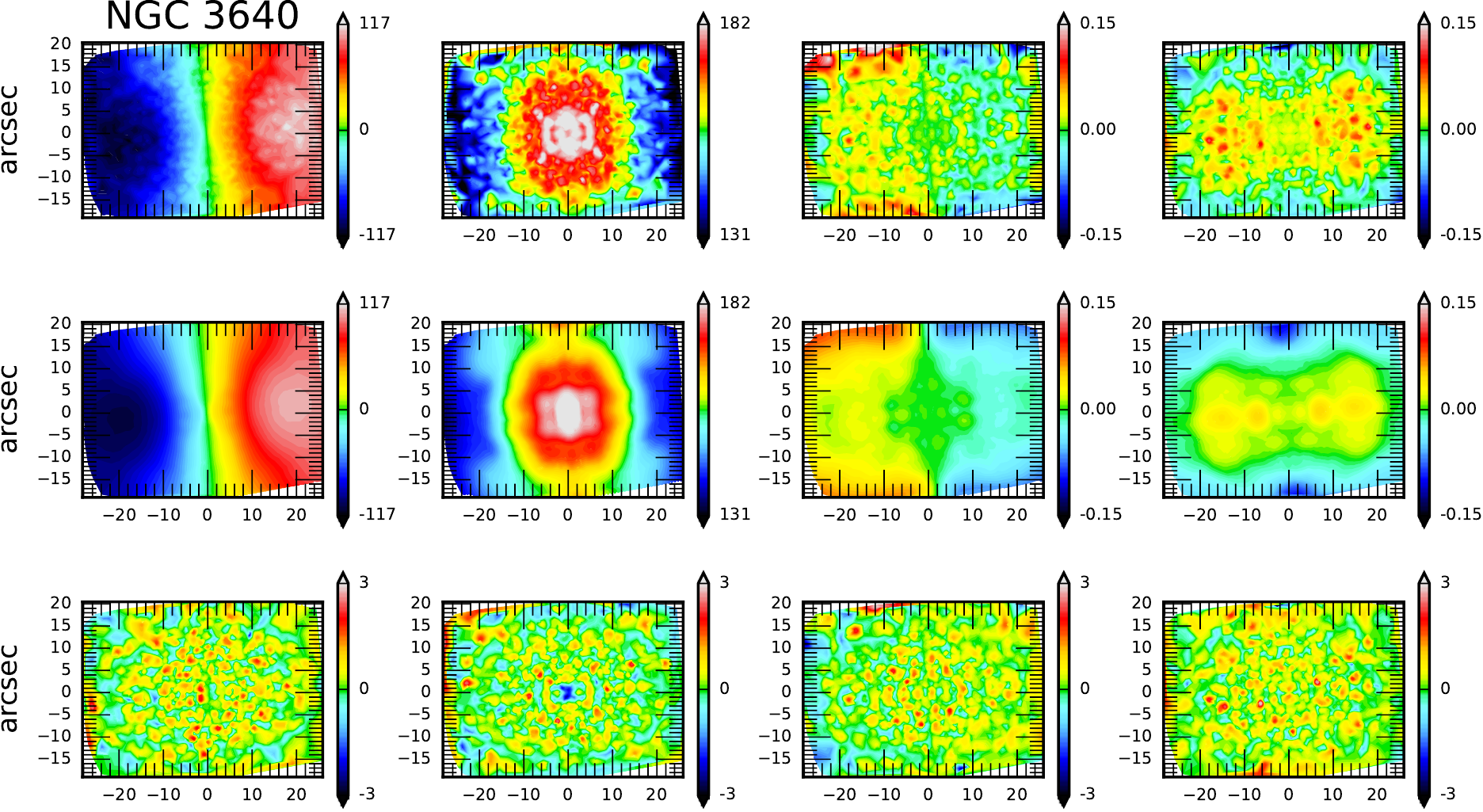}
      \caption{Comparison between symmetrized kinematic data and best-fitting Schwarzschild models for the galaxies NGC 584, NGC 2784 and NGC 3640. For each galaxy we show the SINFONI data on the left side and the large-scale data on the right side. The panels are ordered in the following: From left to right: Mean velocity, velocity dispersion, h3 and h4 Gauss-Hermite moments. From top to bottom: Symmetrized data, model for the best-fitting parameters from Table~\ref{t:results} and residual map defined as difference between the Schwarzschild model and observed kinematics divided by the observational
errors.}
      \label{ff:schwarzschild_comparison1}
\end{figure*}

\begin{figure*}
  \centering
    \includegraphics[width=0.45\textwidth]{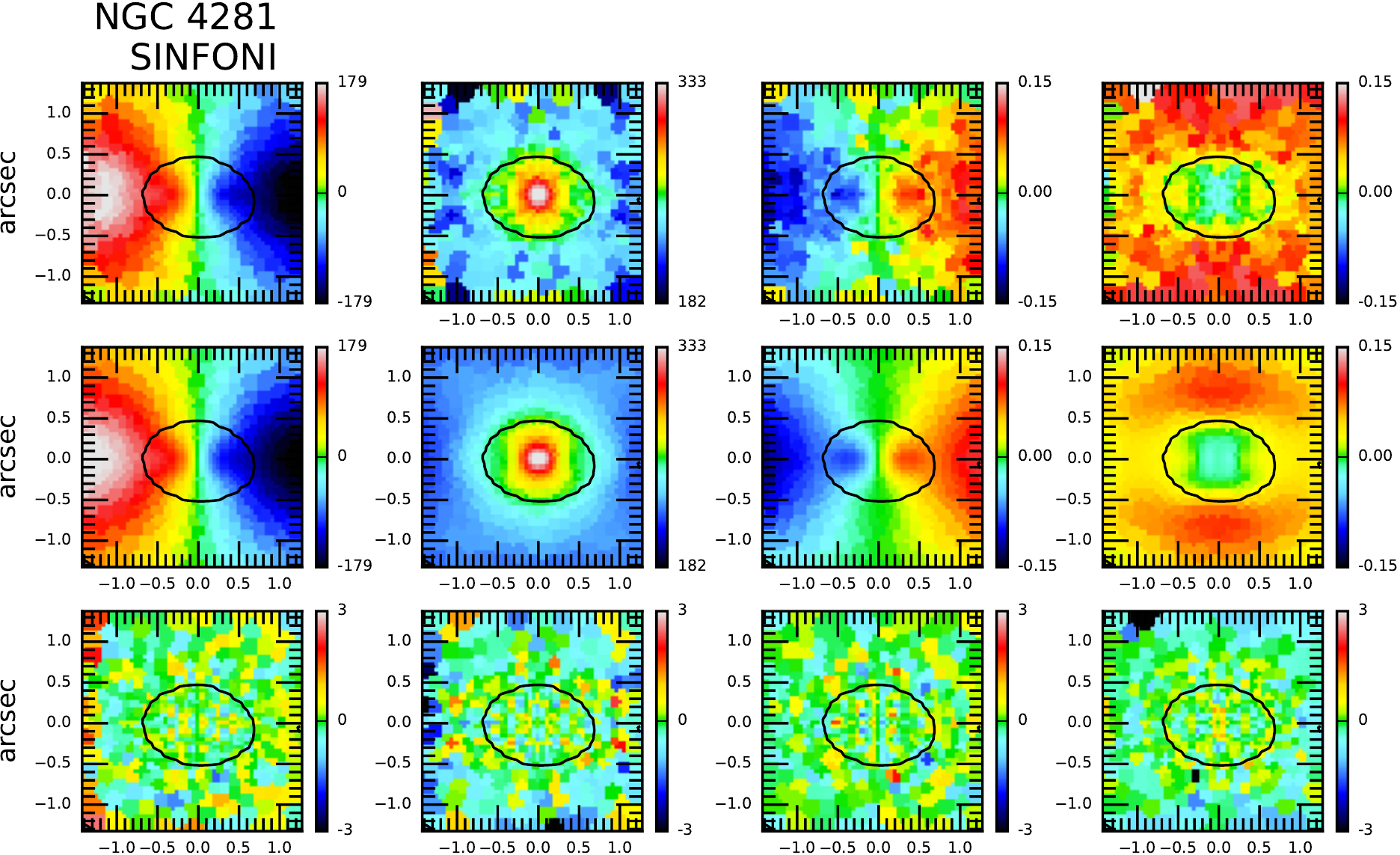}
    \includegraphics[width=0.45\textwidth]{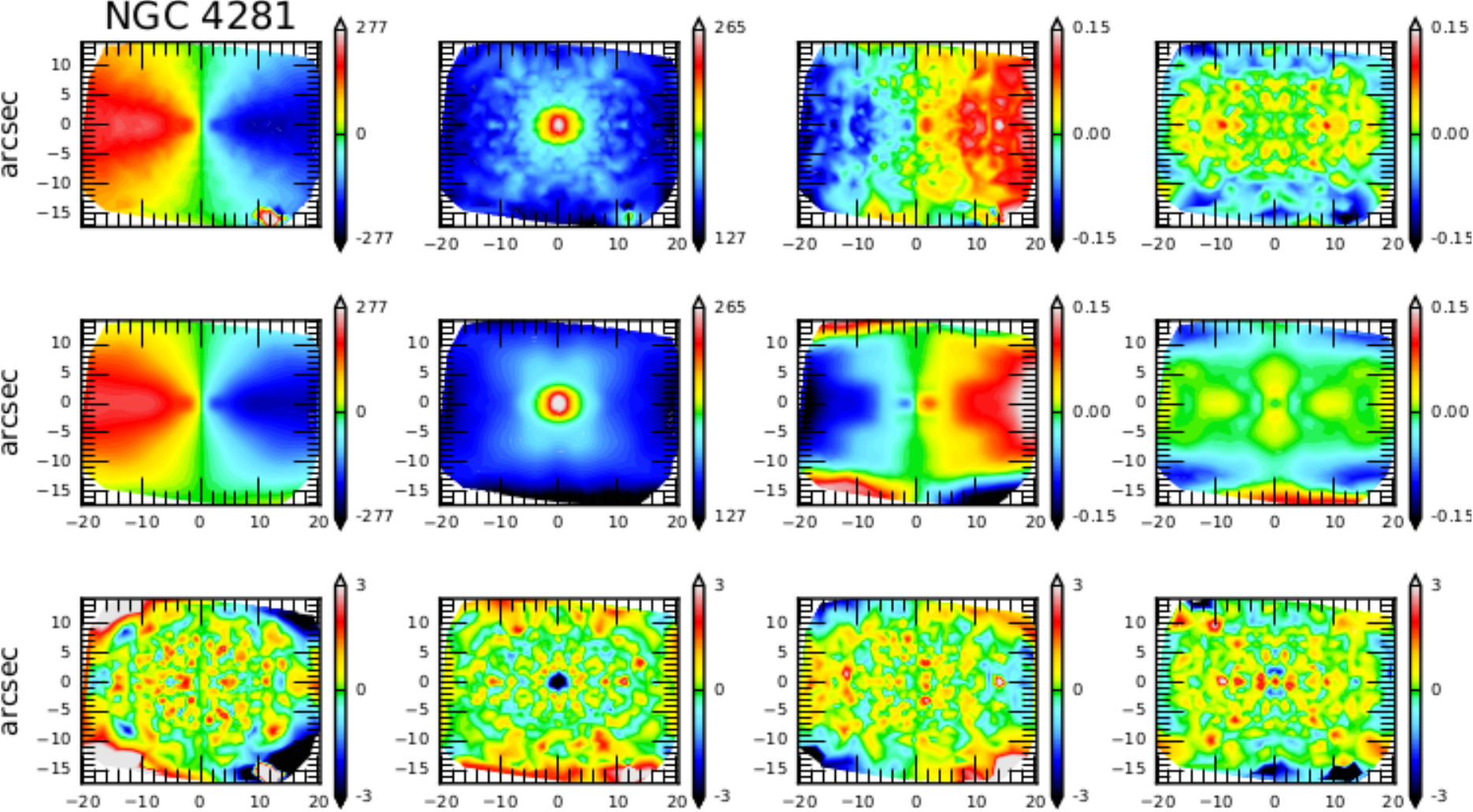}
    \includegraphics[width=0.45\textwidth]{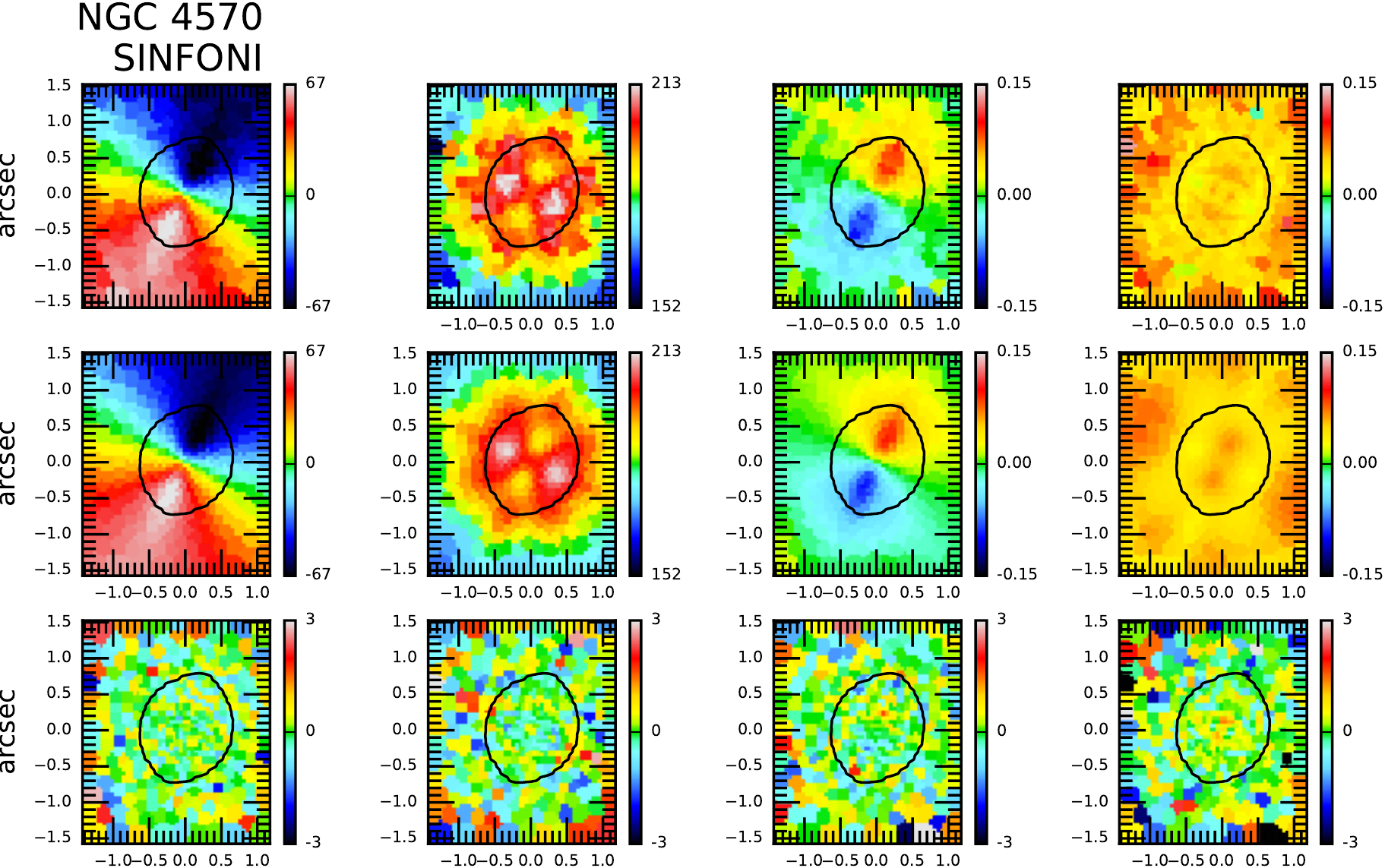}
    \includegraphics[width=0.45\textwidth]{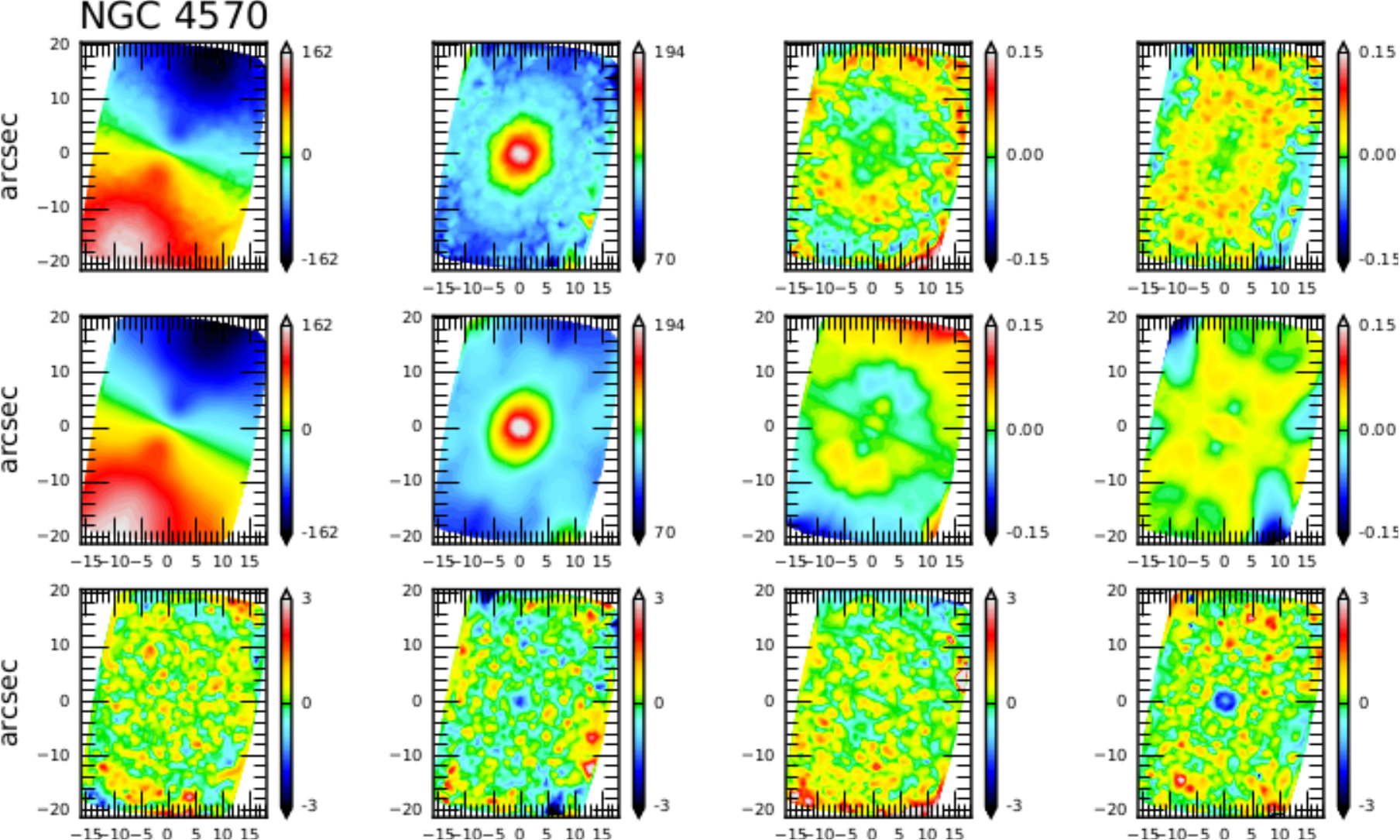}
    \includegraphics[width=0.45\textwidth]{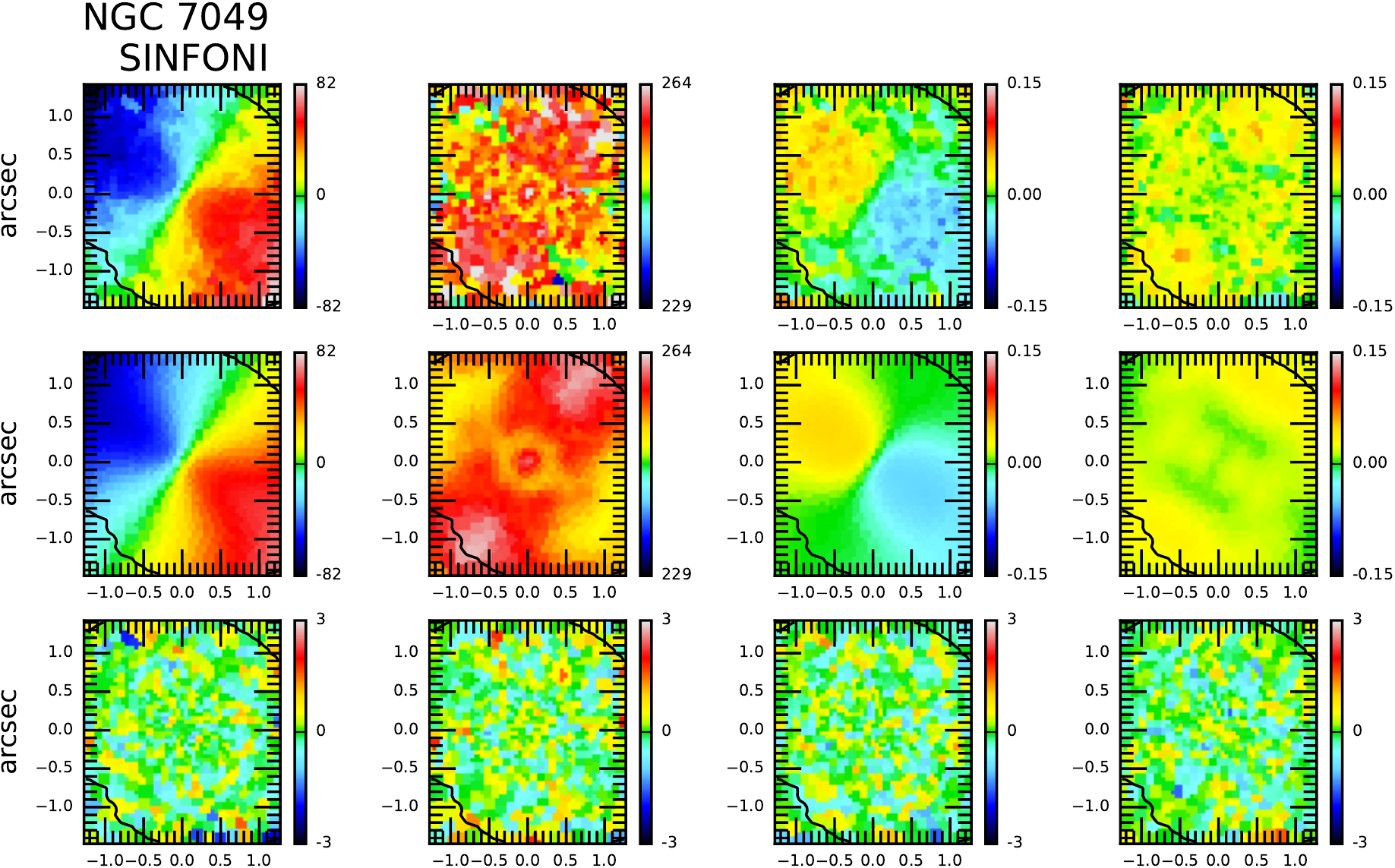}
    \includegraphics[width=0.45\textwidth]{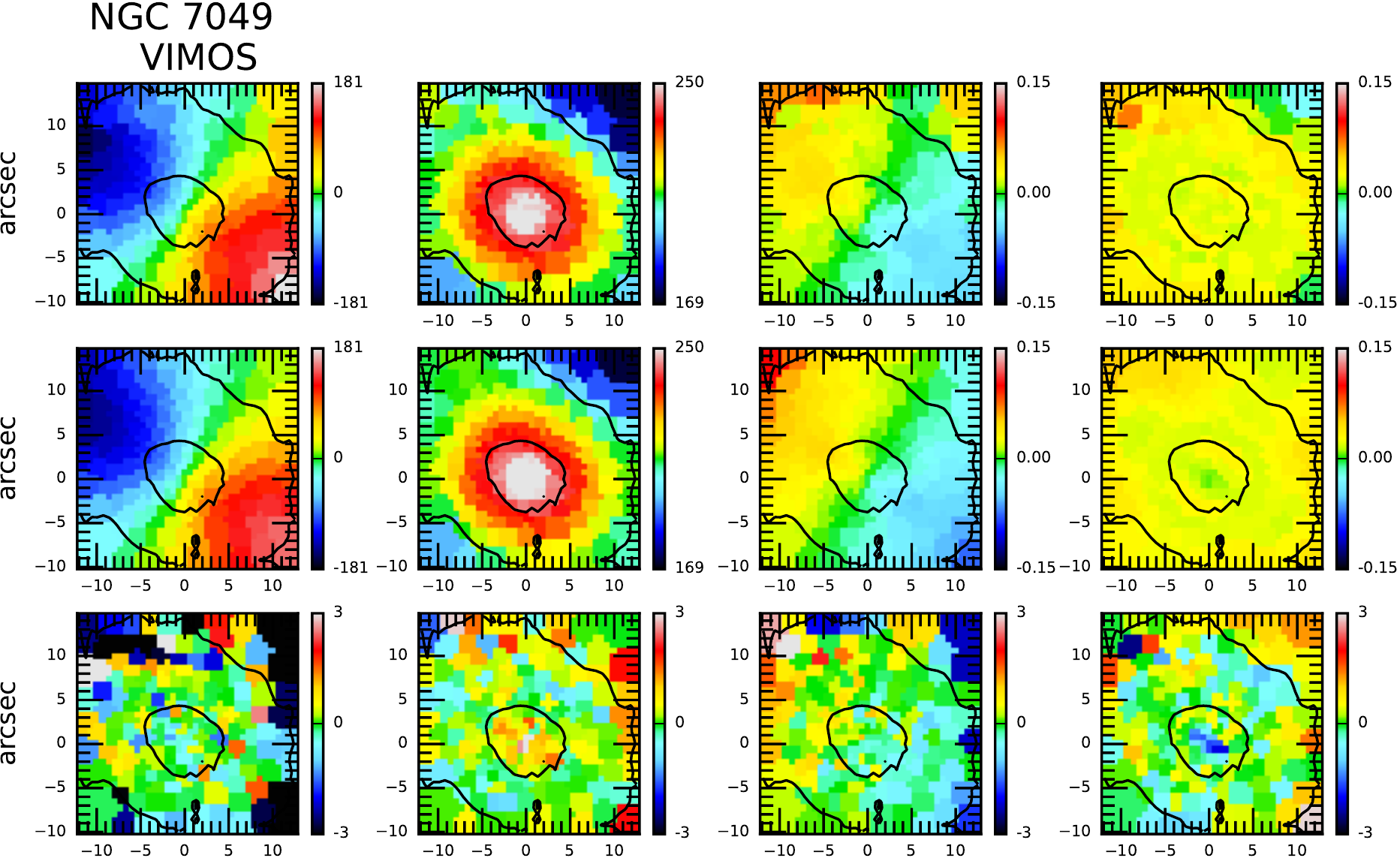}
      \caption{Comparison between symmetrized kinematic data and best-fitting Schwarzschild models for the galaxies NGC 4281, NGC 4570 and NGC 7049. For each galaxy we show the SINFONI data on the left side and the large-scale data on the right side. The panels are ordered in the following: From left to right: Mean velocity, velocity dispersion, h3 and h4 Gauss-Hermite moments. From top to bottom: Symmetrized data, model for the best-fitting parameters from Table~\ref{t:results} and residual map defined as difference between the Schwarzschild model and observed kinematics divided by the observational
errors.}
      \label{ff:schwarzschild_comparison2}
\end{figure*}

\end{appendix}

\end{document}